\documentclass[iop,apj]{emulateapj}
\slugcomment{Received 13-Aug-2021; revised 28-Oct-2022; accepted in ApJ 15-Nov-2022}
\shorttitle{Star Formation-Stellar Mass Relation of COSMOS from $0 < \lowercase{z} < 3.5$}
\shortauthors{Cooke et al. 2022}

\usepackage{natbib}
\usepackage{lineno}

\usepackage{graphicx} 
\usepackage{textcomp}
\usepackage{gensymb}
\usepackage{epsfig}%

\usepackage{verbatim}
\usepackage[outdir=./]{epstopdf}
\usepackage{indentfirst}
\usepackage[colorlinks=true,citecolor=blue,linkcolor=blue]{hyperref}
\usepackage{amssymb}
\usepackage{amsmath}
\usepackage{amsfonts}%
\usepackage{slantsc}
\usepackage{xspace}
\usepackage{fancyhdr}
\usepackage[utf8]{inputenc} 
\newcommand{\galapagosone}{{\scshape Galapagos}\xspace}
\newcommand{\galapagostwo}{{\scshape Galapagos-2}\xspace}

\begin{document}

\title{Morphology \& Environment's Role on the Star Formation Rate -- Stellar Mass Relation in COSMOS from $0 < \lowercase{\emph{z}} < 3.5$}

\author{Kevin C. Cooke$^{1,2,3}$, Jeyhan S. Kartaltepe$^{3}$, Caitlin Rose$^{3}$, K.D. Tyler$^{3}$, Behnam Darvish$^{4}$, Sarah K. Leslie$^{5}$, Ying-jie Peng$^{6}$, Boris H{\"a}u{\upshape{\ss}}ler$^{7}$, Anton M. Koekemoer$^{8}$}
\affil{$^1$AAAS S\&T Policy Fellow hosted at the National Science Foundation,\\ 1200 New York Ave, NW, Washington, DC, US 20005
\\
$^2$Department of Physics \& Astronomy, University of Kansas, Lawrence, KS 66045, USA\\
$^3$Laboratory for Multiwavelength Astrophysics, School of Physics and Astronomy, \\Rochester Institute of Technology, Rochester, NY 14623, USA\\
$^4$Cahill Center for Astrophysics, California Institute of Technology, 1216 East California Boulevard, Pasadena, CA 91125, USA\\
$^5$Leiden Observatory, Leiden University, PO Box 9513, NL-2300 RA Leiden, Netherlands\\
$^6$Kavli Institute for Astronomy and Astrophysics (KIAA), Peking University, Beijing 100871, China\\
$^7$European Southern Observatory, Alonso de Cordova 3107, Vitacura, Casilla 19001, Santiago, Chile\\
$^{8}$Space Telescope Science Institute, 3700 San Martin Dr., Baltimore, MD, 21218, USA}
\email{$^1$ astrokevincooke@gmail.com, $^2$ jeyhan@astro.rit.edu}

\begin{abstract}

We investigate the relationship between environment,  morphology, and the star formation rate -- stellar mass relation derived from a sample of star-forming galaxies (commonly referred to as the `\emph{star formation main sequence}') in the COSMOS field from $0 < z < 3.5$.  We constructed and fit the FUV--FIR SEDs of our stellar mass-selected sample of 111,537 galaxies with stellar and dust emission models using the public packages {\tt MAGPHYS} and {\tt SED3FIT}.  From the best fit parameter estimates, we construct the star formation rate -- stellar mass relation as a function of redshift, local environment, NUVrJ color diagnostics, and morphology.  We find that the shape of the main sequence derived from our color-color and sSFR-selected star forming galaxy population, including the turnover at high stellar mass, does not exhibit an environmental dependence at any redshift from $0 < z < 3.5$.  We investigate the role of morphology in the high mass end of the SFMS to determine whether bulge growth is driving the high mass turnover. We find that star-forming galaxies experience this turnover independent of bulge-to-total ratio, strengthening the case that the turnover is due to the disk component's specific star formation rate evolving with stellar mass rather than bulge growth.   
\end{abstract}



\section{Introduction}\label{sec:intro}
 Toward the goal of understanding the star formation history of today's galaxy population, many surveys have studied the tight correlation between the star formation rate (SFR) and stellar mass (M$_*$) of star-forming galaxies over cosmic time.  Such surveys have found a relatively low dispersion correlation known by many names, such as the star formation main sequence, the star-forming main sequence, the galaxy main sequence, and the star formation -- stellar mass relation \citep[e.g.,][]{Brinchmann:2004aa,Daddi:2007aa,Elbaz:2007aa,Noeske:2007aa,Salim:2007aa,Pannella:2009aa,Magdis:2010aa,Rodighiero:2010aa,Wuyts:2011aa,Whitaker:2012aa,Schreiber:2015aa,Leslie:2019aa}.  Due to its wide use, we will refer to this trend as the star formation main sequence (SFMS).  
	
The SFMS relation is an indication that the star-forming (SF) galaxy population spends a significant amount of its lifetime forming stars at a roughly steady state correlated with its stellar mass, and any deviations in star formation are short-lived until the galaxy is finally quenched, dropping off the SFMS.  Galaxies with SFRs significantly above the SFMS ($> 3 \times$ SFMS) are often classified as `starbursts', with intense star formation triggered through secular processes and/or mergers \citep[e.g.,][]{Kartaltepe:2012aa,Hung:2013aa,Willett:2015aa}.  Galaxies below the SFMS predominantly lie in a cloud of `passive galaxies' exhiting SFRs an order of magnitude or more below the SFMS \citep[e.g.,][]{Morselli:2019aa}.  Between the SFMS and the quiescent population, in a region dubbed the `green valley',  lie an intermediate group of galaxies experiencing a temporary lull or the in-progress quenching of their star formation \citep{Strateva:2001aa,Baldry:2004aa,Baldry:2006aa}, or are experiencing rejuvination \citep{Chauke:2019aa, Mancini:2019aa}.  With rising redshift comes a rising normalization \citep{Noeske:2007aa,Whitaker:2012aa,Rodighiero:2014aa,Johnston:2015aa,Lee:2015aa}, reaching its maximum normalization at cosmic noon \citep[$z\sim2$; ][]{Madau:2014aa}.  This rising normalization is consistent with the higher gas masses of galaxies in the early universe across several decades in stellar mass \citep{Magdis:2012aa,Genzel:2015aa,Schinnerer:2016aa,Scoville:2016aa,Liu:2019aa}.

The shape of the SFMS at $z > 1$ is roughly consistent with low redshift samples \citep[e.g.,][]{Whitaker:2012aa,Lee:2015aa}, however several questions remain at the high mass end.  At stellar masses $>$10$^{10}$ M$_{\odot}$ there is a debated turnover in the shape of the SFMS, where progressively higher masses no longer correspond to higher SFRs.  Recent work found the degree of the turnover is reduced when only considering disk-dominated, or the disk components of, SF galaxies \citep{Abramson:2014aa,Guo:2015aa}.  This was interpreted as an indication that the turnover at high mass is due to the growth of the quiescent bulge component and the star-forming disk components form stars at a consistent level independent of bulge mass.  However, this behavior is contested by studies that find the turnover remains when considering the SFR of disks \citep{Schreiber:2016aa,Catalan-Torrecilla:2017aa,Morselli:2017aa,Belfiore:2018aa,Cook:2020aa}.  Other studies have alternatively interpreted the turnover as a reduction in the star formation per unit stellar mass (specific star formation rate; sSFR) \citep[][]{Whitaker:2012aa,Whitaker:2014aa,Lee:2015aa,Tomczak:2016aa,Lee:2018aa}. Further complicating matters is recent work examining this behavior on an individual star-forming region basis for local galaxies, where ALMA observations find an incredible galaxy to galaxy variation in star formation efficiency at the highest star-forming clump stellar masses \citep{Ellison:2021aa}.  The disagreement between results which observe a turnover in star-forming galaxies' disk components \citep[e.g.,][]{Belfiore:2018aa,Cook:2020aa}, and those that do not \citep[e.g.,]{Abramson:2014aa,Guo:2015aa}, motivate us to determine how the SFMS may change with respect to different morphology-selected samples.

While the role of morphology in the high mass turnover remains under investigation, there are additional questions regarding the role of local environment (i.e., the galaxy number density in the immediate vicinity of a given galaxy) on the shape of the SFMS.  At low redshift, there is a morphology-density relation where quiescent, spheroidal galaxies more commonly inhabit dense environments while star-forming, disk-like galaxies more commonly inhabit the field \citep[e.g.,][]{Dressler:1980aa,Lin:2014aa,Erfanianfar_2015}.  This environmental dependence is thought to be due to the removal of cold gas in satellite galaxies after interacting with the warm intra-cluster medium of the host cluster \citep{Gunn:1972aa}.  As the SFMS is estimated using star-forming galaxies, the shape of the SFMS is dominated by star-forming, disk field galaxies at low redshift.  Measuring the shape of the SFMS across different environments is required to understand whether environment plays a role in the global turnover at high mass.
	
The role of local environment on the shape of the star-formation main sequence becomes uncertain at high redshift (z $>$ 1).  Initially, an inversion of the relation between SFR and local density was observed in at $z \sim 1$ \citep{Elbaz:2007aa,Cooper:2008aa,Hilton:2010aa,Santos:2014aa}, where higher density environments hosted increased star formation rates or galaxies with higher total infrared luminosity. The inversion was hypothesized to be due to the higher gas fractions present in early universe galaxies, enabling elevated levels of star formation in interaction prone environments such as groups and protoclusters.  However, some works have observed a flattening, but not quite an inversion, at $z > 1$ \citep{Feruglio:2010aa,Koyama:2013aa,Scoville:2013aa,Erfanianfar_2015,Alberts:2016aa,Hatfield:2017aa}.  The contrast between the observation of elevated star formation rates in high density regions, such as clusters, by some infrared studies \citep[e.g.,][]{Santos:2014aa} and not others \citep[e.g.,][]{Alberts:2016aa} leads to the question of what trends can be deduced by a study that examines local density surrounding star-forming galaxies across large cosmological volumes.  Addressing this issue as well requires an understand of the shape of the SFMS as a function of environment.

In this paper, we seek to estimate the role of morphology and local environment of star-forming galaxies on the SFMS using consistent methods across a larger span of redshift than examined in previous works, which focused on either low or high-z samples.  Previous works on tracing the SFMS across redshift \citep[e.g., $z < 1.3$,][]{Lee:2015aa} were often limited to redshift ranges constrained by a given observation method's constraints, such as the redshifting out of the optical peak from individual filters.  Additionally, some low-to-high redshift surveys were performed but did not include comparisons to both local environment and morphology \citep[e.g., $0.2<z<2.5$ \& $0.2<z<6$ \& $0<z<9$,][]{Whitaker:2014aa,Pearson:2018aa, Thorne:2021aa}.  Through the use of the k-corrected COSMOS2015 observations across the ultraviolet to far-infrared, we investigate beyond the common stopping points near $z\sim1$ to provide a point of comparison for high-z ($z<3$) studies \citep{Whitaker:2014aa,Schreiber:2015aa} in a self-consistent manner while considering how star-formation correlates with local environment or morphology.

In Section \ref{sec:inst} we describe our sample selection and the photometric and spectroscopic data set we use as input to the SED fitting packages described in Section \ref{sec:methods}.  Our star formation -- stellar mass relations are described in Section \ref{sec:paper4results}  and we consider their implications in Section \ref{sec:disc}.  Finally, we review our findings in Section \ref{sec:conc}. This work assumes a \citet{Chabrier:2003aa} initial mass function (IMF) and the standard cold dark matter ($\Lambda$CDM) cosmological parameters of  H$_{0}$ = 70 Mpc$^{-1}$ km s$^{-1}$, $\Omega_M$ = 0.3, and $\Omega_{vac}$ = 0.7.  All magnitudes are expressed in the AB magnitude system \citep{Oke:1974aa}. 


\section{Sample and Multiwavelength Data}\label{sec:inst}
We select our sample and corresponding photometric data from the public observations of the COSMOS field \citep{Scoville:2007aa}.  COSMOS is the largest contiguous area (2 sq.~deg.) observed with the \emph{Hubble Space Telescope} (HST), located at R.A.~(J2000) = 10:00:28.600, Dec.~(J2000) = +02:12:21.00.  We use the far-ultraviolet (FUV) to far-infrared (FIR) data included in the COSMOS2015 catalog \citep{Laigle:2016aa} to construct spectral energy distributions (SEDs) and estimate galaxy stellar parameters for all galaxies above the stellar mass completeness limit from from $0 < z < 3.5$. We review the photometric data sources and properties below in Section \ref{sec:photometry}; for greater detail, see \citet{Laigle:2016aa}.
We apply galactic foreground corrections to photometry from the FUV-IRAC4 bands using a galactic reddening of R$_v$ = 3.1 \citep{Morrissey:2007aa} and \emph{E(\bv)} values from the dust maps of \citet{Schlegel:1998aa}.


\subsection{Sample Selection}\label{sec:paper4sample}
\begin{figure*}
\begin{center}
\begin{tabular}{lr}
\includegraphics[width=85mm]{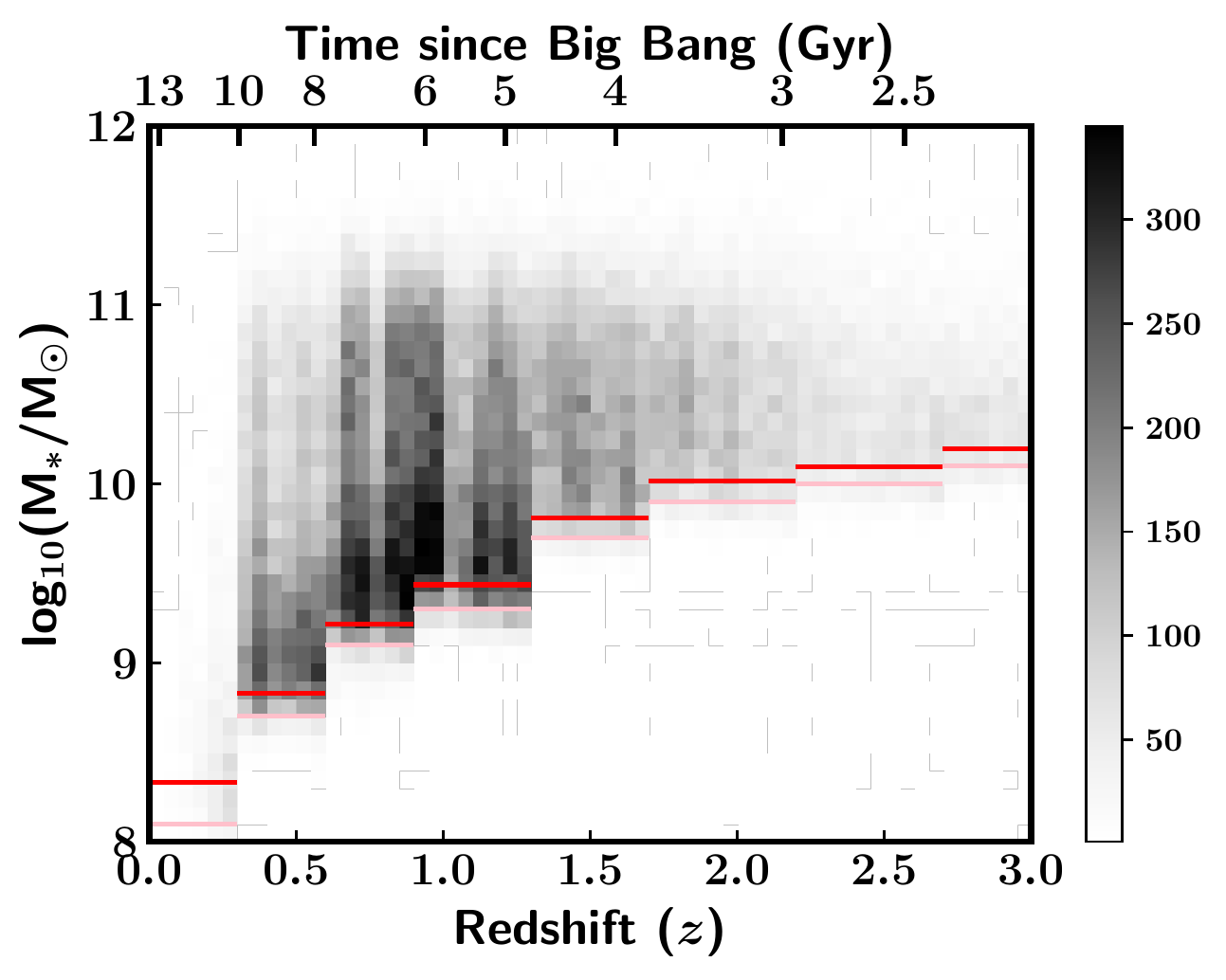} &   \includegraphics[width=90mm]{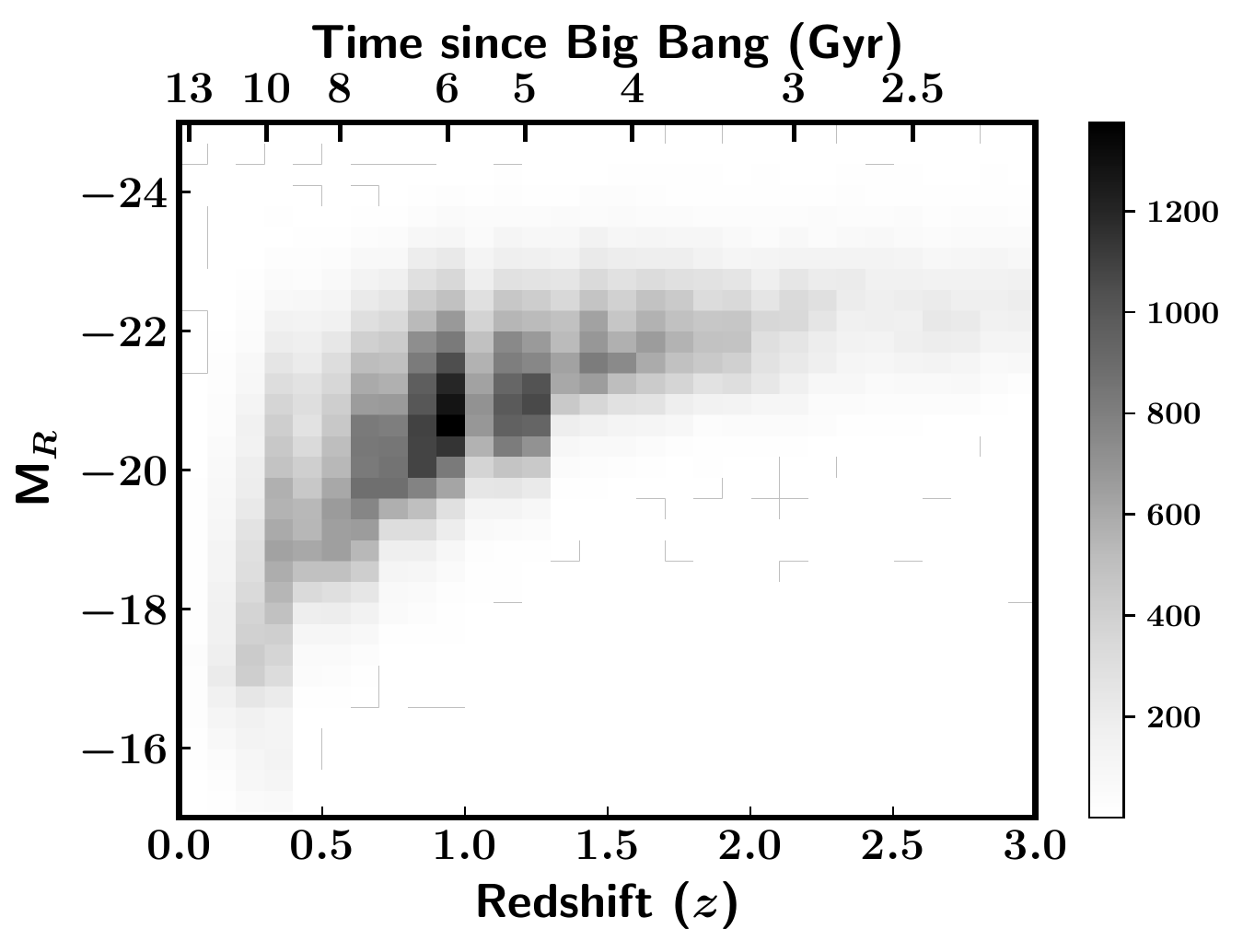}
\end{tabular}
\end{center}
\caption{Left:  The stellar mass distribution of our refit sample.  The original \citet{Laigle:2016aa} stellar mass completeness limits are in pink, and the completeness limits plus median stellar mass error are in red.  Our analysis sample is selected from the completeness plus fitting error limit.  Right:  To enable reference to an observable quantity in parallel to SED-estimated stellar masses, we include the k-corrected M$_{r}$ distribution of our sample with redshift.\label{fig:selection}}
\end{figure*}

For our analysis, we require a mass complete sample across a wide redshift range. To ensure sample completeness across the full redshift range examined here, we select galaxies using the total stellar mass completeness limits from COSMOS2015 \citep{Laigle:2016aa}.  We plot our initial sample in Fig.~\ref{fig:selection} with these stellar mass completeness limits. The \citet{Laigle:2016aa} stellar mass limit was determined using the UltraVISTA DR-2 survey \citep{McCracken:2012aa} depths for the entire COSMOS field.  To maintain sample completeness across the entire COSMOS field, we chose to use the shallower depth of UltraVISTA DR-2 rather than the depth of the UltraVISTA ultra-deep strips of the center of the COSMOS field.  We select galaxies from $0.0 < z < 3.5$ to probe both sides of the global star formation density peak of the universe \citep{Madau:2014aa}.  Photometric redshifts from \citet{Laigle:2016aa} were calculated using {\tt LePhare} \citep{Arnouts:1999aa,Ilbert:2006aa}, yielding a photometric redshift accuracy of 1$\sigma\sim$ 0.01 at $z<3$. 

Our initial sample is only selected based on the COSMOS2015 stellar mass, with no consideration given to previously estimated SFRs, colors, or active galactic nucleus (AGN) activity.  Our initial mass-selected sample from COSMOS2015 totalled 131,842 galaxies. We independently fit the SEDs of our sources because the COSMOS 2015 catalog's original fit estimates presented in \citet{Laigle:2016aa} prioritized photometric redshift and stellar mass determination by using only 12 \citet{Bruzual:2003aa} templates for star-forming galaxies.  To estimate SFRs with high accuracy, refitting with a higher number of models is desired for all objects.  Additionally, the \citet{Laigle:2016aa} fits only included photometry out to the 8$\micron$ band of Spitzer/IRAC Channel 4 as input.  This wavelength range restricts the original fit's capability to estimate obscured star-formation rates, and the resulting fits are primarily used for stellar mass selection of targets.  To estimate SFRs inclusive of obscured and unobscured components, a fitting procedure with both MIR and FIR points should be used.

As described in Section~\ref{sec:sedfitting}, we refit the SEDs of these targets to estimate accurate star formation rates using the full range of IR data available and incorporating spectroscopic redshifts where available. To maintain completeness of the sample, we use SED fits (described in Section~\ref{sec:sedfitting}) from galaxies with best-fit stellar masses above the Laigle completeness mass limit plus the median error of the best-fit SED stellar mass in a given redshift bin.  This removes the parameter space where fits may scatter above or below the original completeness limit due to the re-fitting process. The final mass-selected sample used in our analysis totals 111,537 galaxies.


\subsection{Spectroscopic Redshifts}
Redshift quality plays a great importance in the constraining of stellar models for galaxies at all redshifts.  After identifying our sample described in Section \ref{sec:inst}, we search the COSMOS spectroscopic catalog (Salvato et al., in prep) for matches within 1$''$ of the COSMOS2015 RA and Dec. positions.  Prior to SED fitting, we replace the photometric redshifts provided in the COSMOS2015 catalog with the matched spectroscopic redshifts from the following surveys and observation campaigns: 3D-HST Survey \citep{Brammer:2012aa,Momcheva:2016aa}, DEIMOS 10K Spectroscopic Survey \citep{Hasinger:2018aa}, 2dF Galaxy Redshift Survey \citep{Colless:2001aa}, FMOS-COSMOS Survey \citep{Kartaltepe:2015aa,Silverman:2015aa}, the Gemini GMOS-S spectra of \citet{Balogh:2011aa}, COSMOS Active Galactic Nucleus Spectroscopic Survey \citep{Trump:2007aa,Trump:2009aa}, the KMOS$^{3D}$ Survey \citep{Wisnioski:2015aa}, the AzTEC Millimeter Survey of the COSMOS Field \citep{Scott:2008aa}, the 14$^{th}$ Data Release of the Sloan Digital Sky Survey  \citep[SDSS;][]{Paris:2018aa}, the VLT LEGA-C Spectroscopic Survey \citep{Wel:2016aa}, the ZFIRE KECK/MOSFIRE Spectroscopic Survey \citep{Nanayakkara:2016aa}, the MOSFIRE Deep Evolution Field (MOSDEF) Survey \citep{Kriek:2015aa}, the Keck LRIS spectra of \citet{Casey:2017aa}, MOIRCS Deep Survey \citep{Yoshikawa:2010aa}, the MMT/Hectoscec spectra of \citet{Trump:2009ab}, the Complete Calibration of the Color-Redshift Relation (C3R2) Survey \citep{Masters:2017aa}, the Subaru FOCAS observations of \citet{Trump:2011aa}, the Keck MOSFIRE spectra of \citet{Trakhtenbrot:2016aa}, SCUBA-2 spectroscopic redshifts of \citet{Casey:2013aa}, the Keck DEIMOS spectra of \citet{Kartaltepe:2010aa}, \citet{Capak:2011aa}, and \citet{Mobasher:2016aa}, the SINFONI spectra of \citet{Perna:2015aa}, the zCOSMOS Survey \citep{Lilly:2007aa}, VIMOS Ultra-Deep Survey \citep{Le-Fevre:2015aa}, the HST/WFC3 grism spectra of \citet{Krogager:2014aa}, and the \emph{Spitzer} Infrared Spectrograph spectra of \citet{Fu:2010aa}. Our sample includes 20,959 galaxies with spectroscopic redshifts, a spectroscopic redshift fraction of $\sim$17\%. If a target is observed by more than one program listed above, we select the spectroscopic redshift from the COSMOS catalog with the highest quality flag.  These flags identify the quality of a spectroscopic redshift on the basis of the quality of the original observation, e.g., spectrograph resolution and the number of lines identified during redshift measurement.

\subsection{FUV-MIR Photometry}\label{sec:photometry}

To accurately model the emission from ongoing and unobscured star formation, we include PSF-fit photometric magnitudes from the far-UV (FUV) and near-UV (NUV) bands of the \emph{Galaxy Evolution Explorer} \citep[GALEX;][]{Martin:2005aa}, observed down to a limiting magnitude of m$_{AB} \sim 26$.  For details on the original data reduction and PSF-fitting, see \citet{Zamojski:2007aa}

We also include Canada--France--Hawaii Telescope (CFHT) MegaPrime \citep{Aune:2003aa,Boulade:2003aa} $u^{*}$ observations.  CFHT $u$-band observations reach a depth of m$_{AB} \sim 26.4$ with a seeing of 0.9$''$ \citep{Capak:2007aa}.
		 
Across the optical continuum, we include Subaru/Suprime-Cam observations from five broad filters (B, V, R, $i+$, $z++$) and 11 medium filters (IA427, IA464, IA484, IA505, IA527, IA574, IA624, IA679, IA738, IA767, and IA827), with a $3\sigma$ depth of at least m$_{AB}\sim25.2$.  The maximum PSF FWHM of this filter selection is 1.89$''$ \citep{Taniguchi:2007aa,Taniguchi:2015aa}.  
		
For the near-infrared (NIR) emission sampling the old stellar population, we use Vista/VIRCAM \citep{Sutherland:2015aa} J, H, and K-band observations from the UltraVISTA-DR2 survey \citep{McCracken:2012aa}. The UltraVIST survey observed J, H, and K bands down to $m_{AB} =$ 24.7, 24.3, and 24.0, respectively.
		
Near- to mid-infrared (MIR) observations of the COSMOS field were taken by the SPLASH \citep{Steinhardt:2014aa} and S-COSMOS surveys \citep{Sanders:2007aa} using the \emph{Spitzer Space Telescope} \citep[\textit{Spitzer};][]{Werner:2004aa}
		
We include photometry from \textit{Spitzer}'s Infrared Array Camera (IRAC)'s  3.6, 4.5, 5.7, and 7.9 $\mu$m\; bands \citep[for more information, see][]{Fazio:2004aa}, which have PSF widths of 1\farcs6, 1\farcs6, 1\farcs8, and 1\farcs9 respectively, and are sensitive down to a 3$\sigma$ depth of $m_{AB}$ of 25.5, 25.5, 23.0, and 22.9, respectively.  To bridge the NIR to the FIR and constrain the MIR continuum and polycyclic aromatic hydrocarbon (PAH) features, we include 24 $\mu$m emission observations that were observed down to a 5$\sigma$ depth of 71 $\mu$Jy.

\subsection{Herschel Observations}
The most important wavelength regime in the estimation of star formation in the gas-rich early universe is the FIR, which helps provide an estimate of the obscured star-forming components of a galaxy.  The COSMOS field was imaged using the \emph{Herschel Space Observatory} \cite[\emph{Herschel};][]{Pilbratt:2010aa} Photoconductor Array Camera and Spectrometer \citep[PACS; ][]{Poglitsch:2010aa} 100 and 160 $\mu$m bands and the Spectral and Photometric Imaging Receiver \citep[SPIRE;][]{Griffin:2010aa} using its 250, 350, and 500 $\mu$m bands.  These observations were performed as part of the PACS Evolutionary Probe \citep[PEP:][]{Lutz:2011aa} that observed down to a 3$\sigma$ depth of 5 and 10.2 mJy for 100 and 160 $\mu$m bands, respectively.  In addition to PEP, COSMOS was observed as part of the Herschel Multi-tiered Extragalactic Survey \citep[HerMES:][]{Oliver:2012aa} using SPIRE 250, 350, and 500 $\mu$m bands down to a 3$\sigma$ depth of 8.1, 10.7, 15.4 mJy, respectively.  
		
Unlike the resolved observations in the optical, the FIR observations from $Herschel$ are subject to blending due to the wide beam width (18.1$''$ at 250 $\mu$m, 24.9$''$ at 350 $\mu$m, 36.6$''$ at 500 $\mu$m).   \citet{Lee:2010aa} deblended the sources observed in PEP and HerMES in order of ascending filter FWHM, beginning with Spitzer IRAC observations as priors. Blending is most intense in the SPIRE FIR bands, as they have the widest FWHM of our photometric data set.  We de-select FIR photometry from blended targets using the \citet{Elbaz:2011aa} clean index to ensure that only confidently deblended measurements are included.  These criteria require no neighbors of comparable ($S_{Neighbor}$/$S_{Target} >$ 0.5) flux within 1.1$\times$ the FWHM in 100, 160, 250, 350, and 500 $\mu$m.  For the Herschel 250 and 350 $\micron$ bands, 9095 and 4589 members respectively of our initial mass-selected sample have confident non-blended detections.  Only 1796 targets have non-blended 500 $\mu$m photometry and our results regarding the star formation behavior of the overall sample are consistent within errors when the deblended 500 $\mu$m observations are included or excluded in the SED fitting process.

FIR sources from all bands represent 10\% of our total sample, spread roughly homogeneously from $0.3 < z < 2.0$, with $\sim$300 sources per 0.1 $\Delta$z increment.  This density declines from z $\sim$2 to $\sim$70 per 0.1 $\Delta$z increment by $z\sim3$.  The decline results in the global median SFR becoming more dominated by the rest-frame UV and blue optical emission of the sample at high redshift.  Our median SFR results for the bins, discussed in Section~\ref{sec:sfmsfit}, remain consistent within errors on the median if these FIR detections are excluded. We also find that FIR sources occupy equal density per local environment bin, in the high- and low-density bins (described in Section \ref{sec:mapmatch}) used later in our analysis.  Therefore, we do not observe a bias toward dense or field environments due to the inclusion of the FIR detected sources.


\section{Methods}\label{sec:methods}


\subsection{SED Fitting}\label{sec:sedfitting}

\begin{figure*}
\begin{center}
\includegraphics[width=0.9\textwidth]{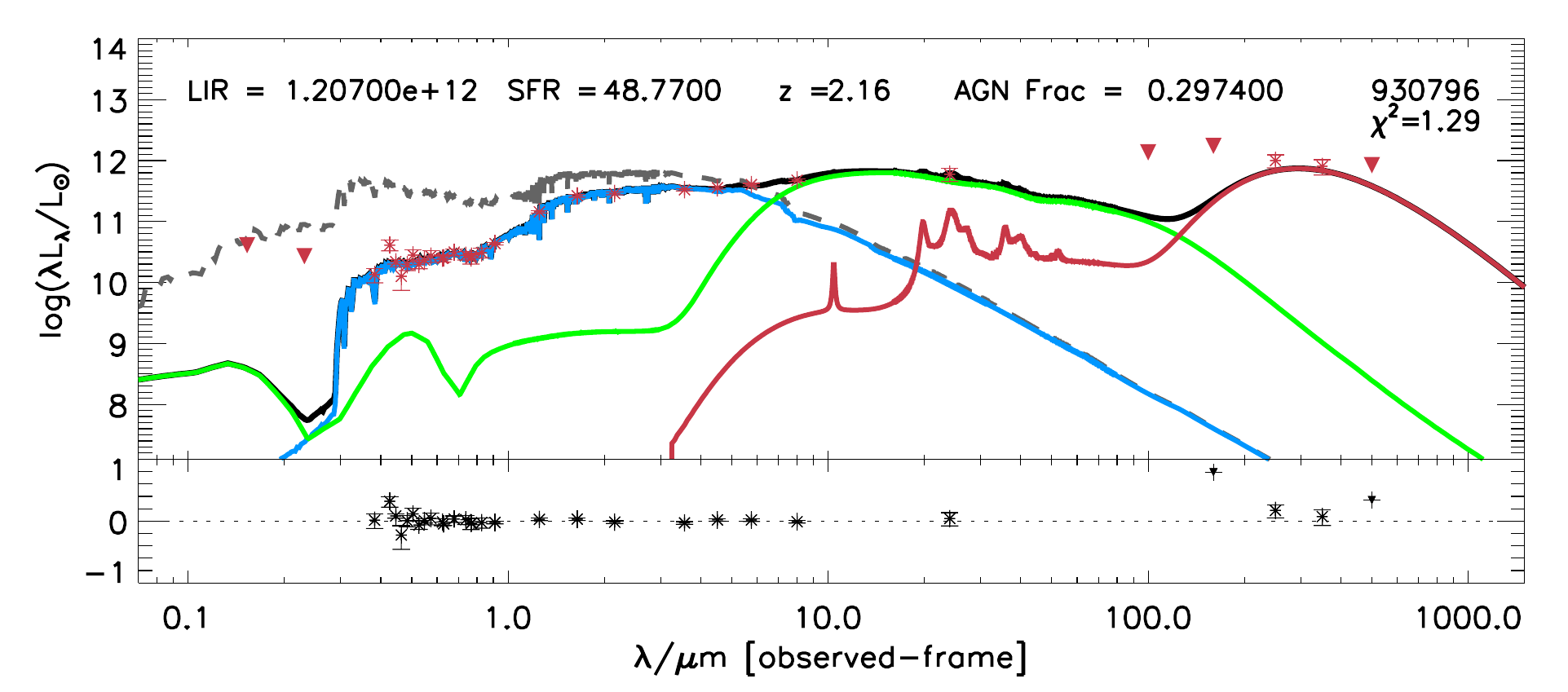}
\end{center}
\caption
{ \label{fig:SEDexample} Example SED from a target requiring the use of SED3FIT to fit its MIR emission.  The best-ft total model (black) is plotted with the unobscured stellar model (dashed grey), obscured stellar model (blue), AGN torus emission (green), and dust model (red). Photometry used in the fit are plotted as red asterisks, with red triangles used in the case of an upper limits in that band.  The lower plot shows the fractional residuals between the best-fit total model and the input photometry.}
\end{figure*}

We estimate the stellar properties of our sample using two SED fitting packages, {\tt MAGPHYS} \citep{da-Cunha:2008aa} for the general population and {\tt SED3FIT} \citep{Berta:2013aa} for {\bf strong} AGN.  These packages were chosen due to their capability to manage energy balance, and their previous successful use with COSMOS2015 data \citep{Cooke:2019aa}. {\tt MAGPHYS} uses photometric measurements and errors from 0.0912 $\mu$m \;$\lesssim \lambda \lesssim$ 1000 $\mu$m and redshift as inputs.  The package fits the data to a library of 50,000 stellar population models \citep{Bruzual:2003aa} and 50,000 dust models \citep{da-Cunha:2008aa} in an energy-balanced manner.  {\tt MAGPHYS} considers the energy radiated from the final stellar model that has been absorbed by the dusty component of the target galaxy and uses this as a prior during the infrared fitting (8--1000 $\mu$m), ensuring energy balance between the two components.  The library models are calculated using an exponentially declining parameterization to model the star-formation history, with starbursts included as constant star-forming episodes \citep{da-Cunha:2008aa}.  For a discussion on the differences between fitting code estimated parameters given a sample population, please see Pacifici et al. (2022, in prep.).    Metallicity models are uniformly included from 0.02-2 times the solar value.  This method produces a marginalized likelihood distribution of each physical parameter (e.g., SFR, stellar mass) by comparing the total FUV-FIR SED with the distribution of models and corresponding solutions in the template library.  Additional details on the computation of the model library can be found in \citet{da-Cunha:2008aa}.  Each target is fit once using this process, yielding SFR and stellar mass estimates with median errors of 13\% and 19\% the estimated value, respectively.

As the public version of {\tt MAGPHYS} does not include an AGN component, we use the AGN-capable {\tt SED3FIT} (Fig.~\ref{fig:SEDexample}) to supplement the results produced by {\tt MAGPHYS}.  {\tt SED3FIT} is a descendent of the {\tt MAGPHYS} package, using the same stellar and dust models but independently changed to include an AGN library.  This is necessary for dominant AGN cases, where the strong UV and/or IR emission from the accretion disk and obscuring torus can cause overestimation of SFRs.  From our original mass-selected sample described in Section \ref{sec:inst}, we identify AGN candidates for refitting with {\tt SED3FIT} using the 8 $\mu$m residuals of the original {\tt MAGPHYS} fit.  If the observed 8 $\mu$m emission is greater than 1.4$\times$ the modeled 8 $\mu$m emission, then the MIR is considered poorly constrained and the source is refit with an AGN component \citep[][Tyler et al. in prep.]{Cooke:2019aa}.  This cutoff identifies galaxies with poorly-fit infrared SED slopes, yielding erroneous SFR and stellar mass estimates. This refitting step is performed rather than fitting the entire sample with SED3FIT because SED3FIT is computationally intensive on large samples and has the inability to determine when an AGN is required at very low AGN fractions. The inclusion of a weak AGN contributing $~$10\% of the bolometric luminosity is easily included in SED3FIT if the fit is degenerate between a weak AGN and no AGN.   In the same manner as {\tt MAGPHYS}, {\tt SED3FIT} \citep{Berta:2013aa} first fits a stellar component and uses this fit as a prior to the infrared component.  However the difference lies in the simultaneous fitting of an AGN model during these steps using the AGN torus and accretion disk models of \citet{Feltre:2012aa}.  Energy balance is considered between the stellar and AGN components when fitting the dust emission.  For more details on this procedure, and the AGN templates, see Tyler et al. in prep. and \citet{Cooke:2019aa}. The number of 8 $\mu$m-selected AGN in each redshift bin is included in Table \ref{tab:sample}, where we find that these 8 $\mu$m-selected AGN candidates are a small fraction ($<1\%$) of the overall sample and are predominantly classified as star-forming galaxies based on their NUVrJ colors (Section \ref{sec:colors}).  The median SFR of an AGN fit by SED3FIT is 40\% below the estimated MAGPHYS value, and the median AGN stellar mass is 13\% lower than the non-AGN MAGPHYS value.  However, due to the rarity of AGN identified using 8 $\mu$m residuals in our sample, their inclusion or exclusion in our SFMS estimates do not move our SFMS trends outside their errors.

\begin{deluxetable*}{lcccccccr}
\tabletypesize{\footnotesize}
\tablecolumns{10}
\tablewidth{\textwidth} 
 \tablecaption{Star-Forming and Quiescent Selected Samples
 \label{tab:sample}}
 \tablehead{\colhead{$z$} & \colhead{log$_{10}$(M$^*_{Limit}$)} & \colhead{Total}& \colhead{SF} & \colhead{Quiescent} & \colhead{Total 8 $\mu$m-bright AGN}& \colhead{SF 8 $\mu$m-bright AGN} & \colhead{Quiescent 8 $\mu$m-bright AGN} } 
 \startdata
 0.0--0.3 & 8.3 & 5925 & 4881 (82.4\%) & 1044 (17.6\%) & 12 & 5 (41.6\%) & 7 (58.3\%) &\\ 
 0.3--0.6 & 8.8 & 17827 & 14087 (79.0\%) & 3740 (21.0\%) & 33 & 28 (84.8\%) & 5 (15.2\%) &\\ 
 0.6--0.9 & 9.2 & 24079 & 18122 (75.2\%) & 5957 (24.8\%) & 106 & 75 (70.7\%) & 31 (29.3\%) &\\ 
 0.9--1.3 & 9.4 & 27805 & 21310 (76.6\%) & 6495 (23.4\%) & 103 & 79 (76.7\%) & 24 (23.3\%) &\\ 
 1.3--1.7 & 9.8 & 15131 & 10483 (69.3\%) & 4648 (30.7\%) & 57 & 33 (57.9\%) & 24 (42.1\%) &\\ 
 1.7--2.2 & 10.0 & 11055 & 8298 (75.1\%) & 2757 (24.9\%) & 89 & 63 (70.8\%) & 26 (29.2\%) &\\ 
 2.2--2.7 & 10.1 & 5215 & 3261 (62.5\%) & 1954 (37.5\%) & 63 & 48 (76.2\%) & 15 (23.8\%) &\\
 2.7--3.5 & 10.2 & 4500 & 3102 (68.9\%) & 1398 (31.1\%) & 23 & 17 (73.9\%) & 6 (26.1\%) &
 \enddata
 \tablecomments{Galaxies selected from the COSMOS2015 catalog above the stellar mass completion limit.  We identify star-forming and quiescent galaxies using NUVrJ rest frame colors \citep{Ilbert:2013aa} and a sSFR cut of 10$^{-11}$(1+z)$^{2.5}$ yr$^{-1}$ \citep{Ilbert:2010aa,Dominguez-Sanchez:2011aa}, and identify 8$\mu$m-bright AGN candidates using 8 $\mu$m fitting residuals from {\tt MAGPHYS}.  These AGN do not represent all AGN types in our sample, only the IR bright sources which require an AGN component to accurately estimate SFRs from their IR luminosity. We chose a NUVrJ selection criteria to be sensitive to $<$Gyr timescale star-forming episodes \citep{Arnouts:2007aa,Salim:2007aa}. We do not find a significant difference in the distribution of quiescent and star--forming galaxies between the MAGPHYS and SED3FIT samples. Note that the star-forming and quiescent fractions are computed over our entire stellar mass range at that redshift.  The difference between star-forming and quiescent fractions here and the distribution shown in Fig.~\ref{fig:SFMSNUVRJ} is due to the sSFR cut used in this table which identifies some blue galaxies as quiescent due to their low sSFR ($< 10^{-11}$(1+z)$^{2.5}$ yr$^{-1}$).  These may represent a subpopulation of galaxies which have just shut down their star formation.}
\end{deluxetable*}

\begin{figure*}
\begin{center}
\begin{tabular}{c}
\includegraphics[width=0.95\textwidth]{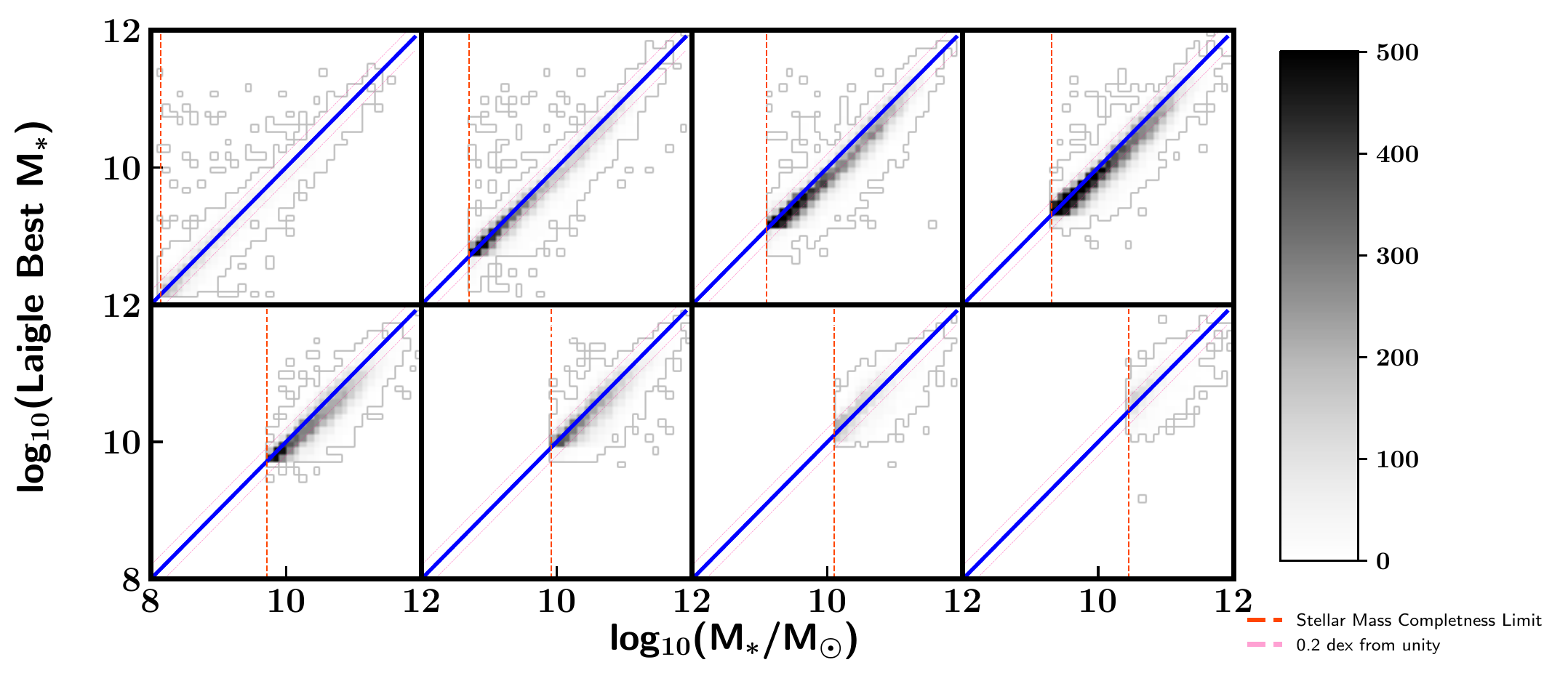}  
\end{tabular}
\begin{tabular}{c}\includegraphics[width=0.95\textwidth]{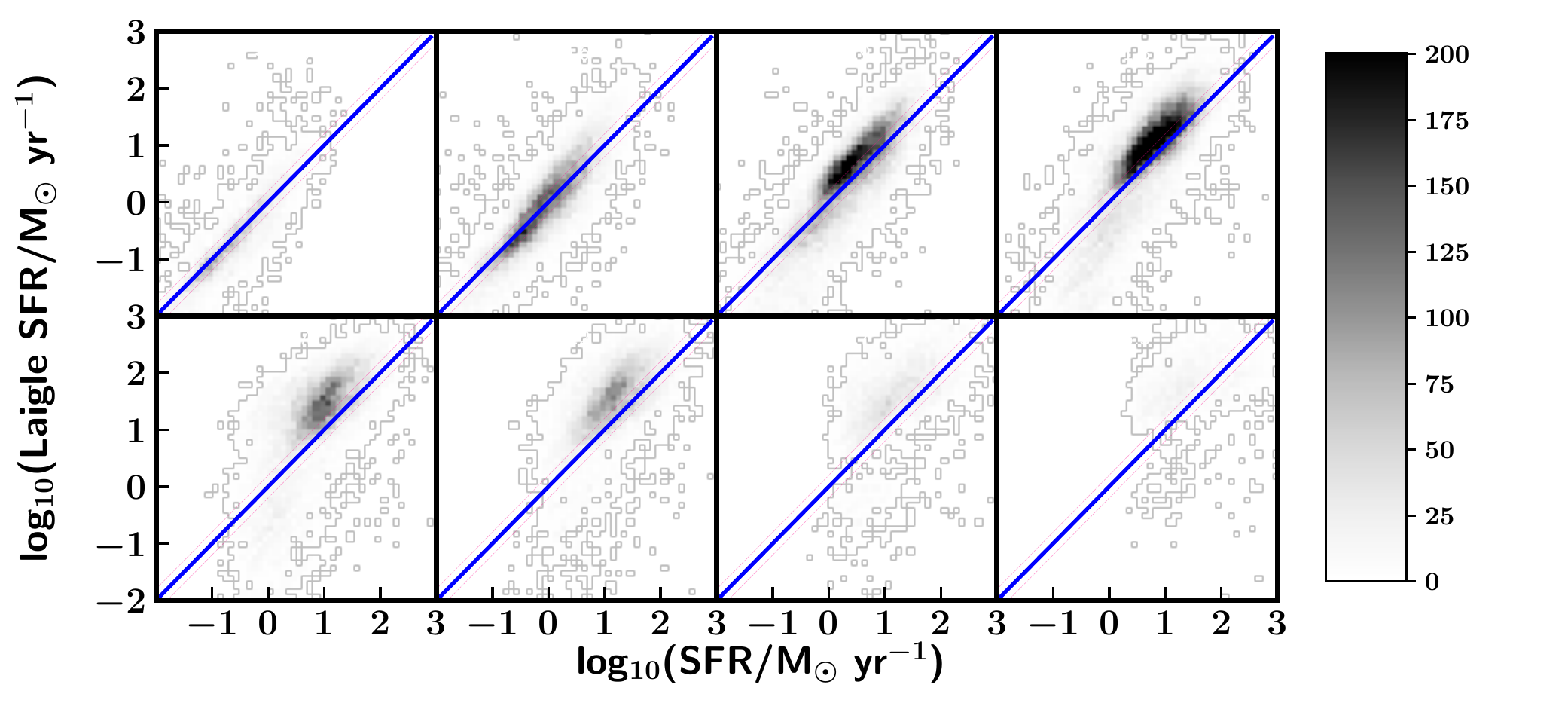}
\end{tabular}
\end{center}
\caption{Top:  The stellar mass distribution of our refit sample in comparison to the original optical-NIR-based fitting results of \citet{Laigle:2016aa}.  We find strong agreement within the two samples, with a dispersion on par with 0.2 dex.  This is consistent with previous comparisons between SED fitting codes (Pacifici et al. In Prep.).  Bottom:  The same Laigle and MAGPHYS-SED3FIT result comparison using total estimated SFR.  We find significantly more scatter, which was an expected result due to our inclusion of MIR-FIR data.  The use of MIR-FIR data in SED fits constrains an other otherwise unconstrained obscured star forming component, yielding lower SFR when included \citep[e.g.,][]{Cooke:2020aa}.\label{fig:fittingcompar}}
\end{figure*}

To test the quality of our fits, we plot a comparison between our final SED fitting estimates with the original optical-NIR-based estimates from the {\tt LePhare} results of COSMOS2015 catalog (Fig.~\ref{fig:fittingcompar}).  We find strong agreement between stellar masses derived from {\tt LePhare} and MAGPHYS/SED3FIT results, with a median difference below 0.1 dex at $z < 1$ and below 0.2 dex at $z < 3$.  Star formation rates are much more variable, especially at higher redshift.  Previous papers comparing SED fitting code results on constant samples have indicated the inclusion of MIR-FIR data points can constrain the obscured SFR component of an SED fit of a star forming galaxy \citep{Cooke:2020aa}, lowering the total SFR estimate (Pacifici et al. 2022 in prep).

\subsection{Color-color Classification}
\label{sec:colors}
\begin{figure*}
\begin{center}
\begin{tabular}{c}
\includegraphics[width=0.95\textwidth]{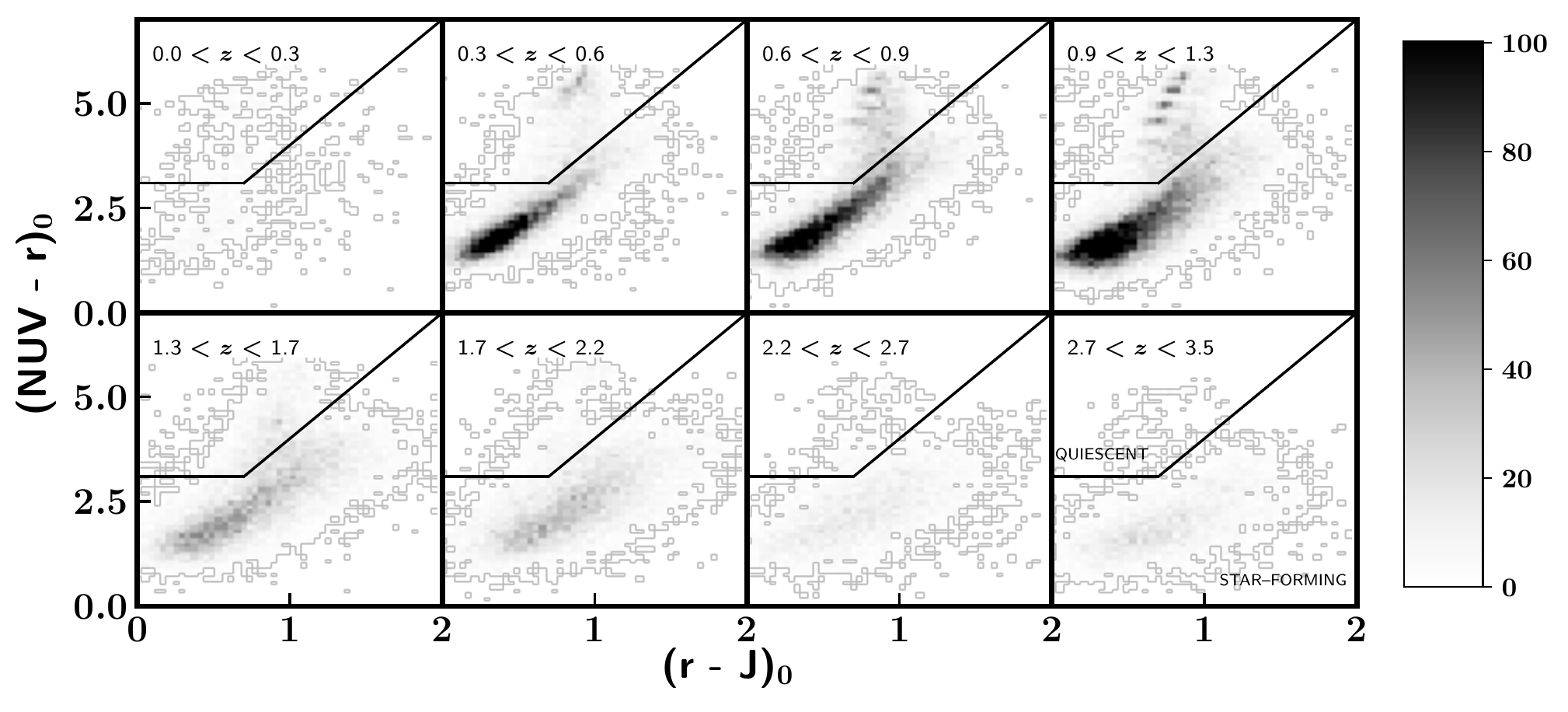}
\end{tabular}
\end{center}
\vspace{-20pt}
\caption
{ \label{fig:SFMSNUVRJ} The distribution of star-forming and quiescent galaxies in our sample as defined by rest-frame NUV-r and r-J colors \citep{Ilbert:2013aa}. We find that the fraction of quiescent galaxies increases by roughly an order of magnitude over our redshift range.  We include the fractions of galaxies in each subsample in Table~\ref{tab:sample}.}
\end{figure*}

For reference of the color distribution of our sample, we plot in Fig.~\ref{fig:SFMSNUVRJ} each galaxy as quiescent or star-forming using rest frame k-corrected NUV-r and r-J colors in the COSMOS2015 catalog \citep{Laigle:2016aa}, but do not apply the sSFR cut used in our SFMS estimation.   We chose these color criteria to be sensitive to the immediate shutdown of star formation \citep{Ilbert:2013aa}.  By using a NUV-sensitive color classification, we probe the distinction between galaxies hosting  O- and B-type stars, and those dominated by lower mass stellar types. The result is that galaxy colors transition to the quiescent parameter space faster than purely optical classification methods \citep{Davidzon:2017aa}, which may classify galaxies dominated by A-type stellar emission as star-forming even when the star formation episode has stopped and the most massive stars born during the latest episode have already died.  

We note that we find a low fraction of quiescent galaxies at $z \sim 2$ ($\sim$20-35\%) across our entire mass range when using a combined sSFR and NUVrJ color-color cut, substantially lower than the ~50\% found in K-band selected spectroscopic programs \citep{Kriek:2006aa,Kriek:2009aa}.  However our values are consistent with the the 35\% found when classifying quiescence using sSFR \citep{Toft:2009aa}. Our results are most similar to the SED fitting results of \citet{Toft:2007aa} and the 20\% quiescent fraction found in the spectroscopic survey of $z \sim 2$ cluster Cl J1449+0856 \citep{Gobat:2013aa}.  The difference in quiescent fractions at $z > 2$ between our results and the K-band program is due to the mass selection inherent in the K-band brightness selection used in \citep{Kriek:2006aa}, which preferentially selected only galaxies more massive than 10$^{11}$ M$_{\odot}$.  If we select only galaxies from our sample with stellar masses greater than 10$^{11}$ M$_{\odot}$, we rederive the 40-50\% quiescence fraction seen in \citet{Kriek:2006aa}.
 
\subsection{Characterizing Morphology}
To determine whether the flattening of the SFMS can be explained by the growth of the bulge component of star-forming galaxies, we require morphological decomposition measurements across a wide range of area and redshift. These classifications enable us to select disk- and bulge-dominated systems, and estimate the SFMS relation of each population.  If the flattening is purely due to bulge-growth, then the disk-dominated systems will exhibit more constant sSFR while the bulge-dominated systems will exhibit a stronger characteristic turnover of SFR above a mass cutoff.  Due to the wide range of redshift covered in our analysis, it is essential to use rest-frame optical morphological measurements. This means that that it is necessary to use an optical classification catalog at $z < 1$ and a near-infrared classification for our high-redshift targets. Therefore we perform the same analyses using two different COSMOS morphological catalogs.

At low redshift ($z < 1$), we retrieve optical classifications from the Zurich Structure $\&$ Morphology Catalog \citep{Sargent:2007aa,Scarlata:2007aa}.  This catalog uses principal component analysis of five nonparametric diagnostics; asymmetry, concentration, Gini coefficient, second-order moment of the
brightest 20\% of galaxy pixels (M20).  In addition, these diagnostics are used in tandem with single S\`ersic fits of the surface brightness distribution to determine a bulgeness parameter correlated with the bulge-to-disk ratio.  This is the value from the catalog used to identify disk dominated galaxies among the full set of disk galaxies, referred to as Type-2 galaxies in the catalog.  The Zurich catalog includes $\sim$130,000 targets across the full 2 square degrees of the COSMOS HST/ACS F814W mosaic down to a limiting magnitude of $I_{AB}\sim24$.  We select Type-2 galaxies with a bulgeness parameter 2 or 3 (disk-dominated composites and purely disk dominated, respectively) for our disk-dominated sample.  Bulge dominated Type-2 galaxies are identified with a bulgeness parameter of 0 or 1.  For full details on this classification system, see \citet{Sargent:2007aa} \& \citet{Scarlata:2007aa}. \citet{Scarlata:2007aa} verified the quality of these fits by visually inspecting the results and removing spurious cases such as overblending and false detections. The median reduced $\chi^2$ of the fits for the targets in both our sample and the full Zurich catalog is 1.04.

At higher redshift ($z > 1$), we estimate rest-frame optical morphological parameters using the infrared HST/WFC3 observations of the COSMOS field taken as part of the Cosmic Assembly Near-IR Deep Extragalactic Legacy Survey (CANDELS) treasury program \citep{Grogin:2011aa,Koekemoer:2011aa} supplemented by the publicly available WFC3/IR imaging available through the Mikulski Archive for Space Telescopes.  We choose to limit our high-z morphology sample to galaxies observed with F125W, F140, or F160W filters, probing the rest frame optical ($\lambda_{observed}>$ 600 nm) for galaxies at $z > 1$. These filter selections, in combination with the partial area of the COSMOS field observed with these filters produces a sample of 4252 galaxies, roughly $5\%$ of the $z > 1$ sample.

\begin{figure*}
\centering
\begin{tabular}{cc}
  \includegraphics[width=70mm]{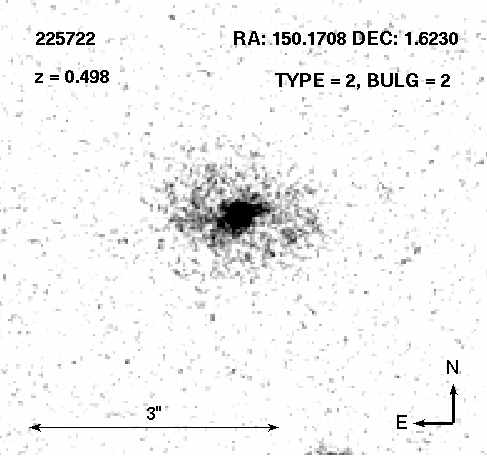} &   \includegraphics[width=65mm]{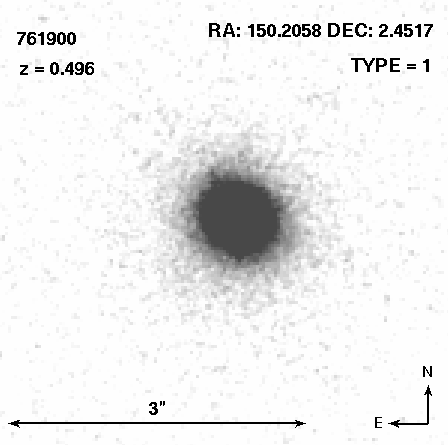} \\
(a) Low-z Disk-dominated (F814W) & (b) Low-z Bulge-dominated (F814W) \\[6pt]
 \includegraphics[width=65mm]{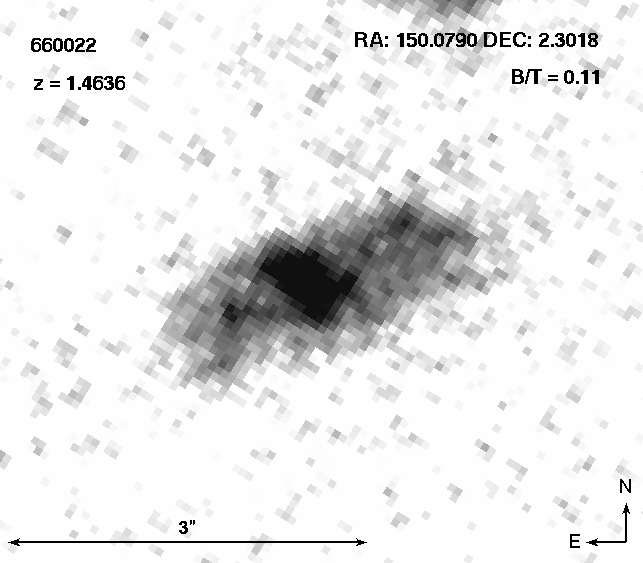} &   \includegraphics[width=65mm]{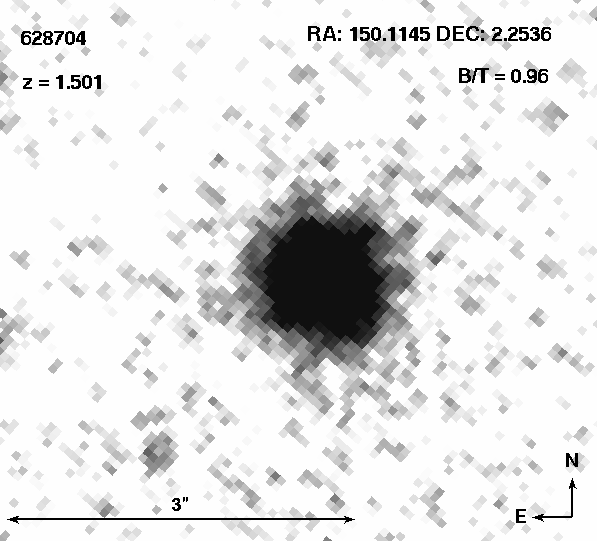} \\
(c) High-z Disk-dominated (F160W) & (d) High-z Bulge-dominated (F140W) 
\end{tabular}
\caption{Examples of disk and bulge dominated galaxies from both redshift samples.  The low-z sample images are taken from the COSMOS field HST ACS/F814W mosaic with morphological type labels from the Zurich morphological catalog.  The high-z sample uses images taken using HST/WFC3 F140W and F160W for these examples and a bulge-to-total ratio calculated using the GALAPAGOS-2 package.}\label{fig:examples}
\end{figure*}

To characterize the morphology of as many galaxies at high redshift ($z > 1$) as possible, we use the MAST Portal to retrieve all public WFC3/IR F125W, F140W, and F160W science observations from the COSMOS field that have not already been used as input to \galapagostwo during the initial CANDELS-COSMOS program (Table~\ref{tab:wfc3table}).  We also do not include data from COSMOS-DASH due to the low surface brightness achieved per WFC3 observation \citep{Mowla:2019aa}. In the case of multiple programs over the same object we use an exposure from the longest exposure program at that location. We also include the HST/ACS F814W images available from the COSMOS field ACS mosaic \cite{Koekemoer:2007aa} in our high-redshift morphology fitting.  We use the same pipeline as the original \galapagostwo CANDELS-COSMOS morphological classification, ensuring we can fairly combine the derived morphological measurements.  Examples of each morphology bin are included in Fig.~\ref{fig:examples}.

The high-redshift fitting was performed using packages from the project MegaMorph \citep{Haussler:2013aa}, which developed a modified version of the IDL code \galapagosone \citep{Barden:2012aa} that takes advantage of a multi-wavelength version of the {\tt Galfit} fitting routine \citep{Peng:2002aa,Peng:2010aa}. This multi-band fitting package is titled \galapagostwo \citep{Haussler:2022aa}, and provides wavelength-dependent two-component S\`ersic fit parameters for bulge-disk decomposition. We input into \galapagostwo postage stamps of any sample member observed by both the HST/ACS F814W COSMOS full-field mosaic and any combination of public HST/WFC3 F125W, F140W, and/or F160W observations. The output provides best-fit parameters for each S\'ersic component.  Specifically, we use the magnitude of each component to estimate bulge-to-total luminosity ratios of each galaxy run through the pipeline. \galapagostwo provides these parameters for each filter an object was observed in, ensuring we can consistently use parameters derived from rest-frame optical continuum emission across our sample.  While \galapagostwo is capable of interpolating between filters, we recognize the difficultly of doing so with so few filters and use the best-fit solution from a target's longest wavelength observation to best estimate the distribution of the stellar population. We find that the B/T ratio from targets with combinations of F125W, F140W, and F160W display consistent B/T ratio solutions when using the magnitudes from each filter.  Shorter wavelength filters such as F110W and F105W were tested but did not consistently reproduce results from longer wavelength observations and were therefore not included in this work. 

\galapagostwo has already been used in two-component mode on the HST/WFC3 images from all five CANDELS fields \citep{Grogin:2011aa,Koekemoer:2011aa} and for each filter any given object has been observed.  We include these values in addition to the estimates made in this work. We note that substructure at high redshift can make the estimation of morphology difficult.

These targets exhibit a median reduced $\chi^2$ $\sim$0.25, a value potentially indicating the presence of overfitting due to the high number of free parameters involved in the two-component S\'ersic fit needed to identify disk-dominated systems.  The magnitudes of each component are well defined, with a median magnitude error of $\sim$0.1 mag. Therefore we make the caveat that the following results from this analysis are most applicable across the population rather than for specific targets.  Additionally, any complex gas and star forming structures of high-z star-forming galaxies would contribute to a more distributed surface brightness distribution, which remains our measure for disk-dominated systems.  While the exact structure of the disk may be different at high-z, the investigation of the distributed star-forming component of high-z galaxies remains our focus and remains roughly identifiable given our data. See \citet{Hashemizadeh:2021aa} for a discussion on visually identified substructure with increasing redshift and the need for a two-component bulge and disk morphological model.

While we caution that these two catalogs both measure morphology, they do so in very different ways and should not be used together.  Therefore we perform any morphological analysis of our sample with each morphological catalog treated independently in Section~\ref{sec:diskdom}, and examples of the classification with images from each data set are available in Figure.~\ref{fig:examples}.  To verify that they produce similar enough results for the sake of discussion however, we compare the Zurich single S\`ersic profile index with the overall index of the two component S\`ersic profile fits from Galapagos and find that for a given target in both catalogs, the median single component S\`ersic index is 12\% lower than the median composite S\`ersic index of the GALAPAGOS result.


\begin{deluxetable}{lcr}
\tabletypesize{\footnotesize}
 \tablecaption{WFC3/IR Input to \galapagostwo
 \label{tab:wfc3table}}
 \tablehead{\colhead{WFC3/IR Filter} & \colhead{Program ID}& \colhead{P.I.}  } 
 \startdata
F125W	&  12461	&  Riess, Adam\\
F125W	&  13046	&  Kirshner, Robert P.\\
F125W	&  13294	&  Karim, Alexander\\
F125W	&  13384	&  Riechers, Dominik A.\\
F125W	&  13641	&  Capak, Peter Lawrence\\
F125W	&  14750	&  Wang, Tao\\
F125W	&  14895	&  Bouwens, Rychard\\
F125W	&  15115	&  Silverman, John David\\
F125W	&  15910	&  Daddi, Emanuele\\
\hline
F140W	&  12190	&  Koekemoer, Anton M.\\
F140W	&  13792	&  Bouwens, Rychard\\
F140W	&  13793	&  Bowler, Rebecca A A\\
F140W	&  14495	&  Sobral, David\\
F140W	&  14719	&  Best, Philip N.\\
F140W	&  14808	&  Suzuki, Nao\\
F140W	&  15115	&  Silverman, John David\\
F140W	&  15363	&  Suzuki, Nao\\
F140W	&  15862	&  Finkelstein, Steven L.\\
\hline
F160W	&  12167	&  Franx, Marijn\\
F160W	&  12440	&  Faber, Sandra M.\\
F160W	&  12461	&  Riess, Adam\\
F160W	&  12578	&  Forster Schreiber, N. M.\\
F160W	&  12990	&  Muzzin, Adam\\
F160W	&  13046	&  Kirshner, Robert P.\\
F160W	&  13294	&  Karim, Alexander\\
F160W	&  13384	&  Riechers, Dominik A.\\
F160W	&  13641	&  Capak, Peter Lawrence\\
F160W	&  13657	&  Kartaltepe, Jeyhan\\
F160W	&  13868	&  Kocevski, Dale D.\\
F160W	&  14596	&  Fan, Xiaohui\\
F160W	&  14699	&  Sobral, David\\
F160W	&  14721	&  Conselice, Christopher J.\\
F160W	&  14750	&  Wang, Tao\\
F160W	&  14895	&  Bouwens, Rychard\\
F160W	&  15229	&  Daddi, Emanuele\\
F160W	&  15692	&  Faisst, Andreas L\\
F160W	&  15910	&  Daddi, Emanuele
 \enddata
 \tablecomments{The programs used as input for \galapagostwo.  Exposures were retrieved using a radial search of 1.2$\deg$ from the COSMOS field center.  For ease of the reader we list the unique program IDs and PIs.}
\end{deluxetable}

\subsection{Quantifying Local Environment}\label{sec:mapmatch}
 Finally, we require a local environment measurement to determine whether star-forming galaxies reside in a given spatially dense region at their redshift, and we can use that measurement to identify any environment-dependent behavior.  We quantify the local environment for each galaxy using the weighted adaptive kernel smoothed maps of the COSMOS field \citep{Darvish:2015aa}.  The spatial density field is calculated in overlapping redshift slices.  Each galaxy within a slice is assigned a weight proportional to the likelihood of the redshift estimate lying within the slice.  This likelihood is the fraction of the galaxy's photo-z probability distribution function that lies within the redshift slice.  With these weights as a prior, \citet{Darvish:2015aa} smooth the spatial density field using a Gaussian kernel with a `global' width of 0.5 Mpc. For reference, this size corresponds to the typical virial radius of X-ray groups in the COSMOS field \citep{Finoguenov:2007aa,George:2011aa}.  Therefore, these maps should be considered most accurate for identifying the cores of clusters and large galaxy groups.  With the parameters used in this specific data set, filaments will be less likely to be detected as this smoothing width is similar to prior widths frequently applied when identifying filaments from density maps \citep[e.g.,][]{Tempel:2014aa}. The density maps provide density values at each location as a multiple of the median density of the map at that given redshift slice.  For further details on this method, see \citet{Darvish:2015aa}.
 
\begin{figure}
\begin{center}
\begin{tabular}{c}
\includegraphics[width=0.48\textwidth]{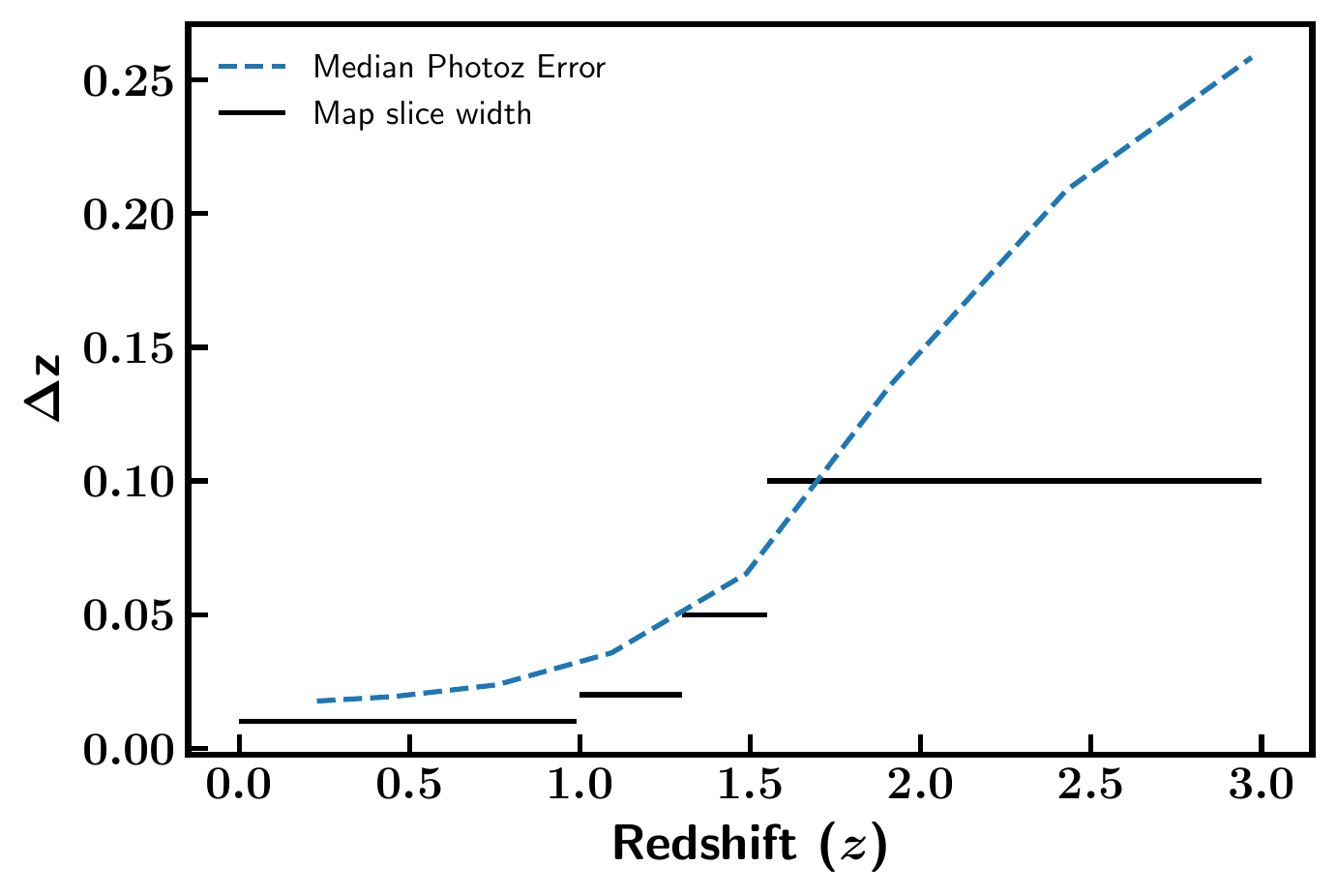}
\end{tabular}
\end{center}
\caption
{ \label{fig:environmentwidth} To characterize the strength of the local environment estimates used here, we compare the median photo-z error of our mass-selected sample as a function of redshift with the redshift slice widths of each map.  Map widths were originally constructed with increasing widths with redshift.  We find that the majority of each target photo-z distributions is well contained by each map slice until $z\sim$2.2, where the photo-z PDF 68th percentile is twice the width of the environment slice.}
\end{figure}

 We consider galaxies in three environment bins: those residing in regions below the median density ($\delta < \delta_{Median}$; field regions), between median density and twice the median density ($\delta_{Median}< \delta < 2\delta_{Median}$; intergroup regions), and regions with any density greater than twice the median density ($\delta > 2\delta_{Median}$; dense regions).  Dense environments are confidently recoverable in the maps below z $\sim$ 2.5, where photo-z errors are $\sim$0.01 \citep{Darvish:2014aa}.  Maps at $z > 2.5$ recover large scale features, with photo-z errors on order $\sim 0.10$ (10s of Mpc), however the intergroup classification will not reliably detect filamentary structures.  To verify that the smoothed maps were sufficiently wide in comparison to our sample's photo-z errors, we calculated the median photo-z for each slice's subsample.  We show in Fig.~\ref{fig:environmentwidth} that the median photo-z error of our mass-selected sample is similar to the width of each map slice until $z > 2.2$.  Above this redshift the maps contain the peak, but no longer the majority, of the median photo-z PDF.
 
\section{Results}\label{sec:paper4results} 

\subsection{The Shape of the SFMS}
After we fit the SEDs of all galaxies selected in Section \ref{sec:paper4sample}, we re-calculate the stellar mass completeness.  Due to the possibility of stellar mass estimates scattering below the original stellar mass cutoff, we redefine our stellar mass selection to include galaxies with stellar mass above the Laigle completeness limit plus the median stellar mass error for that redshift bin.  This removes the parameter space where SED estimate scatter may remove targets and compromise completeness.  The effect of this choice raises the stellar mass limit from \citet{Laigle:2016aa} by 0.1 dex from $0.0 < z < 1.3$ and 0.2 dex between $1.3 < z < 2.2$. 

To estimate the form of the SFMS, we select the star--forming population using NUVrJ color-color criteria \citep{Ilbert:2013aa} and a redshift-dependent sSFR cut corresponding to $> 10^{-11}$ yr$^{-1}$ at present-day \citep{Ilbert:2010aa,Dominguez-Sanchez:2011aa}. NUV, r-, and J-band magnitudes are taken from the k-corrected rest-frame absolute magnitudes provided in COSMOS2015.  The sSFR cut evolves with redshift according to a factor of (1+z)$^{2.5}$, motivated by the results from non-stacked samples of \citet{Speagle:2014aa}. We consider this a less restrictive choice, as observed values for (1+z)$^\gamma$ may vary from $\gamma$ = 2 -- 5 \citep{Salim:2007aa,Karim:2011aa,Whitaker:2014aa,Pearson:2018aa,Popesso:2019aa} and overestimation of $\gamma$ may yield samples biased toward the starbursting side of the SFMS. We include a second cut based on NUVrJ colors to remove potentially rejuvenated quiescent galaxies.  Several mechanisms exist to replenish the cold gas supply of formerly quiescent galaxies, such as the accretion of HI gas from their local environment or the accretion of a gas rich merger companion \citep[e.g.,][]{Kaviraj:2009aa}.  We identify 2258 objects (1.8\% of our total sample) in our original sSFR-selected star-forming sample with red colors which could be these rejuvenated interlopers that are star-forming, yet not undergoing the traditional track of star forming galaxies along the SFMS. For further discussion on our color selection, see Section~\ref{sec:colors}. 

We fit the SFRs and stellar masses of our star--forming sample with a broken power law using the functional form of \citet{Lee:2015aa},
\begin{equation}\label{eq:sfms}\
\log_{10}(SFR) = S_0 - \log_{10}\biggl[1+ \biggl(\frac{M}{M_0}\biggr)^{-\gamma}\biggr],
\end{equation}

\begin{deluxetable}{rrrr}
\tabletypesize{\footnotesize}
 \tablecaption{SFMS Best Fit Function Parameters
 \label{tab:bestfitSFMS}}
 \tablehead{\colhead{$z$} & \colhead{S$_{0}$}& \colhead{M$_{0}$} & \colhead{$\gamma$}  } 
 \startdata
 0.0--0.3 & 1.00 & 11.15 & 0.67\\ 
 0.3--0.6 & 1.01 & 10.68 & 0.77\\ 
 0.6--0.9 & 1.45 & 11.03 & 0.71\\ 
 0.9--1.3 & 2.05 & 11.84 & 0.60\\ 
 1.3--1.7 & 10.26 & 24.51 & 0.64\\ 
 1.7--2.2 & 9.20 & 21.44 & 0.72\\ 
 2.2--2.7 & 9.34 & 21.81 & 0.69\\
 2.7--3.5 & 3.49 & 12.70 & 0.80
 \enddata
 \tablecomments{Best fit \citet{Lee:2015aa} broken power-law parameters using Eq.~\ref{eq:sfms} and the star--forming sample in each of our redshift bins.  S$_{0}$ represents the highest estimated SFR in the sample, M$_{0}$ is the turnover mass for the broken power law, and $\gamma$ is the power law slope at low stellar masses. The rise of M$_{0}$ represents a transition from a SFMS with a turnover at low-z ($z < 1.3$), to without a turnover at high-z ($z > 1.3$)}
\end{deluxetable}

\begin{figure*}
\begin{center}
\begin{tabular}{c}
\includegraphics[width=1\textwidth]{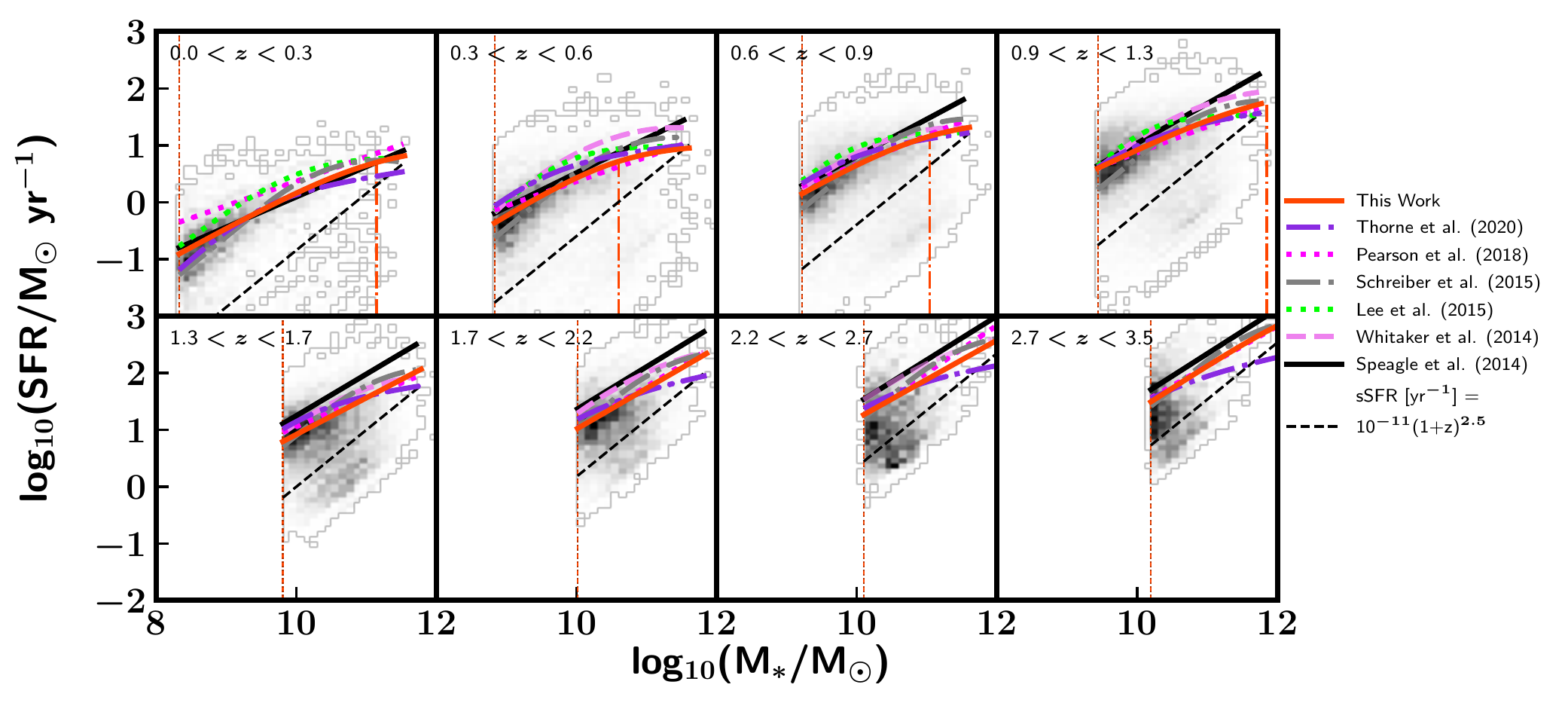}
\\
\includegraphics[width=1\textwidth]{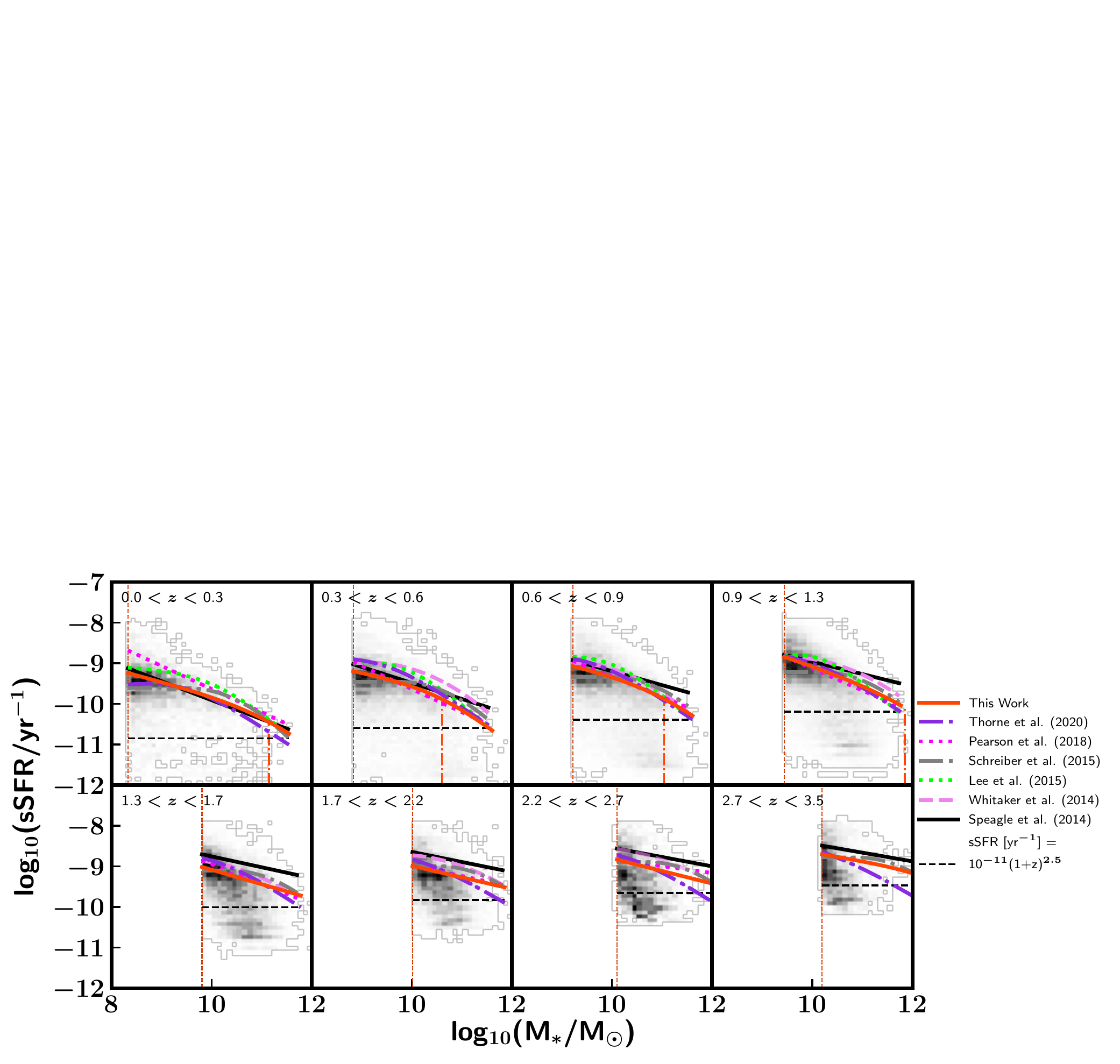}
\end{tabular}
\end{center}
\caption
{ \label{fig:SFMScontours} Top: Density plot of star formation rate versus stellar mass in each redshift bin of our total (star--forming + quiescent) sample.  For the illustrative use of this plot, we have not normalized the underlying density plot to better show the distribution of low-numbered populations, e.g., our lowest redshift bin. The solid orange-red line corresponds to the broken power law fit (described in Equation~\ref{eq:sfms}) to the star--forming population in each redshift selected using NUVrJ color-color criteria (Section \ref{sec:colors}) and sSFR cut of $> 10^{-11}$(1+z)$^{2.5}$ yr$^{-1}$. The estimated turnover mass is indicated using an orange dash-dotted line below the SFMS estimate line.  The SFMS becomes linear at $z > 1.3$, and has not associated turnover mass. Bottom: We move to a specific star formation rate space for the remainder of the paper to better articulate the evolution of star formation efficiency with stellar mass.  For both plots, we plot the SFMS trends observed by \citet{Whitaker:2014aa}, measured from $0.5 < z < 2.5$ as a fuchsia dashed line, \citet{Speagle:2014aa} as a solid black line, \citet{Thorne:2020aa} as a violet dash-dotted line, \citet{Lee:2015aa} as a dotted green line, \citet{Pearson:2018aa} as a dotted fuchsia line, and \citet{Schreiber:2015aa}, renormalized to a \citet{Chabrier:2003aa} IMF, as a grey dot-dashed line. The stellar mass completeness limit for each redshift bin is plotted with a vertical dashed line in orange. The lack of massive, quenched galaxies between $0 < z < 0.6$ is a cosmological effect due to the limited volume of COSMOS at low redshifts (2.26 $\times$ 10$^{6}$ Mpc$^{3}$ in COSMOS below $z < 0.6$). }
\end{figure*}

\noindent where S$_{0}$ represents the highest estimated SFR in the sample, M$_{0}$ is the turnover mass for the broken power law, and $\gamma$ is the power law slope at low stellar masses.  All three parameters are left free during the fitting process to not bias our results towards fits with a turnover or those without. A lack of a turnover is identified if the turnover mass is greater than the maximum stellar mass of our sample. The evolution of higher turnover mass with look-back time has been observed in previous works \citep[e.g.,][]{Thorne:2021aa}. We weight each galaxy by the inverse of the reduced $\chi^2$ of the best-fit SED.  Our sample is well fit with a peak of the reduced $\chi^2$ distribution $\sim$1 and has an evolving median value of $\sim$1.5 to 2.0 with increasing redshift. Confident in the overall quality of our fit sample, this weighting practice minimizes the effect of outliers that may have extremely high SFR estimates or poorly defined SFR estimate probability distribution functions.  Our best-fit parameters are listed in Table~\ref{tab:bestfitSFMS}.  We find that the turnover mass (M$_{0}$) evolves toward higher stellar masses from present-day to z = 1.3. At z $>$ 1.3, the turnover mass is best fit by a value greater than the masses in the sample.  This form is effectively linear, a trend also found in   \citet{Leslie:2019aa} and \citet{Tomczak:2016aa}.

The distribution of our sample's SFR and stellar mass estimates in eight redshift bins from $0 < z < 3.5$ are shown in Fig.~\ref{fig:SFMScontours}, using the same redshift bin definitions as \citet{Laigle:2016aa} with our best-fit SFMS function overplotted in comparison to relevant counterparts in the literature. As observed in previous SFMS studies out to $z \sim 2.5$ \citep{Whitaker:2012aa,Whitaker:2014aa,Lee:2015aa}, the normalization of the SFMS increases with increasing redshift, i.e., galaxies at a fixed stellar mass have higher SFRs at higher redshift.   We also find a turnover of SFR with stellar mass above 10$^{10}$ M$_{\odot}$ for $z < 1.3$, where galaxies of progressively higher stellar mass no longer exhibit higher SFRs.  The turnover mass also evolves to higher stellar mass with increasing redshift.  The high mass turnover has been previously observed in the literature \citep[e.g.,][]{Whitaker:2014aa,Schreiber:2015aa}, and specifically in the COSMOS field \citep{Lee:2015aa, Leslie:2019aa,Thorne:2020aa}.   

\begin{figure}
\begin{center}
\begin{tabular}{c}
\includegraphics[width=0.45\textwidth,angle=0]{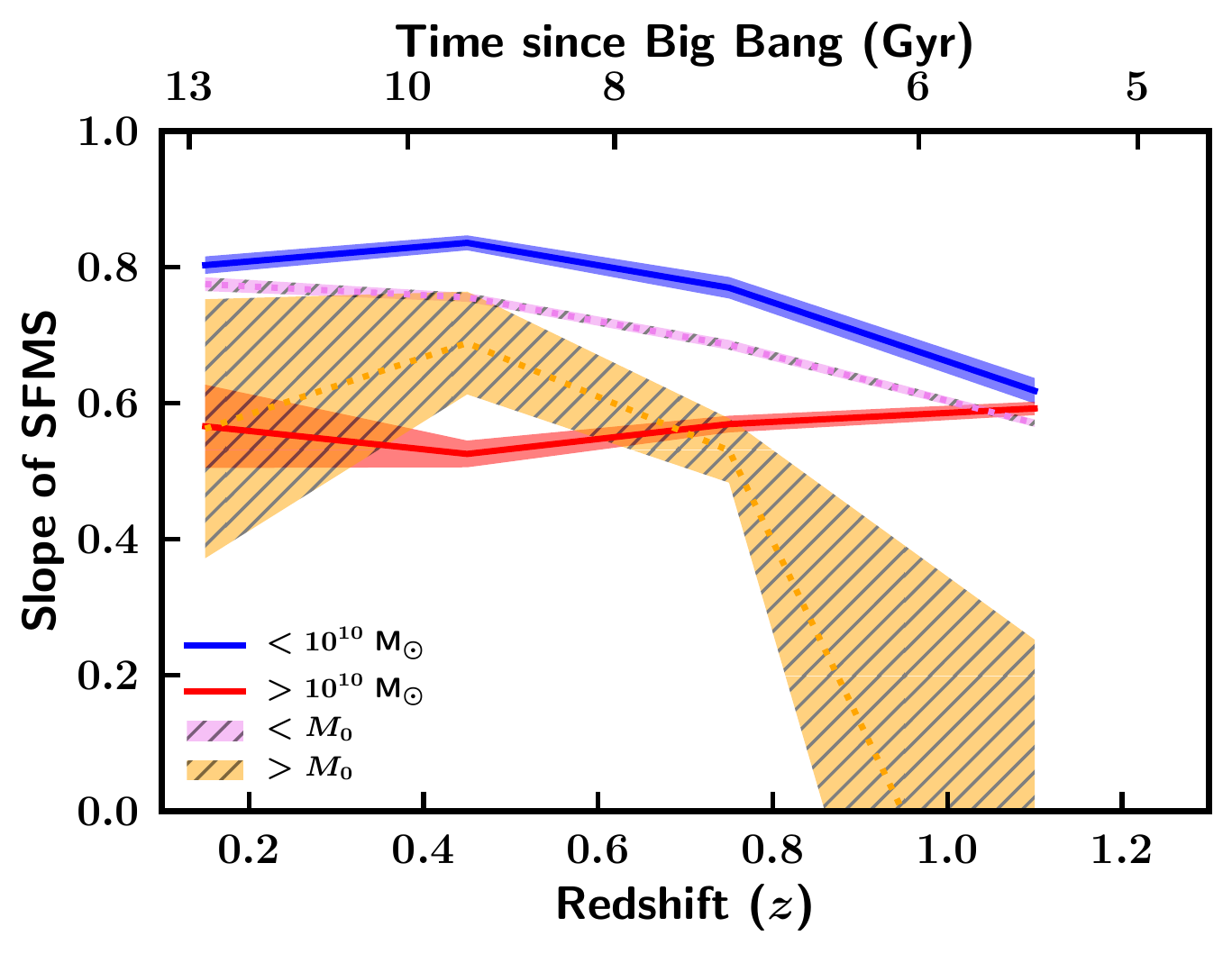}
\end{tabular}
\end{center}
\caption
{ \label{fig:SFMSslope} The slope of the star formation -- stellar mass relation for all star--forming galaxies, calculated using a first order least squares polynomial fit for galaxies above (red) and below (blue) the literature turnover of the relation (10$^{10}$ M$_{\odot}$), weighted by reduced $\chi^2$. We compare this literature-based measurement of our sample to the slopes when measured using our estimates for the `knee' of the SMFS, M$_{0}$ (violet hatched for greater than the knee, violet orange for below the knee). Uncertainty of the fits are plotted as shaded regions. We find that at $z > 1$,  the best fit SFMS M$_0$ becomes higher than the stellar masses of our sample, indicating a linear relation.  We find that the high mass end decreases in slope strongly with redshift while the low mass slope only decreases by ~30\% by z $\sim$ 1.  In comparison, the fixed SFMS knee of 10$^{10}$ used in the literature will include star forming galaxies from 10$^{10}$ to 10$^{10.5}$ in their high mass sample, a mass range that is less sensitive to the decrease in SFR with redshift, decreasing the redshift dependence of the fixed mass selected high-mass slope.
}
\end{figure}

The slope of the global SFMS is a measurement with significant scatter in the literature, with slopes ranging between 0.5 -- 1, the same range exhibited by our sample in  Fig~\ref{fig:SFMSslope}.  This high variation is present regardless of whether the parent sample was chosen via UV, optical, or IR characteristics \citep{Speagle:2014aa}.  In Fig.~\ref{fig:SFMSslope}, we plot the slopes of the SFMS for high- and low- mass bins using both our own turnover mass estimate and the 10$^{10}$ M$_{*}$ value commonly used throughout the literature. Using the literature turnover, we find a convergence in high- and low-mass slope at $z > 1$.  Using our turnover estimates, we find that the high mass end of the SFMS flattens out to $z > 1$.  At high redshift, the low mass end is too loosely constrained to make any conclusions, while the high-mass end rises in slope towards linearity. We tested using an evolving mass cut motivated by our SFMS fits and found that the high mass bin at z $>$ 1 is often too poorly constrained to make any conclusions. 

\subsection{Estimating the Star Formation -- Stellar Mass Relation In Different Environments}\label{sec:sfmsfit}
\begin{figure*}
\begin{center}
\begin{tabular}{c}
\includegraphics[width=0.95\textwidth]{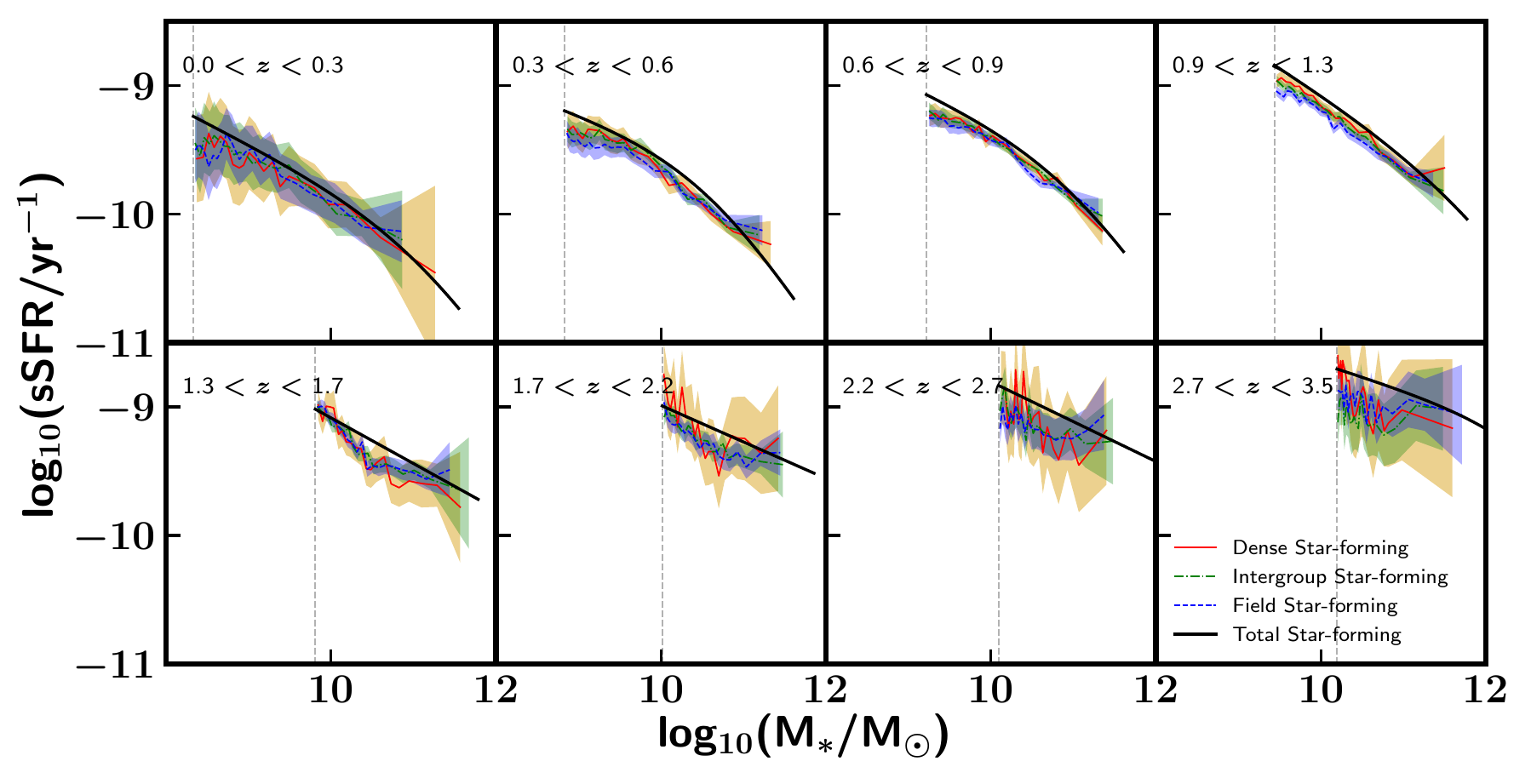}
\end{tabular}
\end{center}
\vspace{-20pt}
\caption
{ \label{fig:SFSMSenvsstarformingfig4} Our SFMS fit compared to the raw median specific star formation rate as a function of stellar mass of star--forming galaxies in field ($\delta < \delta_{Median}$, blue), intergroup ($\delta_{Median} < \delta < 2\times\delta_{Median}$, green), and dense ($\delta > 2\times\delta_{Median}$, red) local environments. We find that star--forming galaxies in all environments are consistent within errors with our global SFMS and do not exhibit any environmental dependence with local environment at any stellar mass examined by our sample. The stellar mass completeness limit for each redshift bin is plotted with a vertical dashed line in black.}
\end{figure*}

We compare the SFMS across three environment bins without consideration of morphology in order to probe local environment's role on the quenching of star-forming galaxies as exhibited by the shape and normalization of the SFMS. In Fig.~\ref{fig:SFSMSenvsstarformingfig4}, we compare the behavior of our fit to the global SFMS to the median SFR per stellar mass bin in field ($\delta < \delta_{Median}$, blue), intergroup ($\delta_{Median} < \delta < 2\times\delta_{Median}$, green), and dense ($\delta > 2\times\delta_{Median}$, red) environments.   When comparing the shape and position of the SFMS within each environment bin, we find no difference in median sSFR with environment. One difference between our environment-binned sample and the overall sample is an offset in sSFR.  This is due to the reduced-$\chi^2$ weighting used in the trend estimation for the total population, which favors higher sSFR galaxies.  With either method, we still observe a decline of sSFR with mass across all redshifts, and across all environments.

The consistent decline of sSFR with stellar mass across environment bin is also observed when moving across redshift bins.  This result leads to the conclusion that star--forming galaxies are forming stars in a self-similar manner independent of the local galaxy density surrounding them, and that star-formation's insensitivity to local environment (previously observed at low redshift) is itself insensitive to redshift out to $z\sim3$.  

\begin{figure*}
\begin{center}
\begin{tabular}{c}
\includegraphics[width=0.95\textwidth]{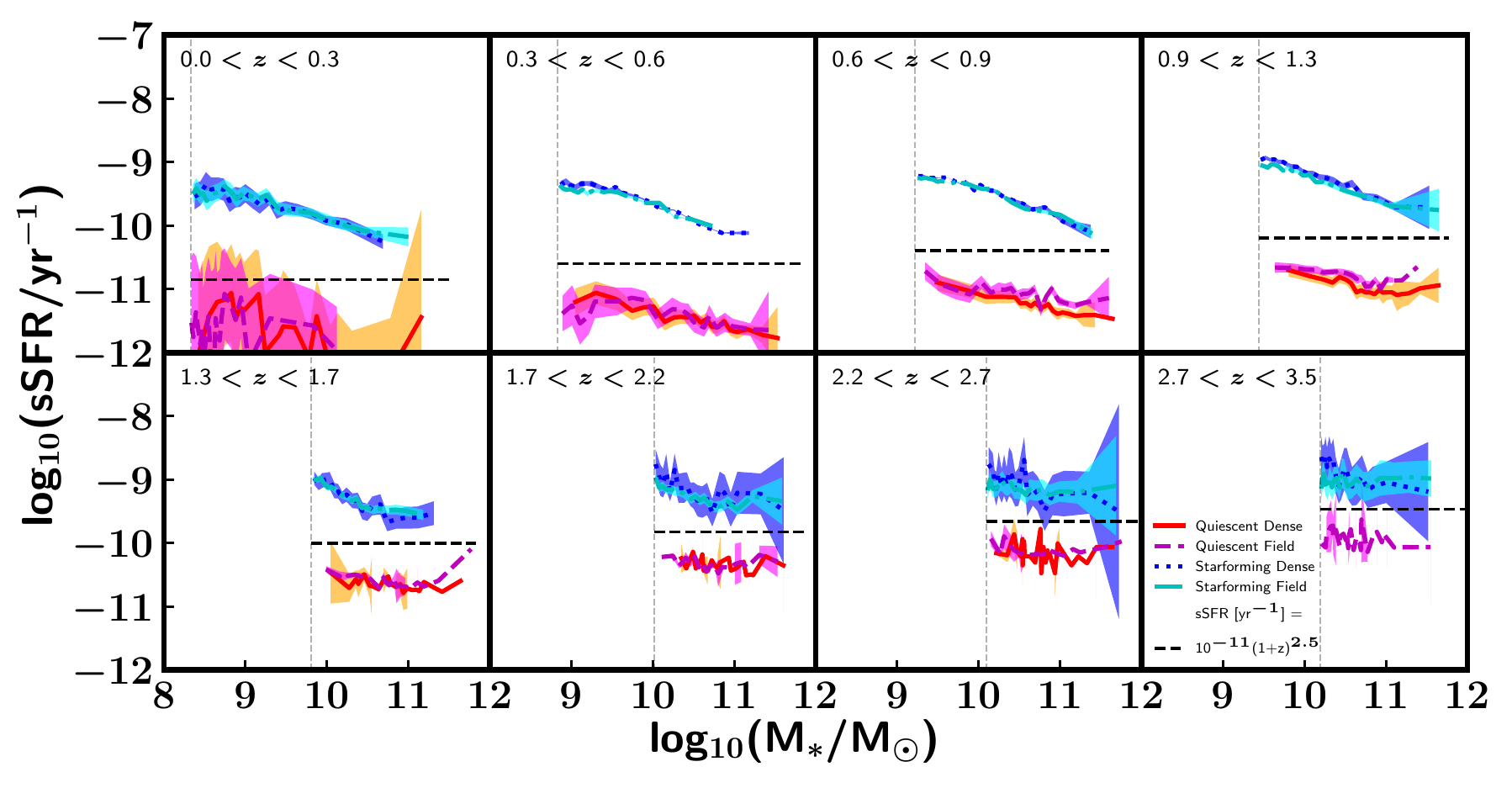}
\end{tabular}
\end{center}
\vspace{-20pt}
\caption
{ \label{fig:SFMScolorcolor} The median specific star formation rate in bins of 150 objects for galaxies in field ($\delta < \delta_{Median}$, blue dashed), and dense ($\delta > 2\delta_{Median}$, red solid) local environments.  For this plot, we classify galaxies as star--forming based on NUVrJ colors only, and plot the sSFR cut used in the final sample selection process.  At $z > 0.9$, we see that the star-forming and quiescent populations show no dependence on local environment. The star--forming populations in dense and field environments are consistent within errors of each other when a sSFR cut is used (see Fig.~\ref{fig:SFSMSenvsstarformingfig4}).  The stellar mass completeness limit for each redshift bin is plotted with a vertical dashed line in black.}
\end{figure*}

To further understand this behavior, we use our NUVrJ color criteria and SFR estimates to examine the behavior of both star-forming and quiescent galaxies in our dense and field environment bins to determine whether the quiescent population shows evidence of environmental dependence and where the quiescent population lies with respect to the star-forming. By binning each environment bin by star-forming or quiescent status, we can answer whether there is any difference in the quenching behavior and star formation end-state, quiescent galaxies, in each environment. In Fig.~\ref{fig:SFMScolorcolor} we plot the median sSFR per stellar mass bin of 150 objects for both star-forming and quiescent galaxies across dense and field environments. To test how our combined sSFR and color selection criteria affect the sample, we only use the NUVrJ color criteria to select the star--forming sample of Fig.~\ref{fig:SFMScolorcolor}, and plot the sSFR delineator of galaxies with sSFR = 10$^{-11}$ yr$^{-1}$(1+z)$^{2.5}$ as a black line for comparison.  We find that at $0.9 < z < 3.5$, both star-forming and quiescent subsamples do not show any difference in median SFR based on local environment.   We note that the median SFR of the purely color-selected star-forming galaxies observed here approach the sSFR cutoff at high-z ($z > 1.3$), however the median remain above the sSFR cutoff within the errors on the median.  Additionally, the shape of the SFMS, as plotted in Fig~\ref{fig:SFMScolorcolor} using stellar-mass binned medians, does not significantly change with redshift.

\subsection{The Role of Morphology in the SFMS}\label{sec:morph}
To address whether the turnover of the SFMS is due to a transition towards bulge-dominated systems, we study the shape of the SFMS for bulge- and disk-dominated SF galaxies.  Due to the wide redshift range used here, any morphological catalog would only measure the distribution of stellar populations within a wavelength-dependent range.  Therefore we perform two independent morphological analyses.  First, a low-z assessment using observer-frame optical data.  Second, a high-z sensitive assessment using observer-frame NIR data.

\subsubsection{Low-z Morphology using Optical Measures}
\begin{figure*}
\begin{center}
\begin{tabular}{c}
\includegraphics[width=1\textwidth]{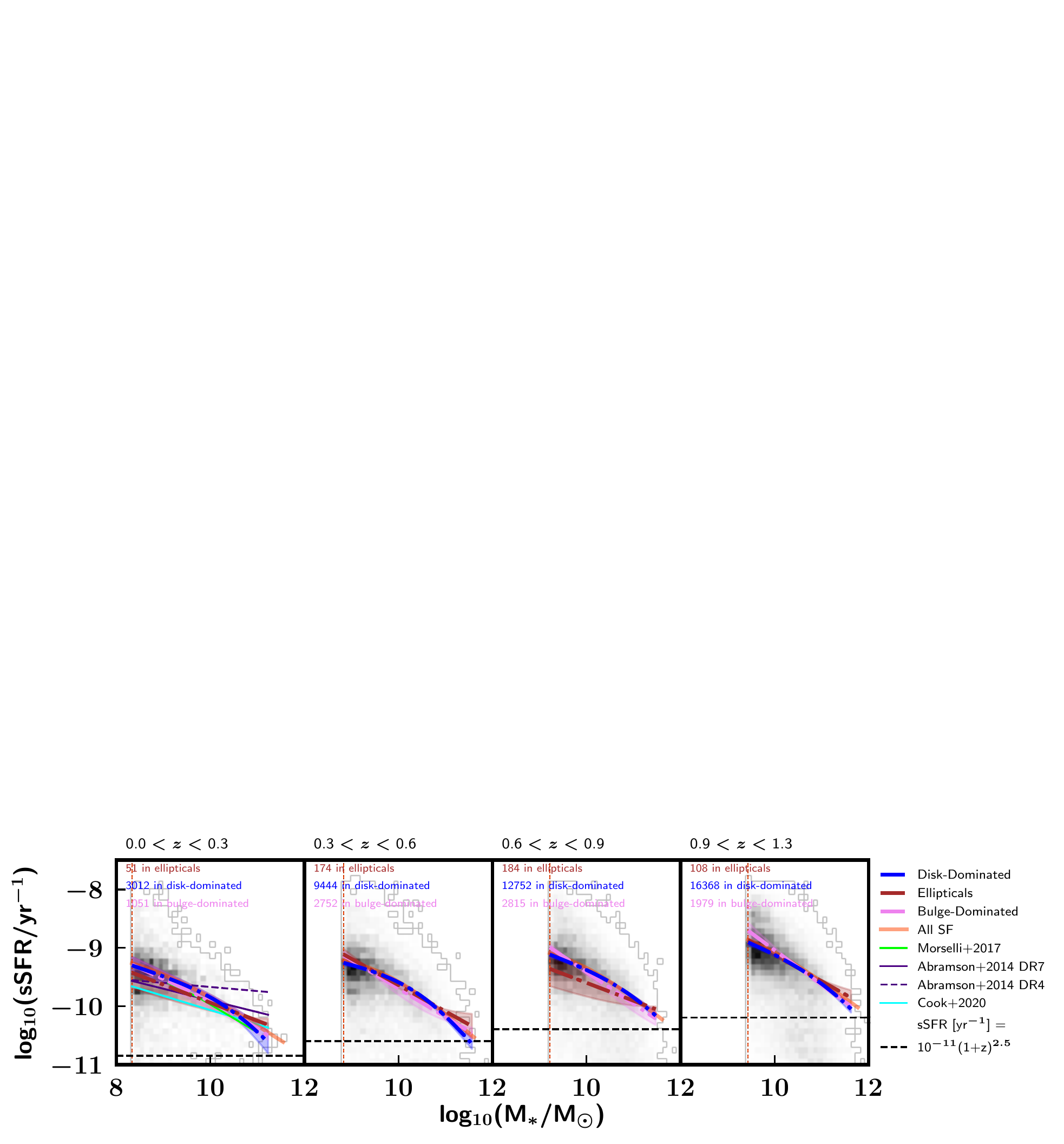}
\end{tabular}
\end{center}
\vspace{-20pt}
\caption
{\label{fig:disksfms} SFMS as estimated using only disk-dominated systems (BULG = 2 or 3, blue dashed-dotted), bulge-dominated disks (BULG = 0 or 1, magenta dashed-dotted), and elliptical systems (TYPE = 1, brown dashed-dotted) from the Zurich Structure $\&$ Morphology Catalog. Finally we also include our global SFMS estimate in orange. We include the full sample as black density bins for reference. In the $z < 0.3$ subplot, we include literature from local studies such as the original \citet{Abramson:2014aa} DR4 results showing a flat disk sSFR with mass (dashed-violet) versus the \citet{Abramson:2014aa} DR7, \citet{Morselli:2017aa}, and \citet{Cook:2020aa} results retaining a change in slope with stellar mass.} 
\end{figure*}

First, in Fig.~\ref{fig:disksfms}, we consider our low-redshift morphological sample using classifications from the Zurich Structure $\&$ Morphology Catalog \citep{Sargent:2007aa,Scarlata:2007aa}. Due to the rest-frame optical wavelengths of this catalog, classifications only accurately probe the bulk of the stellar component out to $z \sim 1$, but are classified across the entire COSMOS field, yielding a uniform data set across the COSMOS field.  For disk-dominated late-types, we use the classification of \verb|TYPE| = 2 and \verb|BULG| = 2 or 3 to only select pure-disks and disk-dominated composites. These classifications are derived using principle component analysis and correlated with, but not limited by, bulge to disk ratios.  For further information on this analysis, please see \citep{Scarlata:2007aa}. These criteria are consistent with previous COSMOS morphological studies using the Zurich catalog \citep[e.g.,][]{Molnar:2018aa}. 

Bulge-dominated disk galaxies are selected using galaxies with a disk morphology but the high bulge dominance labels of \verb|BULG| = 0 or 1. Finally, we also include ellipticals as their own bin using the Zurich morphology \verb|TYPE| = 1, without any overlap with our bulge-dominated disk subsample.  We find that a decline in sSFR with total galaxy mass remains in the disk-dominated sample and is also consistent within the errors on the median trend across all three morphology bins at $z < 1.3$. We consider this as evidence for a declining host sSFR with stellar mass, irrespective of the dominance of the bulge component. 

\begin{figure*}[t]
\begin{center}
\begin{tabular}{c}
\includegraphics[width=0.99\textwidth]{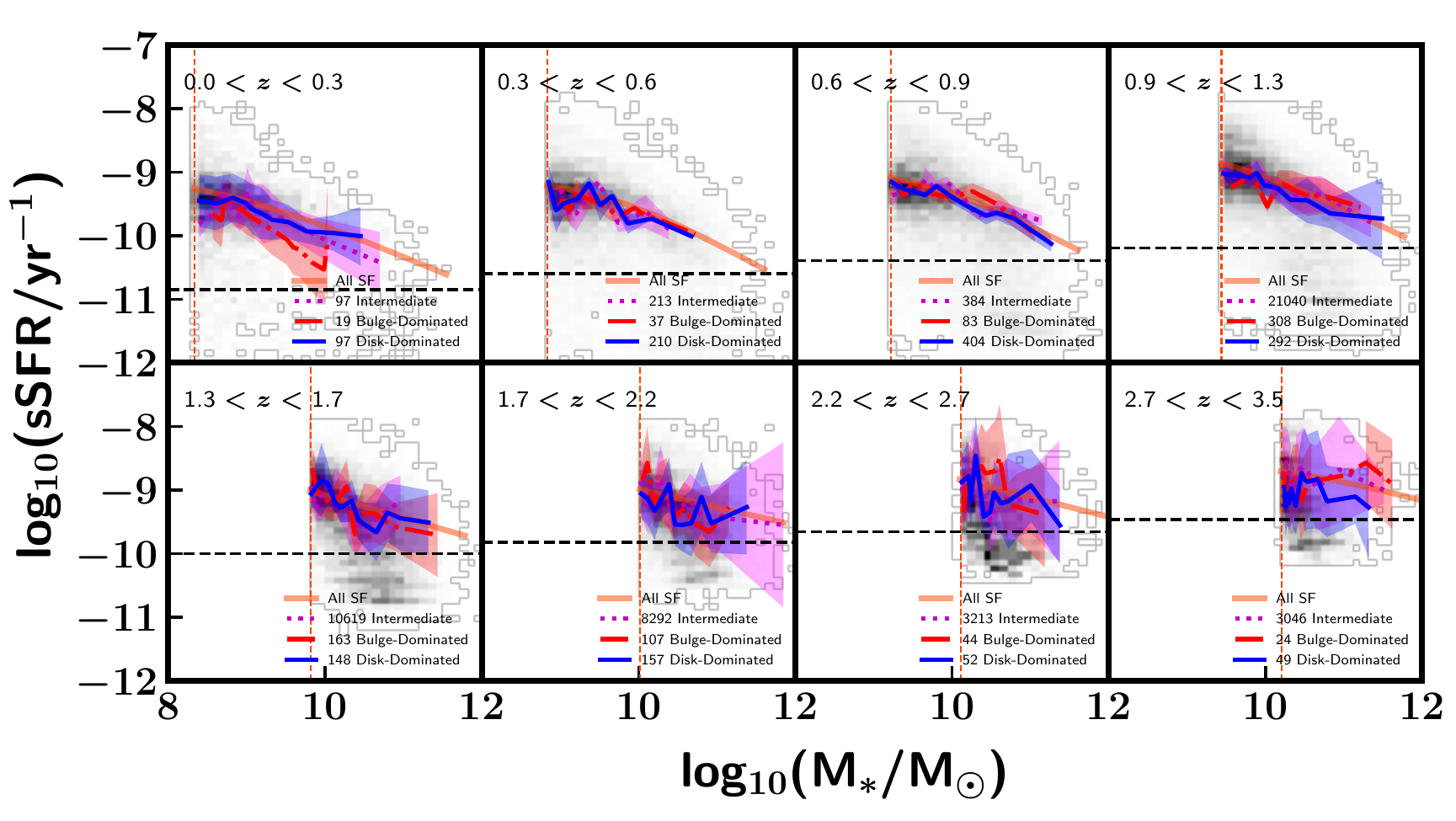}
\end{tabular}
\end{center}
\vspace{-20pt}
\caption
{\label{fig:disksfmscandels} 
Median sSFR of disk-dominated (B/T $<$ 0.25), bulge-dominated (B/T $>$ 0.75), and intermediate (0.25 $<$ B/T $<$ 0.75) star-forming systems using \galapagostwo estimates derived from the CANDELS field and public HST WFC3/IR archival data. Each sample is plotted in 30 stellar mass bins. Star forming galaxies are selected using a NUVrJ color-color criteria and sSFR cut (black dashed line) of $> 10^{-11}$(1+z)$^{2.5}$ yr$^{-1}$ as described in Section~\ref{sec:paper4results} and we include our global SFMS estimate in orange for reference.  Star-forming galaxies from the disk-dominated bin are in blue solid line, and from the bulge-dominated bin in red dashed-dotted line, with intermediates as magenta dotted.  We include the full sample for reference as a black 2D histogram.  We no longer find the turnover in our morphology-selected SFMS estimates at $z > 1.3$, however the errors on the median are too large to determine any difference as a function of B/T ratio.} 
\end{figure*}

\subsubsection{High-z Morphology using NIR Measures}
To determine whether any bulge-dependent effect is driving the turnover at high masses at high redshift, we perform a similar analysis using the two component S\'ersic fits available in the CANDELS MegaMorph catalog supplemented with new two-component decompositions performed using an identical pipeline on the publicly available HST WFC3/IR  F125W, F140W, and F160W images covering the COSMOS field as described in Table~\ref{tab:wfc3table}.  This second analysis enables us to reach higher redshift than the Zurich catalog due to the longer wavelengths observed with HST WFC3/IR filters. The multi-band approach of the \galapagostwo pipeline also enables the separation of bulge and disk components more cleanly \citep{Haussler:2022aa}, enabling analysis to fainter galaxies and greater distances.  The model component magnitudes provided by \galapagostwo can be used to select an explicit bulge-to-total ratio cutoff to select disk-dominated, bulge-dominated, and intermediate galaxies.  

\subsubsection{Results from Low- and High-z Samples}
We limit our disk-dominated sample to those with confidently low B/T ratios, defined as a B/T ratio of $< 0.25$.  Bulge-dominated galaxies are selected with B/T$> 0.75$, and intermediates as $0.25 <$B/T$< 0.75$.  This requirement ensures our disk-dominated sample only considers galaxies in which the bulge is not the dominant environment for the stellar population of a given galaxy. Similar cutoffs are used in the literature to form bulge- and disk-dominated samples to compare against each other \citep[e.g.,][]{Kennedy:2016aa}.  Shown in Fig.~\ref{fig:disksfmscandels}, we find that at $z < 1.3$ the SFMS of disk-dominated systems experiences a weak decline with stellar mass consistent with the turnover exhibited by the overall SF population.  We also see that each morphology bin exhibits the decline within each of the B/T classification within their errors (B/T $<$ 0.25, 0.25 $<$ B/T $<$ 0.75, and B/T $>$ 0.75).  We caution that the results of our $0.9 < z < 1.3$ bin are subject to the F814W filter no longer probing the old stellar population due to the light redshifting out of the rest-frame optical and therefore targets may have under-estimated bulge components in this panel.

In the rarely examined morphology parameter space of $z > 1.3$ plotted in  Fig.~\ref{fig:disksfmscandels}, we work with a significantly less populated sample due to the smaller area of COSMOS observed with F125W, F140W, or F160W bands in the COSMOS-CANDELS subfield (0.05 sq deg.) region and the publicly available WFC3/IR coverage ($\sim$0.05 sq deg.).  For comparison, the low redshift sample is derived from the full 1.7 sq deg.~area covered by ACS.  We note that the distribution of B/T ratios at high-z peaks at an intermediate type (B/T $\sim$ 0.5) due to the difficulty of modelling two morphological components at such a high redshift, even with HST images.  Therefore, we highlight the results of the bulge- and disk-dominated samples.  In Fig.~\ref{fig:disksfmscandels}, we find that at high redshift ($z > 1$), the linearity of the SFMS strengthens with redshift as observed with the general population, however the larger error bars prevent a robust analysis of the turnover mass.  Again, the SFMS estimated using each morphology bin are consistent with each other.


\section{Discussion}\label{sec:disc}


\subsection{Comparison to the SFMS Literature}\label{sec:literature}
In Fig.~\ref{fig:SFMScontours}, we observe that our global star formation rate -- stellar mass relations are consistent with the spread of previous SFMS results at low to intermediate masses.  Our results are most consistent with \citet{Schreiber:2015aa} once re-normalized to a \citet{Chabrier:2003aa} IMF.  Our relations often indicate lower SFRs by 0.1-0.2 dex across stellar mass, but converge at higher stellar masses with other studies that observe a turnover \citep[e.g.,][]{Whitaker:2014aa,Schreiber:2015aa}.  Two systematic effects are present that may be responsible for the offset in median SFR estimates.  First, \citet{Schreiber:2015aa} use a delayed exponentially declining star formation history while our MAGPHYS fits use a non-delayed exponentially declining star formation history.  This delays the decline of SFR at high redshift in the models used in \citet{Schreiber:2015aa}, yielding a higher SFR than our sample fits until both reach a low SFR steady state at low redshift.  Second, the choice of dust law may play a part in such an offset, as \citet{Schreiber:2015aa} use the \citet{Calzetti:2000aa} dust attenuation law while we use the \citet{da-Cunha:2008aa} dust models included in {\tt MAGPHYS} and {\tt SED3FIT} that use a shallower slope, i.e., lower absorption for emission lines, for birth cloud attenuation \citep{Salim:2020aa}.  Overall, our SFMS estimates are within the scatter of previous works on the SFMS and consistent with the scatter about SFR and stellar mass estimates between SED fitting codes identified in Pacifici et al. (2022, in prep.).

 The most similar to our SFMS fit, \citet{Schreiber:2015aa}'s SFMS relation is calculated using galaxies selected from the GOODS-South \citep{Guo:2013ab}, UDS \citep{Galametz:2013aa}, and COSMOS-CANDELS fields \citep{Nayyeri:2017aa}.  Galaxies are selected above the $i$-, K-, or H-band completeness limit of each field.  Similar to \citet{Whitaker:2014aa}, \citet{Schreiber:2015aa} use the UVJ criteria of \citep{Williams:2009aa} to select star forming galaxies and estimate their star formation rates using the FAST package \citep{Kriek:2009aa}.  The far-infrared portion of their SEDs were constrained using median stacking of \emph{Herschel} SPIRE observations.  \citet{Schreiber:2015aa}'s inclusion of NUV to FIR photometry in SED fitting, combined with a similar color-color based selection method may explain our similar results.

 In Fig.~\ref{fig:SFMScontours} we plot the \citet{Whitaker:2014aa} SFMS relation closest to the average redshift of each of our redshift bins. From $0.5 < z < 2.5$, \citet{Whitaker:2014aa}'s results are often offset from this work's to higher SFRs. Their sample is selected from the five CANDELS fields, with stellar mass estimates calculated using the FAST stellar population fitting package \citep{Kriek:2009aa}.  However, their SFR estimates are calculated using two independent components for the unobscured and obscured SFRs.  The bolometric IR (8--1000 $\mu$m) luminosity is calculated assuming the log average of the IR spectrum templates from \citep{Dale:2002aa}.  This total infrared luminosity and the total integrated rest-frame luminosity at 1216--3000 \AA\;are added and converted into a total SFR. The difference in SFRs between their results and ours may be due to the lack of an energy budget conservation, which constrain the height of the FIR peak in our fits and likewise constrain the SFRs estimated.


\subsection{Implications of Environmental Effects}\label{sec:environment}
 The lack of environmental dependence at high redshift seen in Fig.~\ref{fig:SFSMSenvsstarformingfig4} agrees with recent literature that find no environmental dependence on the SFRs of galaxies on the main sequence at high redshift \citep{Erfanianfar_2015,Hatfield:2017aa,Leslie:2019aa}.  \citet{Leslie:2019aa} found no evidence for an environmental dependence of the SFMS out to $z \sim 3$ for a sample of galaxies selected with a comoving number density cut, as well as for X-ray groups out to $z \sim 0.75$, using SFRs derived from 3 GHz radio continuum stacks.  Recently, \citet{Randriamampandry:2020aa} used the VLA-COSMOS 3 GHz catalog to investigate groups using a friend-of-friends algorithm and likewise found no significant difference in SFR behavior as a function of environment at $z > 0.47$. \citet{Erfanianfar_2015} also observed the effect of local environment on the SFMS for galaxies in clusters and groups between $0.15 < z < 0.5$, and found galaxies in groups tended toward quiescence and also hosted a higher bulge to disk ratio than those in field or filament environments, consistent with \citet{Dressler:1980aa}'s earlier findings that clusters hosted a greater frequency of ellipticals.   They also identified a similar cutoff in stellar mass 10$^{10.4}$--10$^{10.6}$ M$_{\odot}$, above which the SFMS flattens and no longer increases in SFR with stellar mass. 

\citet{Erfanianfar_2015}'s sample is drawn from the X-ray groups observed in the ECDFS, CDFN, AEGIS and COSMOS X-ray surveys \citep{Finoguenov:2007aa}, selecting groups with halo masses of 10$^{12.5}$ to 10$^{14.3}$ M$_{\odot}$.  This differs from our local density field estimation method, which classifies density in relation to the median density at a given redshift. However, the majority of their group sample lies in environments with a density $>2$ times the comoving density of their overall galaxy sample, indicating a similarity between their `group' environment and our `dense' environment classifications.  This consistency in environmental measurements may play a part in SFMS environmental results, indicating that massive, dense environments at low redshift support quenching more efficiently than the lower mass halos and environments available at high redshift ($z > 0.9$).  Overall, our results provide a self-consistent estimation of the SFMS across a wide dynamic range of redshift, connecting the individual results of low- and high-z studies into a connected picture that refutes the need for an environmental dependence.

Looking at the quiescent population in Fig.~\ref{fig:SFMScolorcolor}, we compare the median SFRs of both star-forming and quiescent galaxies to examine whether we observe any difference in the SFR of the quiescent sample as a function of environment.  Similar to the star-forming sample, we see little difference between each quiescent subsample.  The consistent SFRs of NUVrJ star-forming or quiescent samples across extreme environments support the model that quenching is a fast process ($<$Gyr timescale) that quickly transfers star-forming galaxies to the quiescent population once cluster-specific quenching begins to take effect.  


\subsection{Interplay of Morphology and Environment}\label{sec:diskdom}

In Fig.~\ref{fig:disksfms}, we show that disk-dominated star-forming galaxies at low redshift experience a consistent decline with stellar mass in the same way as the overall star-forming galaxy population and bulge-dominated galaxies (Fig.~\ref{fig:SFMScontours}).  This consistency indicates that the sSFR of disks in massive star-forming galaxies begin to experience a decline in star formation with stellar mass at $z < 1.3$ and star formation activity per stellar mass bin is insensitive to the bulge fraction. Our estimates are consistent with works that used photometry-based SED fitting to estimate SFR and stellar masses \citep{Schreiber:2016aa,Catalan-Torrecilla:2017aa,Morselli:2017aa,Belfiore:2018aa,Cook:2020aa}. 

In contrast, the papers that inspired our work here found a roughly constant evolution of sSFR with disk M$_{*}$ \citep{Abramson:2014aa,Guo:2015aa}, i.e., a lack of turnover \citep{Popesso:2019aa}.  The \citet{Abramson:2014aa} sample was derived from the Sloan Digital Sky Survey (SDSS) Data Release 4 and 7, and has a median $z \sim 0.08$.
The \citet{Abramson:2014aa} Bulge-to-total (B/T) ratios were estimated using a two component fit of face-on disks and a B/T cutoff of 0.2.  However, one key difference is the use of SFRs from \citet{Brinchmann:2004ab} who use SDSS optical emission lines and classify a galaxy as star-forming only if all four BPT diagram lines are detected.  As shown in Fig.~\ref{fig:disksfms}, \citet{Abramson:2014aa} used the SFRs from both SDSS DR 4 and 7, with the flattest relation being derived from SDSS DR 4. DR 4 spectral-line derived SFRs are warned as erroneously high for quiescent galaxies and this is observed to drive SFRs toward higher values at the high mass, low-SFR end of the star-forming population. \citet{Leslie:2019aa} also used star-forming galaxies in COSMOS and used the Zurich catalog to classify disk-dominated galaxies, but did not observe any turnover.  This is due to a lack of disk-dominated star forming galaxies in their sample above 10$^{11}$ M$_{\odot}$, where our analysis most strongly observed a decline in sSFR with stellar mass.

\citet{Morselli:2017aa} examined disks in a low-z sample from $0.02 < z < 0.1$ using a decomposition of SDSS optical images and SFRs from optical emission lines.  They found that low-z disk SFRs weakly flatten with stellar mass, and can also be correlated linearly within the errors.  Also at low-z, \citet{Cook:2020aa} used the GALEX Arecibo SDSS Survey to examine SFRS of disks using a combined NUV+MIR measurement alongside SDSS $gri$ band images, but found a decline in SFR with mass when bulges were removed.

Building out to intermediate redshift, \citet{Erfanianfar_2015} conducted a similar investigation examining the offset from the SFMS for star-forming galaxies across COSMOS, AEGIS, ECDFS, CDFN at $z < 1.1$.  Across local environments, they found a turnover at 10$^{10.5}$ M$_{\odot}$ but also observe an offset between high and low S\'ersic systems at high stellar masses.
 
Fig.~\ref{fig:disksfmscandels} extends the redshift range that combined morphology and environment studies have estimated the SFMS out to $z < 3$, agreeing with local optical studies that disks and disk-dominated systems experience a turnover at high stellar mass up to $z < 1.3$.  Previous works such as \citet{Lee:2018aa} relied on single S\`ersic profile fits to determine morphology at high-z, so cannot be directly compared to our results.  However \citet{Lee:2018aa} also found a flattening of the SFMS at high mass for their low-bulge ($n < 1.5$) systems, roughly consistent with our conclusions. Once at $z > 1.3$, we find the sSFRs of star-forming galaxies in general, and disk-dominated star-forming galaxies, follows a linear relation that is insensitive to bulge-fraction.  This provides new evidence strengthening the case for a consistently declining specific star formation with stellar mass in star-forming galaxies regardless of bulge fraction.


\section{Conclusions}\label{sec:conc}
 We investigate how the star formation rate -- stellar mass relation evolves with morphology and environment for both the total population and disk-dominated systems.  This is one of the first studies to comprehensively and consistently investigate the evolution of the SFMS as a function of both environment and morphology over such a large fraction of cosmic time ($0 < z < 3.5$).  Our methods enable consistent treatment of environment across our redshift range. To address morphology, we muse use morphology catalogues across two wavelength regimes to measure the rest-frame optical, and care must be taken to compare either or low- or high-z morphology samples to works at other redshifts.  We estimate star formation rates and stellar masses using SED fitting of a stellar mass complete sample from the COSMOS2015 catalog \citep{Laigle:2016aa}.  We consider local environment using the weighted adaptive kernel smoothed maps of the COSMOS field \citep{Darvish:2015aa}, which provide a unique opportunity to discover which morphology populations are driving the previously observed evolution of the SFMS with redshift \citep{Whitaker:2012aa,Whitaker:2014aa,Lee:2015aa}. From our investigation, we identify several key results that provide greater context to our knowledge of the SFMS:
 
\begin{itemize}

\item{A dependence of the SFMS with environment is not observed when either a NUVrJ color-color selection or a combined NUVrJ and sSFR selection function is used to select the star forming sample.  The median star formation rates of star-forming and quiescent galaxies are respectively consistent between their environment bins in all redshift bins.}

\item{We observe that the turnover of the SFMS when considering disk-dominated (B/T $<$ 0.25) systems, is consistent with the global trend as well as more bulge-dominated systems.  This implies that the turnover is due at least in part to a change in sSFR within the disk components and not driven by the transition from disk dominated to bulge-dominated systems.  We also identify a potential cause of disagreements in the literature, as our work supports previous studies based on multiple photometric bands in the SFR estimates.  The flattening of the SFMS may be a feature that requires the ability to identify only moderately star-forming galaxies at the high mass end in addition to starbursts.}

\end{itemize}

From our results in Fig.~\ref{fig:SFMScolorcolor}, we find that the color-color selected star-forming galaxies residing in dense environments host median star formation rates and errors consistent with their equal mass partners in field environments.  As the density maps used in this study estimate the local environment in the COSMOS field as a function of the median density in a given redshift slice, future studies may build upon these results by also investigating the absolute mass of cluster and proto-clusters to confirm whether the environmental effects we observe here are a function of absolute total mass or the increasing dynamic range of environmental densities at low redshift.

When considering these results, we assemble a picture of how similar star-forming galaxies evolve over cosmic time.  At high redshift, galaxies at all stellar masses are forming stars at a rate that is an order of magnitude or more greater than observed at present day.  The most massive galaxies in the universe during this high redshift epoch form stars with a similar sSFR as lower mass galaxies.

From $z \sim 1.3$ to present day, the most massive galaxies undergo a decline in SFR in comparison to low and intermediate mass galaxies at the same redshift.  This is evidence of the process of galactic `downsizing,' where the most massive galaxies conclude their star formation dominated evolutionary eras first.  When considering only disk-dominated systems, we still observe this trend.  This indicates an evolution in specific star formation is experienced by all morphological components with mass, with higher mass bulge- and disk-dominated systems both experiencing a decline in sSFR.

\section*{Acknowledgements}
\acknowledgements{
\linespread{1}
\vspace{-15pt}

  We thank the referee for their useful feedback that provided this paper with a stronger focus.  This paper does not reflect the views or opinions of the American Association for the Advancement of Science (AAAS), National Science Foundation, or any government agency.
  Support for this work was provided by NASA through grants \emph{HST}-GO-13657.010-A,
  \emph{HST}-AR-15802, and \emph{HST}-AR-14298.004-A awarded by the Space Telescope Science Institute, which is operated by the Association of Universities for Research in Astronomy, Inc., under NASA contract NAS 5-26555. Support and observations made with the NASA/ESA \emph{Hubble Space Telescope}, obtained from the Data Archive at the Space Telescope Science Institute, which is operated by the Association of Universities for Research in Astronomy, Inc.,  were also provided by NASA through grant NNX16AB36G as part of the Astrophysics Data Analysis Program.  Spectral energy distribution fitting was performed using the scientific computing resources of the Rochester Institute of Technology ION cluster \citep{https://doi.org/10.34788/0s3g-qd15}.  B.D. acknowledges financial support from NASA through the Astrophysics Data Analysis Program (ADAP), grant No. NNX12AE20G, and the National Science Foundation, grant No. 1716907.  
  
  We thank the generous advice of Elisabete da Cunha and Stefano Berta on the use of MAGPHYS and SED3FIT, respectively. We thank Jorge Zavala and Mattia Vaccari for their comments and discussion on this work that enhanced our discussion on star--forming galaxy selection criteria.  This research made use of Astropy,\footnote{http://www.astropy.org} a community-developed core Python package for Astronomy \citep{Robitaille:2013aa,Price-Whelan:2018aa}.  This research made use of the iPython environment \citep{Perez:2007aa} and the Python packages SciPy \citep{Virtanen:2020aa}, NumPy \citep{Walt:2011aa}, and Matplotlib \citep{Hunter:2007aa}.

 Based on observations made with the NASA Galaxy Evolution Explorer. 
GALEX is operated for NASA by the California Institute of Technology under NASA contract NAS5-98034.  Based on observations made with the NASA/ESA Hubble Space Telescope, obtained from the Data Archive at the Space Telescope Science Institute, which is operated by the Association of Universities for Research in Astronomy, Inc., under NASA contract NAS 5-26555. 
MegaPrime/MegaCam, a joint project of CFHT and CEA/DAPNIA, at the Canada-France-Hawaii Telescope (CFHT) which is operated by the National Research Council (NRC) of Canada, the Institut National des Sciences de l'Univers of the Centre National de la Recherche Scientifique of France, and the University of Hawaii.  Based [in part] on data collected at Subaru Telescope, which is operated by the National Astronomical Observatory of Japan.
Based on observations obtained with WIRCam, a joint project of CFHT, Taiwan, Korea, Canada, France, and the Canada-France-Hawaii Telescope (CFHT) which is operated by the National Research Council (NRC) of Canada, the Institut National des Sciences de l'Univers of the Centre National de la Recherche Scientifique of France, and the University of Hawaii. This work is based [in part] on observations made with the Spitzer Space Telescope, which is operated by the Jet Propulsion Laboratory, California Institute of Technology under a contract with NASA.  Herschel is an ESA space observatory with science instruments provided by European-led Principal Investigator consortia and with important participation from NASA. 

YP acknowledges the National Key R\&D Program of China, Grant 2016YFA0400702 and NSFC Grant No. 11773001. This work was supported by a NASA Keck PI Data Award, administered by the NASA Exoplanet Science Institute. Data presented herein were obtained at the W. M. Keck Observatory from telescope time allocated to the National Aeronautics and Space Administration through the agency's scientific partnership with the California Institute of Technology and the University of California. The Observatory was made possible by the generous financial support of the W. M. Keck Foundation.

The authors wish to recognize and acknowledge the very significant cultural role and reverence that the summit of Mauna Kea has always had within the indigenous Hawaiian community. We are most fortunate to have the opportunity to conduct observations from this mountain.


\newcommand\invisiblesection[1]{%
  \refstepcounter{section}%
  \addcontentsline{toc}{section}{\protect\numberline{\thesection}#1}%
  \sectionmark{#1}}

\invisiblesection{Bibliography}
\bibliography{CookeKartaltepe2019bib}

\begin{thebibliography}{}
\expandafter\ifx\csname natexlab\endcsname\relax\def\natexlab#1{#1}\fi

\bibitem[{Abramson {et~al.}(2014)Abramson, Kelson, Dressler, Poggianti,
  Gladders, Oemler, \& Vulcani}]{Abramson:2014aa}
Abramson, L.~E., Kelson, D.~D., Dressler, A., {et~al.} 2014, \apj, 785, L36

\bibitem[{Alberts {et~al.}(2016)Alberts, Pope, Brodwin, Chung, Cybulski, Dey,
  Eisenhardt, Galametz, Gonzalez, Jannuzi, \& et~al.}]{Alberts:2016aa}
Alberts, S., Pope, A., Brodwin, M., {et~al.} 2016, \apj, 825, 72

\bibitem[{Arnouts {et~al.}(1999)Arnouts, Cristiani, Moscardini, Matarrese,
  Lucchin, Fontana, \& Giallongo}]{Arnouts:1999aa}
Arnouts, S., Cristiani, S., Moscardini, L., {et~al.} 1999, \mnras, 310, 540

\bibitem[{{Arnouts} {et~al.}(2007){Arnouts}, {Walcher}, {Le F{\`e}vre},
  {Zamorani}, {Ilbert}, {Le Brun}, {Pozzetti}, {Bardelli}, {Tresse}, {Zucca},
  {Charlot}, {Lamareille}, {McCracken}, {Bolzonella}, {Iovino}, {Lonsdale},
  {Polletta}, {Surace}, {Bottini}, {Garilli}, {Maccagni}, {Picat},
  {Scaramella}, {Scodeggio}, {Vettolani}, {Zanichelli}, {Adami}, {Cappi},
  {Ciliegi}, {Contini}, {de la Torre}, {Foucaud}, {Franzetti}, {Gavignaud},
  {Guzzo}, {Marano}, {Marinoni}, {Mazure}, {Meneux}, {Merighi}, {Paltani},
  {Pell{\`o}}, {Pollo}, {Radovich}, {Temporin}, \& {Vergani}}]{Arnouts:2007aa}
{Arnouts}, S., {Walcher}, C.~J., {Le F{\`e}vre}, O., {et~al.} 2007, \aap, 476,
  137

\bibitem[{{Aune} {et~al.}(2003){Aune}, {Boulade}, {Charlot}, {Abbon},
  {Borgeaud}, {Carton}, {Carty}, {Da Costa}, {Desforge}, {Deschamps},
  {Eppell{\'e}}, {Gallais}, {Gosset}, {Granelli}, {Gros}, {de Kat}, {Loiseau},
  {Ritou}, {Rouss{\'e}}, {Starzynski}, {Vignal}, \& {Vigroux}}]{Aune:2003aa}
{Aune}, S., {Boulade}, O., {Charlot}, X., {et~al.} 2003, in \procspie, Vol.
  4841, Instrument Design and Performance for Optical/Infrared Ground-based
  Telescopes, ed. M.~{Iye} \& A.~F.~M. {Moorwood}, 513--524

\bibitem[{{Baldry} {et~al.}(2006){Baldry}, {Balogh}, {Bower}, {Glazebrook},
  {Nichol}, {Bamford}, \& {Budavari}}]{Baldry:2006aa}
{Baldry}, I.~K., {Balogh}, M.~L., {Bower}, R.~G., {et~al.} 2006, \mnras, 373,
  469

\bibitem[{{Baldry} {et~al.}(2004){Baldry}, {Glazebrook}, {Brinkmann},
  {Ivezi{\'c}}, {Lupton}, {Nichol}, \& {Szalay}}]{Baldry:2004aa}
{Baldry}, I.~K., {Glazebrook}, K., {Brinkmann}, J., {et~al.} 2004, \apj, 600,
  681

\bibitem[{Balogh {et~al.}(2011)Balogh, McGee, Wilman, Finoguenov, Parker,
  Connelly, Mulchaey, Bower, Tanaka, \& Giodini}]{Balogh:2011aa}
Balogh, M.~L., McGee, S.~L., Wilman, D.~J., {et~al.} 2011, \mnras, 412, 2303

\bibitem[{Barden {et~al.}(2012)Barden, H{\"a}u{\ss}ler, Peng, McIntosh, \&
  Guo}]{Barden:2012aa}
Barden, M., H{\"a}u{\ss}ler, B., Peng, C.~Y., McIntosh, D.~H., \& Guo, Y. 2012,
  \mnras, 422, 449

\bibitem[{Belfiore {et~al.}(2018)Belfiore, Maiolino, Bundy, Masters, Bershady,
  Oyarz{\'u}n, Lin, Cano-Diaz, Wake, Spindler, \& et~al.}]{Belfiore:2018aa}
Belfiore, F., Maiolino, R., Bundy, K., {et~al.} 2018, \mnras, 477, 3014

\bibitem[{{Berta} {et~al.}(2013){Berta}, {Lutz}, {Santini}, {Wuyts}, {Rosario},
  {Brisbin}, {Cooray}, {Franceschini}, {Gruppioni}, {Hatziminaoglou}, {Hwang},
  {Le Floc'h}, {Magnelli}, {Nordon}, {Oliver}, {Page}, {Popesso}, {Pozzetti},
  {Pozzi}, {Riguccini}, {Rodighiero}, {Roseboom}, {Scott}, {Symeonidis},
  {Valtchanov}, {Viero}, \& {Wang}}]{Berta:2013aa}
{Berta}, S., {Lutz}, D., {Santini}, P., {et~al.} 2013, \aap, 551, A100

\bibitem[{{Boulade} {et~al.}(2003){Boulade}, {Charlot}, {Abbon}, {Aune},
  {Borgeaud}, {Carton}, {Carty}, {Da Costa}, {Deschamps}, {Desforge},
  {Eppell{\'e}}, {Gallais}, {Gosset}, {Granelli}, {Gros}, {de Kat}, {Loiseau},
  {Ritou}, {Rouss{\'e}}, {Starzynski}, {Vignal}, \& {Vigroux}}]{Boulade:2003aa}
{Boulade}, O., {Charlot}, X., {Abbon}, P., {et~al.} 2003, in \procspie, Vol.
  4841, Instrument Design and Performance for Optical/Infrared Ground-based
  Telescopes, ed. M.~{Iye} \& A.~F.~M. {Moorwood}, 72--81

\bibitem[{Brammer {et~al.}(2012)Brammer, van Dokkum, Franx, Fumagalli, Patel,
  Rix, Skelton, Kriek, Nelson, Schmidt, \& et~al.}]{Brammer:2012aa}
Brammer, G.~B., van Dokkum, P.~G., Franx, M., {et~al.} 2012, \apjs, 200, 13

\bibitem[{Brinchmann {et~al.}(2004)Brinchmann, Charlot, White, Tremonti,
  Kauffmann, Heckman, \& Brinkmann}]{Brinchmann:2004ab}
Brinchmann, J., Charlot, S., White, S. D.~M., {et~al.} 2004, \mnras, 351, 1151

\bibitem[{{Brinchmann} {et~al.}(2004){Brinchmann}, {Charlot}, {White},
  {Tremonti}, {Kauffmann}, {Heckman}, \& {Brinkmann}}]{Brinchmann:2004aa}
{Brinchmann}, J., {Charlot}, S., {White}, S.~D.~M., {et~al.} 2004, \mnras, 351,
  1151

\bibitem[{{Bruzual} \& {Charlot}(2003)}]{Bruzual:2003aa}
{Bruzual}, G., \& {Charlot}, S. 2003, \mnras, 344, 1000

\bibitem[{{Calzetti} {et~al.}(2000){Calzetti}, {Armus}, {Bohlin}, {Kinney},
  {Koornneef}, \& {Storchi-Bergmann}}]{Calzetti:2000aa}
{Calzetti}, D., {Armus}, L., {Bohlin}, R.~C., {et~al.} 2000, \apj, 533, 682

\bibitem[{{Capak} {et~al.}(2007){Capak}, {Aussel}, {Ajiki}, {McCracken},
  {Mobasher}, {Scoville}, {Shopbell}, {Taniguchi}, {Thompson}, {Tribiano},
  {Sasaki}, {Blain}, {Brusa}, {Carilli}, {Comastri}, {Carollo}, {Cassata},
  {Colbert}, {Ellis}, {Elvis}, {Giavalisco}, {Green}, {Guzzo}, {Hasinger},
  {Ilbert}, {Impey}, {Jahnke}, {Kartaltepe}, {Kneib}, {Koda}, {Koekemoer},
  {Komiyama}, {Leauthaud}, {Le Fevre}, {Lilly}, {Liu}, {Massey}, {Miyazaki},
  {Murayama}, {Nagao}, {Peacock}, {Pickles}, {Porciani}, {Renzini}, {Rhodes},
  {Rich}, {Salvato}, {Sanders}, {Scarlata}, {Schiminovich}, {Schinnerer},
  {Scodeggio}, {Sheth}, {Shioya}, {Tasca}, {Taylor}, {Yan}, \&
  {Zamorani}}]{Capak:2007aa}
{Capak}, P., {Aussel}, H., {Ajiki}, M., {et~al.} 2007, \apjs, 172, 99

\bibitem[{Capak {et~al.}(2011)Capak, Mobasher, Scoville, McCracken, Ilbert,
  Salvato, Men{\'e}ndez-Delmestre, Aussel, Carilli, Civano, \&
  et~al.}]{Capak:2011aa}
Capak, P., Mobasher, B., Scoville, N.~Z., {et~al.} 2011, \apj, 730, 68

\bibitem[{{Casey} {et~al.}(2013){Casey}, {Chen}, {Cowie}, {Barger}, {Capak},
  {Ilbert}, {Koss}, {Lee}, {Le Floc'h}, {Sanders}, \&
  {Williams}}]{Casey:2013aa}
{Casey}, C.~M., {Chen}, C.-C., {Cowie}, L.~L., {et~al.} 2013, \mnras, 436, 1919

\bibitem[{Casey {et~al.}(2017)Casey, Cooray, Killi, Capak, Chen, Hung,
  Kartaltepe, Sanders, \& Scoville}]{Casey:2017aa}
Casey, C.~M., Cooray, A., Killi, M., {et~al.} 2017, \apj, 840, 101

\bibitem[{Catal{\'a}n-Torrecilla {et~al.}(2017)Catal{\'a}n-Torrecilla, Gil~de
  Paz, Castillo-Morales, M{\'e}ndez-Abreu, Falc{\'o}n-Barroso, Bekeraite,
  Costantin, Lorenzo-C{\'a}ceres, Florido, Garc{\'\i}a-Benito, \&
  et~al.}]{Catalan-Torrecilla:2017aa}
Catal{\'a}n-Torrecilla, C., Gil~de Paz, A., Castillo-Morales, A., {et~al.}
  2017, \apj, 848, 87

\bibitem[{{Chabrier}(2003)}]{Chabrier:2003aa}
{Chabrier}, G. 2003, \pasp, 115, 763

\bibitem[{{Chauke} {et~al.}(2019){Chauke}, {van der Wel}, {Pacifici},
  {Bezanson}, {Wu}, {Gallazzi}, {Straatman}, {Franx}, {Bari{\v{s}}i{\'c}},
  {Bell}, {van Houdt}, {Maseda}, {Muzzin}, {Sobral}, \&
  {Spilker}}]{Chauke:2019aa}
{Chauke}, P., {van der Wel}, A., {Pacifici}, C., {et~al.} 2019, \apj, 877, 48

\bibitem[{Colless {et~al.}(2001)Colless, Dalton, Maddox, Sutherland, Norberg,
  Cole, Bland-Hawthorn, Bridges, Cannon, Collins, \& et~al.}]{Colless:2001aa}
Colless, M., Dalton, G., Maddox, S., {et~al.} 2001, \mnras, 328, 1039

\bibitem[{Cook {et~al.}(2020)Cook, Cortese, Catinella, \&
  Robotham}]{Cook:2020aa}
Cook, R. H.~W., Cortese, L., Catinella, B., \& Robotham, A. 2020, \mnras, 493,
  5596

\bibitem[{{Cooke} {et~al.}(2019){Cooke}, {Kartaltepe}, {Tyler}, {Darvish},
  {Casey}, {Le F{\`e}vre}, {Salvato}, \& {Scoville}}]{Cooke:2019aa}
{Cooke}, K.~C., {Kartaltepe}, J.~S., {Tyler}, K.~D., {et~al.} 2019, \apj, 881,
  150

\bibitem[{{Cooke} {et~al.}(2020){Cooke}, {Kirkpatrick}, {Estrada}, {Messias},
  {Peca}, {Cappelluti}, {Ananna}, {Brewster}, {Glikman}, {LaMassa}, {Daisy
  Leung}, {Trump}, {Jane Turner}, \& {Urry}}]{Cooke:2020aa}
{Cooke}, K.~C., {Kirkpatrick}, A., {Estrada}, M., {et~al.} 2020, \apj, 903, 106

\bibitem[{{Cooper} {et~al.}(2008){Cooper}, {Newman}, {Weiner}, {Yan},
  {Willmer}, {Bundy}, {Coil}, {Conselice}, {Davis}, {Faber}, {Gerke},
  {Guhathakurta}, {Koo}, \& {Noeske}}]{Cooper:2008aa}
{Cooper}, M.~C., {Newman}, J.~A., {Weiner}, B.~J., {et~al.} 2008, \mnras, 383,
  1058

\bibitem[{{da Cunha} {et~al.}(2008){da Cunha}, {Charlot}, \&
  {Elbaz}}]{da-Cunha:2008aa}
{da Cunha}, E., {Charlot}, S., \& {Elbaz}, D. 2008, \mnras, 388, 1595

\bibitem[{{Daddi} {et~al.}(2007){Daddi}, {Dickinson}, {Morrison}, {Chary},
  {Cimatti}, {Elbaz}, {Frayer}, {Renzini}, {Pope}, {Alexander}, {Bauer},
  {Giavalisco}, {Huynh}, {Kurk}, \& {Mignoli}}]{Daddi:2007aa}
{Daddi}, E., {Dickinson}, M., {Morrison}, G., {et~al.} 2007, \apj, 670, 156

\bibitem[{Dale \& Helou(2002)}]{Dale:2002aa}
Dale, D.~A., \& Helou, G. 2002, \apj, 576, 159

\bibitem[{Darvish {et~al.}(2015)Darvish, Mobasher, Sobral, Scoville, \&
  Aragon-Calvo}]{Darvish:2015aa}
Darvish, B., Mobasher, B., Sobral, D., Scoville, N., \& Aragon-Calvo, M. 2015,
  \apj, 805, 121

\bibitem[{Darvish {et~al.}(2014)Darvish, Sobral, Mobasher, Scoville, Best,
  Sales, \& Smail}]{Darvish:2014aa}
Darvish, B., Sobral, D., Mobasher, B., {et~al.} 2014, \apj, 1

\bibitem[{Davidzon {et~al.}(2017)Davidzon, Ilbert, Laigle, Coupon, McCracken,
  Delvecchio, Masters, Capak, Hsieh, Le~F{\`e}vre, \& et~al.}]{Davidzon:2017aa}
Davidzon, I., Ilbert, O., Laigle, C., {et~al.} 2017, \aap, 605, A70

\bibitem[{Dom{\'\i}nguez~S{\'a}nchez {et~al.}(2011)Dom{\'\i}nguez~S{\'a}nchez,
  Pozzi, Gruppioni, Cimatti, Ilbert, Pozzetti, McCracken, Capak, Le~Floch,
  Salvato, \& et~al.}]{Dominguez-Sanchez:2011aa}
Dom{\'\i}nguez~S{\'a}nchez, H., Pozzi, F., Gruppioni, C., {et~al.} 2011,
  \mnras, 417, 900

\bibitem[{{Dressler}(1980)}]{Dressler:1980aa}
{Dressler}, A. 1980, \apj, 236, 351

\bibitem[{{Elbaz} {et~al.}(2007){Elbaz}, {Daddi}, {Le Borgne}, {Dickinson},
  {Alexander}, {Chary}, {Starck}, {Brandt}, {Kitzbichler}, {MacDonald},
  {Nonino}, {Popesso}, {Stern}, \& {Vanzella}}]{Elbaz:2007aa}
{Elbaz}, D., {Daddi}, E., {Le Borgne}, D., {et~al.} 2007, \aap, 468, 33

\bibitem[{Elbaz {et~al.}(2011)Elbaz, Dickinson, Hwang, D{\'\i}az-Santos,
  Magdis, Magnelli, Le~Borgne, Galliano, Pannella, Chanial, \&
  et~al.}]{Elbaz:2011aa}
Elbaz, D., Dickinson, M., Hwang, H.~S., {et~al.} 2011, \aap, 533, A119

\bibitem[{{Ellison} {et~al.}(2021){Ellison}, {Lin}, {Thorp}, {Pan}, {Scudder},
  {S{\'a}nchez}, {Bluck}, \& {Maiolino}}]{Ellison:2021aa}
{Ellison}, S.~L., {Lin}, L., {Thorp}, M.~D., {et~al.} 2021, \mnras, 501, 4777

\bibitem[{Erfanianfar {et~al.}(2015)Erfanianfar, Popesso, Finoguenov, Wilman,
  Wuyts, Biviano, Salvato, Mirkazemi, Morselli, Ziparo, \&
  et~al.}]{Erfanianfar_2015}
Erfanianfar, G., Popesso, P., Finoguenov, A., {et~al.} 2015, \mnras, 455, 2839

\bibitem[{{Fazio} {et~al.}(2004){Fazio}, {Hora}, {Allen}, {Ashby}, {Barmby},
  {Deutsch}, {Huang}, {Kleiner}, {Marengo}, {Megeath}, {Melnick}, {Pahre},
  {Patten}, {Polizotti}, {Smith}, {Taylor}, {Wang}, {Willner}, {Hoffmann},
  {Pipher}, {Forrest}, {McMurty}, {McCreight}, {McKelvey}, {McMurray}, {Koch},
  {Moseley}, {Arendt}, {Mentzell}, {Marx}, {Losch}, {Mayman}, {Eichhorn},
  {Krebs}, {Jhabvala}, {Gezari}, {Fixsen}, {Flores}, {Shakoorzadeh}, {Jungo},
  {Hakun}, {Workman}, {Karpati}, {Kichak}, {Whitley}, {Mann}, {Tollestrup},
  {Eisenhardt}, {Stern}, {Gorjian}, {Bhattacharya}, {Carey}, {Nelson},
  {Glaccum}, {Lacy}, {Lowrance}, {Laine}, {Reach}, {Stauffer}, {Surace},
  {Wilson}, {Wright}, {Hoffman}, {Domingo}, \& {Cohen}}]{Fazio:2004aa}
{Fazio}, G.~G., {Hora}, J.~L., {Allen}, L.~E., {et~al.} 2004, \apjs, 154, 10

\bibitem[{{Feltre} {et~al.}(2012){Feltre}, {Hatziminaoglou}, {Fritz}, \&
  {Franceschini}}]{Feltre:2012aa}
{Feltre}, A., {Hatziminaoglou}, E., {Fritz}, J., \& {Franceschini}, A. 2012,
  \mnras, 426, 120

\bibitem[{Feruglio {et~al.}(2010)Feruglio, Aussel, Le~Floc'h, Ilbert, Salvato,
  Capak, Fiore, Kartaltepe, Sanders, Scoville, \& et~al.}]{Feruglio:2010aa}
Feruglio, C., Aussel, H., Le~Floc'h, E., {et~al.} 2010, \apj, 721, 607

\bibitem[{{Finoguenov} {et~al.}(2007){Finoguenov}, {Guzzo}, {Hasinger},
  {Scoville}, {Aussel}, {B{\"o}hringer}, {Brusa}, {Capak}, {Cappelluti},
  {Comastri}, {Giodini}, {Griffiths}, {Impey}, {Koekemoer}, {Kneib},
  {Leauthaud}, {Le F{\`e}vre}, {Lilly}, {Mainieri}, {Massey}, {McCracken},
  {Mobasher}, {Murayama}, {Peacock}, {Sakelliou}, {Schinnerer}, {Silverman},
  {Smol{\v c}i{\'c}}, {Taniguchi}, {Tasca}, {Taylor}, {Trump}, \&
  {Zamorani}}]{Finoguenov:2007aa}
{Finoguenov}, A., {Guzzo}, L., {Hasinger}, G., {et~al.} 2007, \apjs, 172, 182

\bibitem[{Fu {et~al.}(2010)Fu, Yan, Scoville, Capak, Aussel, Le~Floc'h, Ilbert,
  Salvato, Kartaltepe, Frayer, \& et~al.}]{Fu:2010aa}
Fu, H., Yan, L., Scoville, N.~Z., {et~al.} 2010, \apj, 722, 653

\bibitem[{Galametz {et~al.}(2013)Galametz, Grazian, Fontana, Ferguson, Ashby,
  Barro, Castellano, Dahlen, Donley, Faber, \& et~al.}]{Galametz:2013aa}
Galametz, A., Grazian, A., Fontana, A., {et~al.} 2013, \apjs, 206, 10

\bibitem[{{Genzel} {et~al.}(2015){Genzel}, {Tacconi}, {Lutz}, {Saintonge},
  {Berta}, {Magnelli}, {Combes}, {Garc{\'{\i}}a-Burillo}, {Neri}, {Bolatto},
  {Contini}, {Lilly}, {Boissier}, {Boone}, {Bouch{\'e}}, {Bournaud}, {Burkert},
  {Carollo}, {Colina}, {Cooper}, {Cox}, {Feruglio}, {F{\"o}rster Schreiber},
  {Freundlich}, {Gracia-Carpio}, {Juneau}, {Kovac}, {Lippa}, {Naab}, {Salome},
  {Renzini}, {Sternberg}, {Walter}, {Weiner}, {Weiss}, \&
  {Wuyts}}]{Genzel:2015aa}
{Genzel}, R., {Tacconi}, L.~J., {Lutz}, D., {et~al.} 2015, \apj, 800, 20

\bibitem[{{George} {et~al.}(2011){George}, {Leauthaud}, {Bundy}, {Finoguenov},
  {Tinker}, {Lin}, {Mei}, {Kneib}, {Aussel}, {Behroozi}, {Busha}, {Capak},
  {Coccato}, {Covone}, {Faure}, {Fiorenza}, {Ilbert}, {Le Floc'h}, {Koekemoer},
  {Tanaka}, {Wechsler}, \& {Wolk}}]{George:2011aa}
{George}, M.~R., {Leauthaud}, A., {Bundy}, K., {et~al.} 2011, \apj, 742, 125

\bibitem[{{Gobat} {et~al.}(2013){Gobat}, {Strazzullo}, {Daddi}, {Onodera},
  {Carollo}, {Renzini}, {Finoguenov}, {Cimatti}, {Scarlata}, \&
  {Arimoto}}]{Gobat:2013aa}
{Gobat}, R., {Strazzullo}, V., {Daddi}, E., {et~al.} 2013, \apj, 776, 9

\bibitem[{Griffin {et~al.}(2010)Griffin, Abergel, Abreu, Ade, Andr{\'e},
  Augueres, Babbedge, Bae, Baillie, Baluteau, \& et~al.}]{Griffin:2010aa}
Griffin, M.~J., Abergel, A., Abreu, A., {et~al.} 2010, \aap, 518, L3

\bibitem[{Grogin {et~al.}(2011)Grogin, Kocevski, Faber, Ferguson, Koekemoer,
  Riess, Acquaviva, Alexander, Almaini, Ashby, \& et~al.}]{Grogin:2011aa}
Grogin, N.~A., Kocevski, D.~D., Faber, S.~M., {et~al.} 2011, \apjs, 197, 35

\bibitem[{Gunn \& Gott(1972)}]{Gunn:1972aa}
Gunn, J.~E., \& Gott, J.~Richard, I. 1972, \apj, 176, 1

\bibitem[{Guo {et~al.}(2015)Guo, Zheng, Wang, \& Fu}]{Guo:2015aa}
Guo, K., Zheng, X.~Z., Wang, T., \& Fu, H. 2015, \apj, 808, L49

\bibitem[{Guo {et~al.}(2013)Guo, Ferguson, Giavalisco, Barro, Willner, Ashby,
  Dahlen, Donley, Faber, Fontana, \& et~al.}]{Guo:2013ab}
Guo, Y., Ferguson, H.~C., Giavalisco, M., {et~al.} 2013, \apjs, 207, 24

\bibitem[{{Hashemizadeh} {et~al.}(2021){Hashemizadeh}, {Driver}, {Davies},
  {Robotham}, {Bellstedt}, {Windhorst}, {Bremer}, {Phillipps}, {Jarvis},
  {Holwerda}, {Lagos}, {Koushan}, {Siudek}, {Maddox}, {Thorne}, \&
  {Elahi}}]{Hashemizadeh:2021aa}
{Hashemizadeh}, A., {Driver}, S.~P., {Davies}, L. J.~M., {et~al.} 2021, \mnras,
  505, 136

\bibitem[{Hasinger {et~al.}(2018)Hasinger, Capak, Salvato, Barger, Cowie,
  Faisst, Hemmati, Kakazu, Kartaltepe, Masters, \& et~al.}]{Hasinger:2018aa}
Hasinger, G., Capak, P., Salvato, M., {et~al.} 2018, \apj, 858, 77

\bibitem[{Hatfield \& Jarvis(2017)}]{Hatfield:2017aa}
Hatfield, P.~W., \& Jarvis, M.~J. 2017, \mnras, 472, 3570

\bibitem[{H{\"a}u{\ss}ler {et~al.}(2013)H{\"a}u{\ss}ler, Bamford, Vika, Rojas,
  Barden, Kelvin, Alpaslan, Robotham, Driver, Baldry, \&
  et~al.}]{Haussler:2013aa}
H{\"a}u{\ss}ler, B., Bamford, S.~P., Vika, M., {et~al.} 2013, \mnras, 430, 330

\bibitem[{{H{\"a}u{\ss}ler} {et~al.}(2022){H{\"a}u{\ss}ler}, {Vika}, {Bamford},
  {Johnston}, {Brough}, {Casura}, {Holwerda}, {Kelvin}, \&
  {Popescu}}]{Haussler:2022aa}
{H{\"a}u{\ss}ler}, B., {Vika}, M., {Bamford}, S.~P., {et~al.} 2022, arXiv
  e-prints, arXiv:2204.05907

\bibitem[{Hilton {et~al.}(2010)Hilton, Lloyd-Davies, Stanford, Stott, Collins,
  Romer, Hosmer, Hoyle, Kay, Liddle, \& et~al.}]{Hilton:2010aa}
Hilton, M., Lloyd-Davies, E., Stanford, S.~A., {et~al.} 2010, \apj, 718, 133

\bibitem[{Hung {et~al.}(2013)Hung, Sanders, Casey, Lee, Barnes, Capak,
  Kartaltepe, Koss, Larson, Le~Floc'h, \& et~al.}]{Hung:2013aa}
Hung, C.-L., Sanders, D.~B., Casey, C.~M., {et~al.} 2013, \apj, 778, 129

\bibitem[{Hunter(2007)}]{Hunter:2007aa}
Hunter, J.~D. 2007, Computing in Science \& Engineering, 9, 90

\bibitem[{Ilbert {et~al.}(2006)Ilbert, Arnouts, McCracken, Bolzonella, Bertin,
  Le~F{\`e}vre, Mellier, Zamorani, Pell{\`o}, Iovino, \&
  et~al.}]{Ilbert:2006aa}
Ilbert, O., Arnouts, S., McCracken, H.~J., {et~al.} 2006, \aap, 457, 841

\bibitem[{{Ilbert} {et~al.}(2010){Ilbert}, {Salvato}, {Le Floc'h}, {Aussel},
  {Capak}, {McCracken}, {Mobasher}, {Kartaltepe}, {Scoville}, {Sanders},
  {Arnouts}, {Bundy}, {Cassata}, {Kneib}, {Koekemoer}, {Le F{\`e}vre}, {Lilly},
  {Surace}, {Taniguchi}, {Tasca}, {Thompson}, {Tresse}, {Zamojski}, {Zamorani},
  \& {Zucca}}]{Ilbert:2010aa}
{Ilbert}, O., {Salvato}, M., {Le Floc'h}, E., {et~al.} 2010, \apj, 709, 644

\bibitem[{Ilbert {et~al.}(2013)Ilbert, McCracken, Le~Fevre, Capak, Dunlop,
  Karim, Renzini, Caputi, Boissier, Arnouts, \& et~al.}]{Ilbert:2013aa}
Ilbert, O., McCracken, H.~J., Le~Fevre, O., {et~al.} 2013, \aap, 556, A55

\bibitem[{Johnston {et~al.}(2015)Johnston, Vaccari, Jarvis, Smith, Giovannoli,
  H{\"a}u{\ss}ler, \& Prescott}]{Johnston:2015aa}
Johnston, R., Vaccari, M., Jarvis, M., {et~al.} 2015, \mnras, 453, 2541

\bibitem[{Karim {et~al.}(2011)Karim, Schinnerer, Mart{\'\i}nez-Sansigre,
  Sargent, van~der Wel, Rix, Ilbert, Smol{\v c}i{\'c}, Carilli, Pannella, \&
  et~al.}]{Karim:2011aa}
Karim, A., Schinnerer, E., Mart{\'\i}nez-Sansigre, A., {et~al.} 2011, \apj,
  730, 61

\bibitem[{Kartaltepe {et~al.}(2010)Kartaltepe, Sanders, Le~Floc'h, Frayer,
  Aussel, Arnouts, Ilbert, Salvato, Scoville, Surace, \&
  et~al.}]{Kartaltepe:2010aa}
Kartaltepe, J.~S., Sanders, D.~B., Le~Floc'h, E., {et~al.} 2010, \apj, 709, 572

\bibitem[{Kartaltepe {et~al.}(2012)Kartaltepe, Dickinson, Alexander, Bell,
  Dahlen, Elbaz, Faber, Lotz, McIntosh, Wiklind, \& et~al.}]{Kartaltepe:2012aa}
Kartaltepe, J.~S., Dickinson, M., Alexander, D.~M., {et~al.} 2012, \apj, 757,
  23

\bibitem[{Kartaltepe {et~al.}(2015)Kartaltepe, Mozena, Kocevski, McIntosh,
  Lotz, Bell, Faber, Ferguson, Koo, Bassett, \& et~al.}]{Kartaltepe:2015aa}
Kartaltepe, J.~S., Mozena, M., Kocevski, D., {et~al.} 2015, \apjs, 221, 11

\bibitem[{{Kaviraj} {et~al.}(2009){Kaviraj}, {Peirani}, {Khochfar}, {Silk}, \&
  {Kay}}]{Kaviraj:2009aa}
{Kaviraj}, S., {Peirani}, S., {Khochfar}, S., {Silk}, J., \& {Kay}, S. 2009,
  \mnras, 394, 1713

\bibitem[{{Kennedy} {et~al.}(2016){Kennedy}, {Bamford}, {H{\"a}u{\ss}ler},
  {Baldry}, {Bremer}, {Brough}, {Brown}, {Driver}, {Duncan}, {Graham},
  {Holwerda}, {Hopkins}, {Kelvin}, {Lange}, {Phillipps}, {Vika}, \&
  {Vulcani}}]{Kennedy:2016aa}
{Kennedy}, R., {Bamford}, S.~P., {H{\"a}u{\ss}ler}, B., {et~al.} 2016, \mnras,
  460, 3458

\bibitem[{{Koekemoer} {et~al.}(2007){Koekemoer}, {Aussel}, {Calzetti}, {Capak},
  {Giavalisco}, {Kneib}, {Leauthaud}, {Le F{\`e}vre}, {McCracken}, {Massey},
  {Mobasher}, {Rhodes}, {Scoville}, \& {Shopbell}}]{Koekemoer:2007aa}
{Koekemoer}, A.~M., {Aussel}, H., {Calzetti}, D., {et~al.} 2007, \apjs, 172,
  196

\bibitem[{Koekemoer {et~al.}(2011)Koekemoer, Faber, Ferguson, Grogin, Kocevski,
  Koo, Lai, Lotz, Lucas, McGrath, \& et~al.}]{Koekemoer:2011aa}
Koekemoer, A.~M., Faber, S.~M., Ferguson, H.~C., {et~al.} 2011, \apjs, 197, 36

\bibitem[{Koyama {et~al.}(2013)Koyama, Smail, Kurk, Geach, Sobral, Kodama,
  Nakata, Swinbank, Best, Hayashi, \& et~al.}]{Koyama:2013aa}
Koyama, Y., Smail, I., Kurk, J., {et~al.} 2013, \mnras, 434, 423

\bibitem[{Kriek {et~al.}(2009)Kriek, van Dokkum, Labb{\'e}, Franx, Illingworth,
  Marchesini, \& Quadri}]{Kriek:2009aa}
Kriek, M., van Dokkum, P.~G., Labb{\'e}, I., {et~al.} 2009, \apj, 700, 221

\bibitem[{{Kriek} {et~al.}(2006){Kriek}, {van Dokkum}, {Franx}, {Quadri},
  {Gawiser}, {Herrera}, {Illingworth}, {Labb{\'e}}, {Lira}, {Marchesini},
  {Rix}, {Rudnick}, {Taylor}, {Toft}, {Urry}, \& {Wuyts}}]{Kriek:2006aa}
{Kriek}, M., {van Dokkum}, P.~G., {Franx}, M., {et~al.} 2006, \apjl, 649, L71

\bibitem[{Kriek {et~al.}(2015)Kriek, Shapley, Reddy, Siana, Coil, Mobasher,
  Freeman, Groot, Price, Sanders, \& et~al.}]{Kriek:2015aa}
Kriek, M., Shapley, A.~E., Reddy, N.~A., {et~al.} 2015, \apjs, 218, 15

\bibitem[{Krogager {et~al.}(2014)Krogager, Zirm, Toft, Man, \&
  Brammer}]{Krogager:2014aa}
Krogager, J.-K., Zirm, A.~W., Toft, S., Man, A., \& Brammer, G. 2014, \apj,
  797, 17

\bibitem[{{Laigle} {et~al.}(2016){Laigle}, {McCracken}, {Ilbert}, {Hsieh},
  {Davidzon}, {Capak}, {Hasinger}, {Silverman}, {Pichon}, {Coupon}, {Aussel},
  {Le Borgne}, {Caputi}, {Cassata}, {Chang}, {Civano}, {Dunlop}, {Fynbo},
  {Kartaltepe}, {Koekemoer}, {Le F{\`e}vre}, {Le Floc'h}, {Leauthaud}, {Lilly},
  {Lin}, {Marchesi}, {Milvang-Jensen}, {Salvato}, {Sanders}, {Scoville},
  {Smolcic}, {Stockmann}, {Taniguchi}, {Tasca}, {Toft}, {Vaccari}, \&
  {Zabl}}]{Laigle:2016aa}
{Laigle}, C., {McCracken}, H.~J., {Ilbert}, O., {et~al.} 2016, \apjs, 224, 24

\bibitem[{Le~F{\`e}vre {et~al.}(2015)Le~F{\`e}vre, Tasca, Cassata, Garilli,
  Le~Brun, Maccagni, Pentericci, Thomas, Vanzella, Zamorani, \&
  et~al.}]{Le-Fevre:2015aa}
Le~F{\`e}vre, O., Tasca, L. A.~M., Cassata, P., {et~al.} 2015, \aap, 576, A79

\bibitem[{Lee {et~al.}(2018)Lee, Giavalisco, Whitaker, Williams, Ferguson,
  Acquaviva, Koekemoer, Straughn, Guo, Kartaltepe, \& et~al.}]{Lee:2018aa}
Lee, B., Giavalisco, M., Whitaker, K., {et~al.} 2018, \apj, 853, 131

\bibitem[{{Lee} {et~al.}(2010){Lee}, {Le Floc'h}, {Sanders}, {Frayer},
  {Arnouts}, {Ilbert}, {Aussel}, {Salvato}, {Scoville}, \&
  {Kartaltepe}}]{Lee:2010aa}
{Lee}, N., {Le Floc'h}, E., {Sanders}, D.~B., {et~al.} 2010, \apj, 717, 175

\bibitem[{{Lee} {et~al.}(2015){Lee}, {Sanders}, {Casey}, {Toft}, {Scoville},
  {Hung}, {Le Floc'h}, {Ilbert}, {Zahid}, {Aussel}, {Capak}, {Kartaltepe},
  {Kewley}, {Li}, {Schawinski}, {Sheth}, \& {Xiao}}]{Lee:2015aa}
{Lee}, N., {Sanders}, D.~B., {Casey}, C.~M., {et~al.} 2015, \apj, 801, 80

\bibitem[{{Leslie} {et~al.}(2020){Leslie}, {Schinnerer}, {Liu}, {Magnelli},
  {Algera}, {Karim}, {Davidzon}, {Gozaliasl}, {Jim{\'e}nez-Andrade}, {Lang},
  {Sargent}, {Novak}, {Groves}, {Smol{\v{c}}i{\'c}}, {Zamorani}, {Vaccari},
  {Battisti}, {Vardoulaki}, {Peng}, \& {Kartaltepe}}]{Leslie:2019aa}
{Leslie}, S.~K., {Schinnerer}, E., {Liu}, D., {et~al.} 2020, \apj, 899, 58

\bibitem[{Lilly {et~al.}(2007)Lilly, Fevre, Renzini, Zamorani, Scodeggio,
  Contini, Carollo, Hasinger, Kneib, Iovino, \& et~al.}]{Lilly:2007aa}
Lilly, S.~J., Fevre, O.~L., Renzini, A., {et~al.} 2007, \apjs, 172, 70

\bibitem[{Lin {et~al.}(2014)Lin, Jian, Foucaud, Norberg, Bower, Cole,
  Arnalte-Mur, Chen, Coupon, Hsieh, \& et~al.}]{Lin:2014aa}
Lin, L., Jian, H.-Y., Foucaud, S., {et~al.} 2014, \apj, 782, 33

\bibitem[{Liu {et~al.}(2019)Liu, Lang, Magnelli, Schinnerer, Leslie, Fudamoto,
  Bondi, Groves, Jim{\'e}nez-Andrade, Harrington, \& et~al.}]{Liu:2019aa}
Liu, D., Lang, P., Magnelli, B., {et~al.} 2019, \apjs, 244, 40

\bibitem[{{Lutz} {et~al.}(2011){Lutz}, {Poglitsch}, {Altieri}, {Andreani},
  {Aussel}, {Berta}, {Bongiovanni}, {Brisbin}, {Cava}, {Cepa}, {Cimatti},
  {Daddi}, {Dominguez-Sanchez}, {Elbaz}, {F{\"o}rster Schreiber}, {Genzel},
  {Grazian}, {Gruppioni}, {Harwit}, {Le Floc'h}, {Magdis}, {Magnelli},
  {Maiolino}, {Nordon}, {P{\'e}rez Garc{\'{\i}}a}, {Popesso}, {Pozzi},
  {Riguccini}, {Rodighiero}, {Saintonge}, {Sanchez Portal}, {Santini}, {Shao},
  {Sturm}, {Tacconi}, {Valtchanov}, {Wetzstein}, \& {Wieprecht}}]{Lutz:2011aa}
{Lutz}, D., {Poglitsch}, A., {Altieri}, B., {et~al.} 2011, \aap, 532, A90

\bibitem[{{Madau} \& {Dickinson}(2014)}]{Madau:2014aa}
{Madau}, P., \& {Dickinson}, M. 2014, \araa, 52, 415

\bibitem[{Magdis {et~al.}(2010)Magdis, Elbaz, Daddi, Morrison, Dickinson,
  Rigopoulou, Gobat, \& Hwang}]{Magdis:2010aa}
Magdis, G.~E., Elbaz, D., Daddi, E., {et~al.} 2010, \apj, 714, 1740

\bibitem[{{Magdis} {et~al.}(2012){Magdis}, {Daddi}, {B{\'e}thermin}, {Sargent},
  {Elbaz}, {Pannella}, {Dickinson}, {Dannerbauer}, {da Cunha}, {Walter},
  {Rigopoulou}, {Charmandaris}, {Hwang}, \& {Kartaltepe}}]{Magdis:2012aa}
{Magdis}, G.~E., {Daddi}, E., {B{\'e}thermin}, M., {et~al.} 2012, \apj, 760, 6

\bibitem[{{Mancini} {et~al.}(2019){Mancini}, {Daddi}, {Juneau}, {Renzini},
  {Rodighiero}, {Cappellari}, {Rodr{\'\i}guez-Mu{\~n}oz}, {Liu}, {Pannella},
  {Baronchelli}, {Franceschini}, {Bergamini}, {D'Eugenio}, \&
  {Puglisi}}]{Mancini:2019aa}
{Mancini}, C., {Daddi}, E., {Juneau}, S., {et~al.} 2019, in The Art of
  Measuring Galaxy Physical Properties, 49

\bibitem[{{Martin} {et~al.}(2005){Martin}, {Fanson}, {Schiminovich},
  {Morrissey}, {Friedman}, {Barlow}, {Conrow}, {Grange}, {Jelinsky},
  {Milliard}, {Siegmund}, {Bianchi}, {Byun}, {Donas}, {Forster}, {Heckman},
  {Lee}, {Madore}, {Malina}, {Neff}, {Rich}, {Small}, {Surber}, {Szalay},
  {Welsh}, \& {Wyder}}]{Martin:2005aa}
{Martin}, D.~C., {Fanson}, J., {Schiminovich}, D., {et~al.} 2005, \apjl, 619,
  L1

\bibitem[{{Masters} {et~al.}(2017){Masters}, {Stern}, {Cohen}, {Capak},
  {Rhodes}, {Castander}, \& {Paltani}}]{Masters:2017aa}
{Masters}, D.~C., {Stern}, D.~K., {Cohen}, J.~G., {et~al.} 2017, \apj, 841, 111

\bibitem[{{McCracken} {et~al.}(2012){McCracken}, {Milvang-Jensen}, {Dunlop},
  {Franx}, {Fynbo}, {Le F{\`e}vre}, {Holt}, {Caputi}, {Goranova}, {Buitrago},
  {Emerson}, {Freudling}, {Hudelot}, {L{\'o}pez-Sanjuan}, {Magnard}, {Mellier},
  {M{\o}ller}, {Nilsson}, {Sutherland}, {Tasca}, \& {Zabl}}]{McCracken:2012aa}
{McCracken}, H.~J., {Milvang-Jensen}, B., {Dunlop}, J., {et~al.} 2012, \aap,
  544, A156

\bibitem[{{Mobasher}(2016)}]{Mobasher:2016aa}
{Mobasher}, B. 2016, {Spectroscopic Properties of Galaxies in the Cosmic Web.},
  Keck Observatory Archive U200D

\bibitem[{{Moln{\'a}r} {et~al.}(2018){Moln{\'a}r}, {Sargent}, {Delhaize},
  {Delvecchio}, {Smol{\v{c}}i{\'c}}, {Novak}, {Schinnerer}, {Zamorani},
  {Bondi}, {Herrera-Ruiz}, {Murphy}, {Vardoulaki}, {Karim}, {Leslie},
  {Magnelli}, {Carollo}, \& {Middelberg}}]{Molnar:2018aa}
{Moln{\'a}r}, D.~C., {Sargent}, M.~T., {Delhaize}, J., {et~al.} 2018, \mnras,
  475, 827

\bibitem[{Momcheva {et~al.}(2016)Momcheva, Brammer, van Dokkum, Skelton,
  Whitaker, Nelson, Fumagalli, Maseda, Leja, Franx, \&
  et~al.}]{Momcheva:2016aa}
Momcheva, I.~G., Brammer, G.~B., van Dokkum, P.~G., {et~al.} 2016, \apjs, 225,
  27

\bibitem[{{Morrissey} {et~al.}(2007){Morrissey}, {Conrow}, {Barlow}, {Small},
  {Seibert}, {Wyder}, {Budav{\'a}ri}, {Arnouts}, {Friedman}, {Forster},
  {Martin}, {Neff}, {Schiminovich}, {Bianchi}, {Donas}, {Heckman}, {Lee},
  {Madore}, {Milliard}, {Rich}, {Szalay}, {Welsh}, \& {Yi}}]{Morrissey:2007aa}
{Morrissey}, P., {Conrow}, T., {Barlow}, T.~A., {et~al.} 2007, \apjs, 173, 682

\bibitem[{Morselli {et~al.}(2019)Morselli, Popesso, Cibinel, Oesch, Montes,
  Atek, Illingworth, \& Holden}]{Morselli:2019aa}
Morselli, L., Popesso, P., Cibinel, A., {et~al.} 2019, \aap, 626, A61

\bibitem[{Morselli {et~al.}(2017)Morselli, Popesso, Erfanianfar, \&
  Concas}]{Morselli:2017aa}
Morselli, L., Popesso, P., Erfanianfar, G., \& Concas, A. 2017, \aap, 597, A97

\bibitem[{{Mowla} {et~al.}(2019){Mowla}, {van Dokkum}, {Brammer}, {Momcheva},
  {van der Wel}, {Whitaker}, {Nelson}, {Bezanson}, {Muzzin}, {Franx},
  {MacKenty}, {Leja}, {Kriek}, \& {Marchesini}}]{Mowla:2019aa}
{Mowla}, L.~A., {van Dokkum}, P., {Brammer}, G.~B., {et~al.} 2019, \apj, 880,
  57

\bibitem[{{Nanayakkara} {et~al.}(2016){Nanayakkara}, {Glazebrook}, {Kacprzak},
  {Yuan}, {Tran}, {Spitler}, {Kewley}, {Straatman}, {Cowley}, {Fisher},
  {Labbe}, {Tomczak}, {Allen}, \& {Alcorn}}]{Nanayakkara:2016aa}
{Nanayakkara}, T., {Glazebrook}, K., {Kacprzak}, G.~G., {et~al.} 2016, \apj,
  828, 21

\bibitem[{Nayyeri {et~al.}(2017)Nayyeri, Hemmati, Mobasher, Ferguson, Cooray,
  Barro, Faber, Dickinson, Koekemoer, Peth, \& et~al.}]{Nayyeri:2017aa}
Nayyeri, H., Hemmati, S., Mobasher, B., {et~al.} 2017, \apjs, 228, 7

\bibitem[{{Noeske} {et~al.}(2007){Noeske}, {Faber}, {Weiner}, {Koo}, {Primack},
  {Dekel}, {Papovich}, {Conselice}, {Le Floc'h}, {Rieke}, {Coil}, {Lotz},
  {Somerville}, \& {Bundy}}]{Noeske:2007aa}
{Noeske}, K.~G., {Faber}, S.~M., {Weiner}, B.~J., {et~al.} 2007, \apjl, 660,
  L47

\bibitem[{of~Technology(2019)}]{https://doi.org/10.34788/0s3g-qd15}
of~Technology, R.~I. 2019, Research Computing Services, doi:10.34788/0S3G-QD15

\bibitem[{Oke(1974)}]{Oke:1974aa}
Oke, J.~B. 1974, \apjs, 27, 21

\bibitem[{{Oliver} {et~al.}(2012){Oliver}, {Bock}, {Altieri}, {Amblard},
  {Arumugam}, {Aussel}, {Babbedge}, {Beelen}, {B{\'e}thermin}, {Blain},
  {Boselli}, {Bridge}, {Brisbin}, {Buat}, {Burgarella},
  {Castro-Rodr{\'{\i}}guez}, {Cava}, {Chanial}, {Cirasuolo}, {Clements},
  {Conley}, {Conversi}, {Cooray}, {Dowell}, {Dubois}, {Dwek}, {Dye}, {Eales},
  {Elbaz}, {Farrah}, {Feltre}, {Ferrero}, {Fiolet}, {Fox}, {Franceschini},
  {Gear}, {Giovannoli}, {Glenn}, {Gong}, {Gonz{\'a}lez Solares}, {Griffin},
  {Halpern}, {Harwit}, {Hatziminaoglou}, {Heinis}, {Hurley}, {Hwang}, {Hyde},
  {Ibar}, {Ilbert}, {Isaak}, {Ivison}, {Lagache}, {Le Floc'h}, {Levenson},
  {Faro}, {Lu}, {Madden}, {Maffei}, {Magdis}, {Mainetti}, {Marchetti},
  {Marsden}, {Marshall}, {Mortier}, {Nguyen}, {O'Halloran}, {Omont}, {Page},
  {Panuzzo}, {Papageorgiou}, {Patel}, {Pearson}, {P{\'e}rez-Fournon}, {Pohlen},
  {Rawlings}, {Raymond}, {Rigopoulou}, {Riguccini}, {Rizzo}, {Rodighiero},
  {Roseboom}, {Rowan-Robinson}, {S{\'a}nchez Portal}, {Schulz}, {Scott},
  {Seymour}, {Shupe}, {Smith}, {Stevens}, {Symeonidis}, {Trichas}, {Tugwell},
  {Vaccari}, {Valtchanov}, {Vieira}, {Viero}, {Vigroux}, {Wang}, {Ward},
  {Wardlow}, {Wright}, {Xu}, \& {Zemcov}}]{Oliver:2012aa}
{Oliver}, S.~J., {Bock}, J., {Altieri}, B., {et~al.} 2012, \mnras, 424, 1614

\bibitem[{Pannella {et~al.}(2009)Pannella, Carilli, Daddi, McCracken, Owen,
  Renzini, Strazzullo, Civano, Koekemoer, Schinnerer, \&
  et~al.}]{Pannella:2009aa}
Pannella, M., Carilli, C.~L., Daddi, E., {et~al.} 2009, \apj, 698, L116

\bibitem[{{P{\^a}ris} {et~al.}(2018){P{\^a}ris}, {Petitjean}, {Aubourg},
  {Myers}, {Streblyanska}, {Lyke}, {Anderson}, {Armengaud}, {Bautista},
  {Blanton}, {Blomqvist}, {Brinkmann}, {Brownstein}, {Brandt}, {Burtin},
  {Dawson}, {de la Torre}, {Georgakakis}, {Gil-Mar{\'{\i}}n}, {Green}, {Hall},
  {Kneib}, {LaMassa}, {Le Goff}, {MacLeod}, {Mariappan}, {McGreer}, {Merloni},
  {Noterdaeme}, {Palanque-Delabrouille}, {Percival}, {Ross}, {Rossi},
  {Schneider}, {Seo}, {Tojeiro}, {Weaver}, {Weijmans}, {Y{\`e}che}, {Zarrouk},
  \& {Zhao}}]{Paris:2018aa}
{P{\^a}ris}, I., {Petitjean}, P., {Aubourg}, {\'E}., {et~al.} 2018, \aap, 613,
  A51

\bibitem[{Pearson {et~al.}(2018)Pearson, Wang, Hurley, Ma{\l}ek, Buat,
  Burgarella, Farrah, Oliver, Smith, \& van~der Tak}]{Pearson:2018aa}
Pearson, W.~J., Wang, L., Hurley, P.~D., {et~al.} 2018, \aap, 615, A146

\bibitem[{Peng {et~al.}(2002)Peng, Ho, Impey, \& Rix}]{Peng:2002aa}
Peng, C.~Y., Ho, L.~C., Impey, C.~D., \& Rix, H.-W. 2002, \aj, 124, 266

\bibitem[{Peng {et~al.}(2010)Peng, Ho, Impey, \& Rix}]{Peng:2010aa}
---. 2010, \aj, 139, 2097

\bibitem[{Perez \& Granger(2007)}]{Perez:2007aa}
Perez, F., \& Granger, B.~E. 2007, Computing in Science \& Engineering, 9, 21

\bibitem[{Perna {et~al.}(2015)Perna, Brusa, Salvato, Cresci, Lanzuisi, Berta,
  Delvecchio, Fiore, Lutz, Le~Floc'h, \& et~al.}]{Perna:2015aa}
Perna, M., Brusa, M., Salvato, M., {et~al.} 2015, \aap, 583, A72

\bibitem[{{Pilbratt} {et~al.}(2010){Pilbratt}, {Riedinger}, {Passvogel},
  {Crone}, {Doyle}, {Gageur}, {Heras}, {Jewell}, {Metcalfe}, {Ott}, \&
  {Schmidt}}]{Pilbratt:2010aa}
{Pilbratt}, G.~L., {Riedinger}, J.~R., {Passvogel}, T., {et~al.} 2010, \aap,
  518, L1

\bibitem[{{Poglitsch} {et~al.}(2010){Poglitsch}, {Waelkens}, {Geis},
  {Feuchtgruber}, {Vandenbussche}, {Rodriguez}, {Krause}, {Renotte}, {van
  Hoof}, {Saraceno}, {Cepa}, {Kerschbaum}, {Agn{\`e}se}, {Ali}, {Altieri},
  {Andreani}, {Augueres}, {Balog}, {Barl}, {Bauer}, {Belbachir}, {Benedettini},
  {Billot}, {Boulade}, {Bischof}, {Blommaert}, {Callut}, {Cara}, {Cerulli},
  {Cesarsky}, {Contursi}, {Creten}, {De Meester}, {Doublier}, {Doumayrou},
  {Duband}, {Exter}, {Genzel}, {Gillis}, {Gr{\"o}zinger}, {Henning},
  {Herreros}, {Huygen}, {Inguscio}, {Jakob}, {Jamar}, {Jean}, {de Jong},
  {Katterloher}, {Kiss}, {Klaas}, {Lemke}, {Lutz}, {Madden}, {Marquet},
  {Martignac}, {Mazy}, {Merken}, {Montfort}, {Morbidelli}, {M{\"u}ller},
  {Nielbock}, {Okumura}, {Orfei}, {Ottensamer}, {Pezzuto}, {Popesso},
  {Putzeys}, {Regibo}, {Reveret}, {Royer}, {Sauvage}, {Schreiber}, {Stegmaier},
  {Schmitt}, {Schubert}, {Sturm}, {Thiel}, {Tofani}, {Vavrek}, {Wetzstein},
  {Wieprecht}, \& {Wiezorrek}}]{Poglitsch:2010aa}
{Poglitsch}, A., {Waelkens}, C., {Geis}, N., {et~al.} 2010, \aap, 518, L2

\bibitem[{Popesso {et~al.}(2019)Popesso, Morselli, Concas, Schreiber,
  Rodighiero, Cresci, Belli, Ilbert, Erfanianfar, Mancini, \&
  et~al.}]{Popesso:2019aa}
Popesso, P., Morselli, L., Concas, A., {et~al.} 2019, \mnras,
  doi:10.1093/mnras/stz2635

\bibitem[{Price-Whelan {et~al.}(2018)Price-Whelan, Sip{\H o}cz, G{\"u}nther,
  Lim, Crawford, Conseil, Shupe, Craig, Dencheva, \&
  et~al.}]{Price-Whelan:2018aa}
Price-Whelan, A.~M., Sip{\H o}cz, B.~M., G{\"u}nther, H.~M., {et~al.} 2018,
  \aj, 156, 123

\bibitem[{{Randriamampandry} {et~al.}(2020){Randriamampandry}, {Vaccari}, \&
  {Hess}}]{Randriamampandry:2020aa}
{Randriamampandry}, S.~M., {Vaccari}, M., \& {Hess}, K.~M. 2020, \mnras, 499,
  948

\bibitem[{Robitaille {et~al.}(2013)Robitaille, Tollerud, Greenfield,
  Droettboom, Bray, Aldcroft, Davis, Ginsburg, Price-Whelan, \&
  et~al.}]{Robitaille:2013aa}
Robitaille, T.~P., Tollerud, E.~J., Greenfield, P., {et~al.} 2013, \aap, 558,
  A33

\bibitem[{Rodighiero {et~al.}(2010)Rodighiero, Cimatti, Gruppioni, Popesso,
  Andreani, Altieri, Aussel, Berta, Bongiovanni, Brisbin, \&
  et~al.}]{Rodighiero:2010aa}
Rodighiero, G., Cimatti, A., Gruppioni, C., {et~al.} 2010, \aap, 518, L25

\bibitem[{Rodighiero {et~al.}(2014)Rodighiero, Renzini, Daddi, Baronchelli,
  Berta, Cresci, Franceschini, Gruppioni, Lutz, Mancini, \&
  et~al.}]{Rodighiero:2014aa}
Rodighiero, G., Renzini, A., Daddi, E., {et~al.} 2014, \mnras, 443, 19

\bibitem[{{Salim} \& {Narayanan}(2020)}]{Salim:2020aa}
{Salim}, S., \& {Narayanan}, D. 2020, \araa, 58, 529

\bibitem[{Salim {et~al.}(2007)Salim, Rich, Charlot, Brinchmann, Johnson,
  Schiminovich, Seibert, Mallery, Heckman, Forster, \& et~al.}]{Salim:2007aa}
Salim, S., Rich, R.~M., Charlot, S., {et~al.} 2007, \apjs, 173, 267

\bibitem[{{Sanders} {et~al.}(2007){Sanders}, {Salvato}, {Aussel}, {Ilbert},
  {Scoville}, {Surace}, {Frayer}, {Sheth}, {Helou}, {Brooke}, {Bhattacharya},
  {Yan}, {Kartaltepe}, {Barnes}, {Blain}, {Calzetti}, {Capak}, {Carilli},
  {Carollo}, {Comastri}, {Daddi}, {Ellis}, {Elvis}, {Fall}, {Franceschini},
  {Giavalisco}, {Hasinger}, {Impey}, {Koekemoer}, {Le F{\`e}vre}, {Lilly},
  {Liu}, {McCracken}, {Mobasher}, {Renzini}, {Rich}, {Schinnerer}, {Shopbell},
  {Taniguchi}, {Thompson}, {Urry}, \& {Williams}}]{Sanders:2007aa}
{Sanders}, D.~B., {Salvato}, M., {Aussel}, H., {et~al.} 2007, \apjs, 172, 86

\bibitem[{Santos {et~al.}(2014)Santos, Altieri, Tanaka, Valtchanov, Saintonge,
  Dickinson, Foucaud, Kodama, Rawle, \& Tadaki}]{Santos:2014aa}
Santos, J.~S., Altieri, B., Tanaka, M., {et~al.} 2014, \mnras, 438, 2565

\bibitem[{Sargent {et~al.}(2007)Sargent, Carollo, Lilly, Scarlata, Feldmann,
  Kampczyk, Koekemoer, Scoville, Kneib, Leauthaud, \& et~al.}]{Sargent:2007aa}
Sargent, M.~T., Carollo, C.~M., Lilly, S.~J., {et~al.} 2007, \apjs, 172, 434

\bibitem[{Scarlata {et~al.}(2007)Scarlata, Carollo, Lilly, Sargent, Feldmann,
  Kampczyk, Porciani, Koekemoer, Scoville, Kneib, \& et~al.}]{Scarlata:2007aa}
Scarlata, C., Carollo, C.~M., Lilly, S., {et~al.} 2007, \apjs, 172, 406

\bibitem[{{Schinnerer} {et~al.}(2016){Schinnerer}, {Groves}, {Sargent},
  {Karim}, {Oesch}, {Magnelli}, {LeFevre}, {Tasca}, {Civano}, {Cassata}, \&
  {Smol{\v c}i{\'c}}}]{Schinnerer:2016aa}
{Schinnerer}, E., {Groves}, B., {Sargent}, M.~T., {et~al.} 2016, \apj, 833, 112

\bibitem[{{Schlegel} {et~al.}(1998){Schlegel}, {Finkbeiner}, \&
  {Davis}}]{Schlegel:1998aa}
{Schlegel}, D.~J., {Finkbeiner}, D.~P., \& {Davis}, M. 1998, \apj, 500, 525

\bibitem[{Schreiber {et~al.}(2016)Schreiber, Elbaz, Pannella, Ciesla, Wang,
  Koekemoer, Rafelski, \& Daddi}]{Schreiber:2016aa}
Schreiber, C., Elbaz, D., Pannella, M., {et~al.} 2016, \aap, 589, A35

\bibitem[{Schreiber {et~al.}(2015)Schreiber, Pannella, Elbaz, B{\'e}thermin,
  Inami, Dickinson, Magnelli, Wang, Aussel, Daddi, \&
  et~al.}]{Schreiber:2015aa}
Schreiber, C., Pannella, M., Elbaz, D., {et~al.} 2015, \aap, 575, A74

\bibitem[{{Scott} {et~al.}(2008){Scott}, {Austermann}, {Perera}, {Wilson},
  {Aretxaga}, {Bock}, {Hughes}, {Kang}, {Kim}, {Mauskopf}, {Sanders},
  {Scoville}, \& {Yun}}]{Scott:2008aa}
{Scott}, K.~S., {Austermann}, J.~E., {Perera}, T.~A., {et~al.} 2008, \mnras,
  385, 2225

\bibitem[{{Scoville} {et~al.}(2007){Scoville}, {Aussel}, {Brusa}, {Capak},
  {Carollo}, {Elvis}, {Giavalisco}, {Guzzo}, {Hasinger}, {Impey}, {Kneib},
  {LeFevre}, {Lilly}, {Mobasher}, {Renzini}, {Rich}, {Sanders}, {Schinnerer},
  {Schminovich}, {Shopbell}, {Taniguchi}, \& {Tyson}}]{Scoville:2007aa}
{Scoville}, N., {Aussel}, H., {Brusa}, M., {et~al.} 2007, \apjs, 172, 1

\bibitem[{Scoville {et~al.}(2013)Scoville, Arnouts, Aussel, Benson, Bongiorno,
  Bundy, Calvo, Capak, Carollo, Civano, \& et~al.}]{Scoville:2013aa}
Scoville, N., Arnouts, S., Aussel, H., {et~al.} 2013, \apjs, 206, 3

\bibitem[{{Scoville} {et~al.}(2016){Scoville}, {Sheth}, {Aussel}, {Vanden
  Bout}, {Capak}, {Bongiorno}, {Casey}, {Murchikova}, {Koda},
  {{\'A}lvarez-M{\'a}rquez}, {Lee}, {Laigle}, {McCracken}, {Ilbert}, {Pope},
  {Sanders}, {Chu}, {Toft}, {Ivison}, \& {Manohar}}]{Scoville:2016aa}
{Scoville}, N., {Sheth}, K., {Aussel}, H., {et~al.} 2016, \apj, 820, 83

\bibitem[{Silverman {et~al.}(2015)Silverman, Kashino, Sanders, Kartaltepe,
  Arimoto, Renzini, Rodighiero, Daddi, Zahid, Nagao, \&
  et~al.}]{Silverman:2015aa}
Silverman, J.~D., Kashino, D., Sanders, D., {et~al.} 2015, \apjs, 220, 12

\bibitem[{Speagle {et~al.}(2014)Speagle, Steinhardt, Capak, \&
  Silverman}]{Speagle:2014aa}
Speagle, J.~S., Steinhardt, C.~L., Capak, P.~L., \& Silverman, J.~D. 2014,
  \apjs, 214, 15

\bibitem[{{Steinhardt} {et~al.}(2014){Steinhardt}, {Speagle}, {Capak},
  {Silverman}, {Carollo}, {Dunlop}, {Hashimoto}, {Hsieh}, {Ilbert}, {Le Fevre},
  {Le Floc'h}, {Lee}, {Lin}, {Lin}, {Masters}, {McCracken}, {Nagao}, {Petric},
  {Salvato}, {Sanders}, {Scoville}, {Sheth}, {Strauss}, \&
  {Taniguchi}}]{Steinhardt:2014aa}
{Steinhardt}, C.~L., {Speagle}, J.~S., {Capak}, P., {et~al.} 2014, \apjl, 791,
  L25

\bibitem[{{Strateva} {et~al.}(2001){Strateva}, {Ivezi{\'c}}, {Knapp},
  {Narayanan}, {Strauss}, {Gunn}, {Lupton}, {Schlegel}, {Bahcall}, {Brinkmann},
  {Brunner}, {Budav{\'a}ri}, {Csabai}, {Castander}, {Doi}, {Fukugita}, {Gy{\H
  o}ry}, {Hamabe}, {Hennessy}, {Ichikawa}, {Kunszt}, {Lamb}, {McKay},
  {Okamura}, {Racusin}, {Sekiguchi}, {Schneider}, {Shimasaku}, \&
  {York}}]{Strateva:2001aa}
{Strateva}, I., {Ivezi{\'c}}, {\v Z}., {Knapp}, G.~R., {et~al.} 2001, \aj, 122,
  1861

\bibitem[{{Sutherland} {et~al.}(2015){Sutherland}, {Emerson}, {Dalton},
  {Atad-Ettedgui}, {Beard}, {Bennett}, {Bezawada}, {Born}, {Caldwell}, {Clark},
  {Craig}, {Henry}, {Jeffers}, {Little}, {McPherson}, {Murray}, {Stewart},
  {Stobie}, {Terrett}, {Ward}, {Whalley}, \& {Woodhouse}}]{Sutherland:2015aa}
{Sutherland}, W., {Emerson}, J., {Dalton}, G., {et~al.} 2015, \aap, 575, A25

\bibitem[{{Taniguchi} {et~al.}(2007){Taniguchi}, {Scoville}, {Murayama},
  {Sanders}, {Mobasher}, {Aussel}, {Capak}, {Ajiki}, {Miyazaki}, {Komiyama},
  {Shioya}, {Nagao}, {Sasaki}, {Koda}, {Carilli}, {Giavalisco}, {Guzzo},
  {Hasinger}, {Impey}, {LeFevre}, {Lilly}, {Renzini}, {Rich}, {Schinnerer},
  {Shopbell}, {Kaifu}, {Karoji}, {Arimoto}, {Okamura}, \&
  {Ohta}}]{Taniguchi:2007aa}
{Taniguchi}, Y., {Scoville}, N., {Murayama}, T., {et~al.} 2007, \apjs, 172, 9

\bibitem[{{Taniguchi} {et~al.}(2015){Taniguchi}, {Kajisawa}, {Kobayashi},
  {Shioya}, {Nagao}, {Capak}, {Aussel}, {Ichikawa}, {Murayama}, {Scoville},
  {Ilbert}, {Salvato}, {Sanders}, {Mobasher}, {Miyazaki}, {Komiyama}, {Le
  F{\`e}vre}, {Tasca}, {Lilly}, {Carollo}, {Renzini}, {Rich}, {Schinnerer},
  {Kaifu}, {Karoji}, {Arimoto}, {Okamura}, {Ohta}, {Shimasaku}, \&
  {Hayashino}}]{Taniguchi:2015aa}
{Taniguchi}, Y., {Kajisawa}, M., {Kobayashi}, M.~A.~R., {et~al.} 2015, \pasj,
  67, 104

\bibitem[{{Tempel} {et~al.}(2014){Tempel}, {Kipper}, {Saar}, {Bussov},
  {Hektor}, \& {Pelt}}]{Tempel:2014aa}
{Tempel}, E., {Kipper}, R., {Saar}, E., {et~al.} 2014, \aap, 572, A8

\bibitem[{{Thorne} {et~al.}(2020){Thorne}, {Robotham}, {Davies}, {Bellstedt},
  {Driver}, {Bravo}, {Bremer}, {Holwerda}, {Hopkins}, {Lagos}, {Phillipps},
  {Siudek}, {Taylor}, \& {Wright}}]{Thorne:2020aa}
{Thorne}, J.~E., {Robotham}, A. S.~G., {Davies}, L. J.~M., {et~al.} 2020, arXiv
  e-prints, arXiv:2011.13605

\bibitem[{{Thorne} {et~al.}(2021){Thorne}, {Robotham}, {Davies}, {Bellstedt},
  {Driver}, {Bravo}, {Bremer}, {Holwerda}, {Hopkins}, {Lagos}, {Phillipps},
  {Siudek}, {Taylor}, \& {Wright}}]{Thorne:2021aa}
---. 2021, \mnras, 505, 540

\bibitem[{{Toft} {et~al.}(2009){Toft}, {Franx}, {van Dokkum}, {F{\"o}rster
  Schreiber}, {Labbe}, {Wuyts}, \& {Marchesini}}]{Toft:2009aa}
{Toft}, S., {Franx}, M., {van Dokkum}, P., {et~al.} 2009, \apj, 705, 255

\bibitem[{{Toft} {et~al.}(2007){Toft}, {van Dokkum}, {Franx}, {Labbe},
  {F{\"o}rster Schreiber}, {Wuyts}, {Webb}, {Rudnick}, {Zirm}, {Kriek}, {van
  der Werf}, {Blakeslee}, {Illingworth}, {Rix}, {Papovich}, \&
  {Moorwood}}]{Toft:2007aa}
{Toft}, S., {van Dokkum}, P., {Franx}, M., {et~al.} 2007, \apj, 671, 285

\bibitem[{Tomczak {et~al.}(2016)Tomczak, Quadri, Tran, Labb{\'e}, Straatman,
  Papovich, Glazebrook, Allen, Brammer, Cowley, \& et~al.}]{Tomczak:2016aa}
Tomczak, A.~R., Quadri, R.~F., Tran, K.-V.~H., {et~al.} 2016, \apj, 817, 118

\bibitem[{Trakhtenbrot {et~al.}(2016)Trakhtenbrot, Civano, Urry, Schawinski,
  Marchesi, Elvis, Rosario, Suh, Mejia-Restrepo, Simmons, \&
  et~al.}]{Trakhtenbrot:2016aa}
Trakhtenbrot, B., Civano, F., Urry, C.~M., {et~al.} 2016, \apj, 825, 4

\bibitem[{Trump {et~al.}(2007)Trump, Impey, McCarthy, Elvis, Huchra, Brusa,
  Hasinger, Schinnerer, Capak, Lilly, \& et~al.}]{Trump:2007aa}
Trump, J.~R., Impey, C.~D., McCarthy, P.~J., {et~al.} 2007, \apjs, 172, 383

\bibitem[{{Trump} {et~al.}(2009){Trump}, {Impey}, {Elvis}, {McCarthy},
  {Huchra}, {Brusa}, {Salvato}, {Capak}, {Cappelluti}, {Civano}, {Comastri},
  {Gabor}, {Hao}, {Hasinger}, {Jahnke}, {Kelly}, {Lilly}, {Schinnerer},
  {Scoville}, \& {Smol{\v c}i{\'c}}}]{Trump:2009ab}
{Trump}, J.~R., {Impey}, C.~D., {Elvis}, M., {et~al.} 2009, \apj, 696, 1195

\bibitem[{Trump {et~al.}(2009)Trump, Impey, Elvis, McCarthy, Huchra, Brusa,
  Salvato, Capak, Cappelluti, Civano, \& et~al.}]{Trump:2009aa}
Trump, J.~R., Impey, C.~D., Elvis, M., {et~al.} 2009, \apj, 696, 1195

\bibitem[{Trump {et~al.}(2011)Trump, Nagao, Ikeda, Murayama, Impey, Stocke,
  Civano, Elvis, Jahnke, Kelly, \& et~al.}]{Trump:2011aa}
Trump, J.~R., Nagao, T., Ikeda, H., {et~al.} 2011, \apj, 732, 23

\bibitem[{van~der Walt {et~al.}(2011)van~der Walt, Colbert, \&
  Varoquaux}]{Walt:2011aa}
van~der Walt, S., Colbert, S.~C., \& Varoquaux, G. 2011, Computing in Science
  \& Engineering, 13, 22

\bibitem[{van~der Wel {et~al.}(2016)van~der Wel, Noeske, Bezanson, Pacifici,
  Gallazzi, Franx, Mu{\~n}oz-Mateos, Bell, Brammer, Charlot, \&
  et~al.}]{Wel:2016aa}
van~der Wel, A., Noeske, K., Bezanson, R., {et~al.} 2016, \apjs, 223, 29

\bibitem[{Virtanen {et~al.}(2020)Virtanen, Gommers, Oliphant, Haberland, Reddy,
  Cournapeau, Burovski, Peterson, Weckesser, Bright, {van der Walt}, Brett,
  Wilson, Millman, Mayorov, Nelson, Jones, Kern, Larson, Carey, Polat, Feng,
  Moore, {VanderPlas}, Laxalde, Perktold, Cimrman, Henriksen, Quintero, Harris,
  Archibald, Ribeiro, Pedregosa, {van Mulbregt}, \& {SciPy 1.0
  Contributors}}]{Virtanen:2020aa}
Virtanen, P., Gommers, R., Oliphant, T.~E., {et~al.} 2020, Nature Methods, 17,
  261

\bibitem[{{Werner} {et~al.}(2004){Werner}, {Roellig}, {Low}, {Rieke}, {Rieke},
  {Hoffmann}, {Young}, {Houck}, {Brandl}, {Fazio}, {Hora}, {Gehrz}, {Helou},
  {Soifer}, {Stauffer}, {Keene}, {Eisenhardt}, {Gallagher}, {Gautier}, {Irace},
  {Lawrence}, {Simmons}, {Van Cleve}, {Jura}, {Wright}, \&
  {Cruikshank}}]{Werner:2004aa}
{Werner}, M.~W., {Roellig}, T.~L., {Low}, F.~J., {et~al.} 2004, \apjs, 154, 1

\bibitem[{Whitaker {et~al.}(2012)Whitaker, van Dokkum, Brammer, \&
  Franx}]{Whitaker:2012aa}
Whitaker, K.~E., van Dokkum, P.~G., Brammer, G., \& Franx, M. 2012, \apj, 754,
  L29

\bibitem[{Whitaker {et~al.}(2014)Whitaker, Franx, Leja, van Dokkum, Henry,
  Skelton, Fumagalli, Momcheva, Brammer, Labb{\'e}, \&
  et~al.}]{Whitaker:2014aa}
Whitaker, K.~E., Franx, M., Leja, J., {et~al.} 2014, \apj, 795, 104

\bibitem[{Willett {et~al.}(2015)Willett, Schawinski, Simmons, Masters, Skibba,
  Kaviraj, Melvin, Wong, Nichol, Cheung, \& et~al.}]{Willett:2015aa}
Willett, K.~W., Schawinski, K., Simmons, B.~D., {et~al.} 2015, \mnras, 449, 820

\bibitem[{Williams {et~al.}(2009)Williams, Quadri, Franx, van Dokkum, \&
  Labb{\'e}}]{Williams:2009aa}
Williams, R.~J., Quadri, R.~F., Franx, M., van Dokkum, P., \& Labb{\'e}, I.
  2009, \apj, 691, 1879

\bibitem[{{Wisnioski} {et~al.}(2015){Wisnioski}, {F{\"o}rster Schreiber},
  {Wuyts}, {Wuyts}, {Bandara}, {Wilman}, {Genzel}, {Bender}, {Davies},
  {Fossati}, {Lang}, {Mendel}, {Beifiori}, {Brammer}, {Chan}, {Fabricius},
  {Fudamoto}, {Kulkarni}, {Kurk}, {Lutz}, {Nelson}, {Momcheva}, {Rosario},
  {Saglia}, {Seitz}, {Tacconi}, \& {van Dokkum}}]{Wisnioski:2015aa}
{Wisnioski}, E., {F{\"o}rster Schreiber}, N.~M., {Wuyts}, S., {et~al.} 2015,
  \apj, 799, 209

\bibitem[{{Wuyts} {et~al.}(2011){Wuyts}, {F{\"o}rster Schreiber}, {van der
  Wel}, {Magnelli}, {Guo}, {Genzel}, {Lutz}, {Aussel}, {Barro}, {Berta},
  {Cava}, {Graci{\'a}-Carpio}, {Hathi}, {Huang}, {Kocevski}, {Koekemoer},
  {Lee}, {Le Floc'h}, {McGrath}, {Nordon}, {Popesso}, {Pozzi}, {Riguccini},
  {Rodighiero}, {Saintonge}, \& {Tacconi}}]{Wuyts:2011aa}
{Wuyts}, S., {F{\"o}rster Schreiber}, N.~M., {van der Wel}, A., {et~al.} 2011,
  \apj, 742, 96

\bibitem[{Yoshikawa {et~al.}(2010)Yoshikawa, Akiyama, Kajisawa, Alexander,
  Ohta, Suzuki, Tokoku, Uchimoto, Konishi, Yamada, \&
  et~al.}]{Yoshikawa:2010aa}
Yoshikawa, T., Akiyama, M., Kajisawa, M., {et~al.} 2010, \apj, 718, 112

\bibitem[{{Zamojski} {et~al.}(2007){Zamojski}, {Schiminovich}, {Rich},
  {Mobasher}, {Koekemoer}, {Capak}, {Taniguchi}, {Sasaki}, {McCracken},
  {Mellier}, {Bertin}, {Aussel}, {Sanders}, {Le F{\`e}vre}, {Ilbert},
  {Salvato}, {Thompson}, {Kartaltepe}, {Scoville}, {Barlow}, {Forster},
  {Friedman}, {Martin}, {Morrissey}, {Neff}, {Seibert}, {Small}, {Wyder},
  {Bianchi}, {Donas}, {Heckman}, {Lee}, {Madore}, {Milliard}, {Szalay},
  {Welsh}, \& {Yi}}]{Zamojski:2007aa}
{Zamojski}, M.~A., {Schiminovich}, D., {Rich}, R.~M., {et~al.} 2007, \apjs,
  172, 468

\end{thebibliography}

\end{document}